\documentclass[12pt,preprint]{aastex}

\usepackage{amsmath}
\usepackage{graphics,graphicx}
\usepackage{multirow}
\usepackage{array}
\usepackage{rotating}
\usepackage[T1]{fontenc}
\usepackage{textcomp}
\usepackage[hyphens]{url}
\usepackage{hyperref}

\begin{document}
%

\title{Characterizing Rocky and Gaseous Exoplanets with 2-meter Class Space-based Coronagraphs}

\author{Tyler D. Robinson\altaffilmark{1,2}}
\affil{Department of Astronomy and Astrophysics, University of California, Santa Cruz, CA 95064, USA}
\email{tyler.d.robinson@nasa.gov}

\author{Karl R. Stapelfeldt}
\affil{Exoplanets and Stellar Astrophysics Laboratory, Code 667, NASA Goddard Space Flight Center, Greenbelt, MD 20771, USA}

\and

\author{Mark S. Marley}
\affil{NASA Ames Research Center, MS 245-3, Moffett Field, CA 94035, USA}

\altaffiltext{1}{Sagan Fellow}
\altaffiltext{2}{NASA Astrobiology Institute's Virtual Planetary Laboratory}

\begin{abstract}

Several concepts now exist for small, space-based missions to directly characterize 
exoplanets in reflected light.  While studies have been performed that investigate the 
potential detection yields of such missions, little work has been done to understand how 
instrumental and astrophysical parameters will affect the ability of these missions to 
obtain spectra that are useful for characterizing their planetary targets.  Here, we develop 
an instrument noise model suitable for studying the spectral characterization potential of a 
coronagraph-equipped, space-based telescope.  We adopt a baseline set of telescope 
and instrument parameters appropriate for near-future planned missions like 
WFIRST-AFTA, including a 2~m diameter primary aperture, an operational wavelength 
range of 0.4--1.0~$\mu$m, and an instrument spectral resolution of 
$\lambda/\Delta\lambda=70$, and apply our baseline model to a variety of spectral models 
of different planet types, including Earth twins, Jupiter twins, and warm and cool Jupiters 
and Neptunes.  With our exoplanet spectral models, we explore wavelength-dependent 
planet-star flux ratios for main sequence stars of various effective temperatures, and 
discuss how coronagraph inner and outer working angle constraints will influence the 
potential to study different types of planets.  For planets most favorable to spectroscopic 
characterization---cool Jupiters and Neptunes as well as nearby  
super-Earths---we study the integration times required to achieve moderate signal-to-noise 
ratio spectra.  We also explore the sensitivity of the integration times required to either 
detect the bottom or presence of key absorption bands (for methane, water vapor, and 
molecular oxygen) to coronagraph raw contrast performance, exozodiacal light levels, 
and the distance to the planetary system.  Decreasing detector quantum efficiency at 
longer visible wavelengths makes the detection of water vapor in the atmospheres of 
Earth-like planets extremely challenging, and also hinders detections of the 0.89~$\mu$m 
methane band.  Additionally, most modeled observations have 
noise dominated by dark current, indicating that improving CCD performance could 
substantially drive down requisite integration times.  Finally, we briefly discuss the 
extension of our models to a more distant future Large UV-Optical-InfraRed (LUVOIR) 
mission.

\end{abstract}
%


%
\section{Introduction}

Exoplanet atmospheric characterization is a rapidly progressing field, and will likely 
continue on this trajectory as the detection and study of nearby exoplanets was 
highlighted as one of three main science objectives in the 2010 National Research 
Council decadal survey of astronomy and astrophysics\footnotemark[1].  Following 
the first detection of an exoplanet atmosphere \citep{charbonneauetal2002}, 
secondary eclipse and transit observations have been used to characterize the 
atmospheres of a number of hot Jupiters \citep{grillmairetal2008,swainetal2008,
pontetal2008,swainetal2009,singetal2009,madhusudhan&seager2009}.  More 
recently,  observations have begun to probe new categories of planets, including  
mini-Neptune and super-Earth exoplanets \citep{stevensonetal2010,beanetal2010,
kreidbergetal2014,knutsonetal2014a,knutsonetal2014b,ehrenreichetal2014,
fraineetal2014}.  Of course, and as with any young field, some findings remain 
controversial or have undergone substantial revision \citep{lineetal2014,
hansenetal2014,dimaondloweetal2014}.

\footnotetext[1]{\scriptsize{\url{www.nationalacademies.org/astro2010}}}

Anticipating future exoplanet direct detection missions, several studies have 
investigated the detection capabilities of coronagraph or starshade instruments.  
\citet{agol2007} presented analytic models of the detection efficiency of coronagraph 
searches and investigated dependencies on key parameters, including survey duration, 
telescope diameter, and telescope inner working angle (IWA).  \citet{brown&soummer2010} 
discussed methods for optimizing survey ``completeness'' \citep{brown2004,brown2005}, 
while other studies \citep{savranskyetal2010,turnbulletal2012} have emphasized 
techniques for evaluating the effectiveness of proposed missions.  \citet{mawetetal2014} 
studied the statistics of high-contrast imaging at small angles, and found that the limited 
number of resolution elements near to the star (i.e., at $\lesssim2\lambda/D$) can lead to large 
errors in contrast estimations and false alarm probabilities computed by less sophisticated 
methods.  \citet{starketal2014} presented an improved technique for optimizing exoplanet 
detection completeness, and explored the dependencies of exo-Earth yield on a number of 
astrophysical, mission, and instrument parameters.  In a follow-up paper, 
\citet{starketal2015} used their completeness model to compute lower limits on the aperture 
size required to find and characterize a significant number of Earth-twins.  Very recently, 
\citet{greco&burrows2015} studied the detection capabilities of a coronagraph-equipped 
WFIRST-AFTA mission\footnotemark[2] \citep{spergeletal2013}.

\footnotetext[2]{\scriptsize{\url{http://wfirst.gsfc.nasa.gov/science/sdt_public/WFIRST-AFTA_SDT_Report_150310_Final.pdf}}}

Technologies and ideas for direct detection and characterization of exoplanets 
have long been discussed, and there exists a body of white papers and summary 
documents that outline strategies for direct characterization missions 
\citep{beichmanetal1999,desmaraisetal2002,schneideretal2008,lunineetal2008,
cockelletal2009}.  Recently, following a growing interest in exoplanet direct 
detection\footnotemark[3]$^{,}$\footnotemark[4], NASA's Exoplanet Exploration Program 
commissioned a number of ``quick study'' reports focused on exoplanet characterization 
at visible wavelengths.  \citet{marleyetal2014} discussed how reflected light observations 
could be used to characterize existing radial velocity-detected exoplanets.  \citet{hu2014} 
studied how direct imaging could constrain methane abundances and cloud properties 
using low resolution spectra.  Finally, \citet{burrows2014} explored the diversity of giant 
exoplanet spectra that may be accessible to a coronagraph-equiped WFIRST-AFTA 
mission.

\footnotetext[3]{\scriptsize{\url{http://exep.jpl.nasa.gov/stdt/Exo-C_Final_Report_for_Unlimited_Release_150323.pdf}}}
\footnotetext[4]{\scriptsize{\url{http://exep.jpl.nasa.gov/stdt/Exo-S_Starshade_Probe_Class_Final_Report_150312_URS250118.pdf}}}

Only a subset of the aforementioned quick studies considered the influence of noise on 
spectral observations, which, if included, is commonly applied by assuming a constant 
signal-to-noise ratio (SNR) across the entire modeled ``observation''.  While some 
studies have done extensive modeling of the influence of speckle noise (and techniques 
for minimizing such noise) on direct exoplanet observations \citep{maroisetal2000,
sparks&ford2002}, few published studies exist that comprehensively examine how spectral 
observations are influenced by stellar leakage, zodiacal light, exozodiacal light, and other 
noise sources.  \citet{maireetal2012} investigated how noise would 
influence the characterization capabilities of the proposed the Spectro-Polarimetric Imaging 
and Characterization of Exoplanetary Systems (SPICES) mission 
\citep{boccalettietal2012}.  This work included noise due to speckles, exozodiacal 
light, zodiacal light, and read-out, resulting in a handful of spectral simulations for 
planet-star systems for a specified distance and integration time.  Recently, 
\citet{legeretal2015} studied design missions for a 2-meter class space-based coronagraph 
mission and a four-spacecraft formation flying interferometer mission, assuming noise 
from only stellar leakage and exozodiacal light, with a primary goal of  
determining how  effective such missions would be in studying exo-Earths given new 
estimates of $\eta_{\oplus}$.

Here, we build on the limited previous studies of the influence of noise on the ability of 
space-based coronagraphs to characterize exoplanets.  We begin by outlining a framework 
for modeling noise for such missions, including stellar leakage, exozodiacal light, zodiacal 
light, dark current, and read noise.  We then investigate the capabilities of a 2-meter class 
mission to observe and study a variety of Jupiter- and Neptune-sized worlds as well as 
terrestrial exoplanets.  Using a characteristic set of mission parameters, we show how 
expected integration times will scale with wavelength and host star spectral type, and, 
for key spectral features, we show the sensitivity of these integration times to distance to 
the star-planet system, exozodiacal light levels, and coronagraph raw contrast performance.  
Finally, we discuss how our framework extends to the characterization of Earth twins with 
Large UV-Optical-InfraRed (LUVOIR) telescope missions.

\section{Model Description}

We consider a space-based telescope of diameter $D$ designed to detect and 
characterize exoplanets in reflected light.  The telescope is equipped with a 
coronagraph that achieves a raw contrast $C$ (which can, in general, be 
wavelength-dependent) and has inner and outer working angles of $\theta_{\rm{IWA}}$ 
and $\theta_{\rm{OWA}}$, respectively.  The planetary light is is distributed into a point 
spread function (PSF), where an aperture of width (or diameter) $X\lambda/D$ is used in 
the image plane to extract the  planetary signal.  However, a number of other sources will 
contribute to the detector counts within the defined aperture, including Solar System zodiacal 
light, exozodiacal light, leaked stellar light, dark current, and read noise.

The Appendix contains a presentation of our models of the planet count rate and 
count rates from key noise sources, which are similar to those of \citet{brown2005}.  
For convenience, a synopsis of all of the model parameters, variables, and outputs are 
given in Table~1.  Briefly, our planet count rate ($c_{\rm{p}}$), zodiacal light count rate 
($c_{\rm{z}}$), exozodiacal light count rate ($c_{\rm{ez}}$), stellar leakage count rate 
($c_{\rm{lk}}$), dark current count rate ($c_{\rm{D}}$), and read noise count rate 
($c_{\rm{R}}$) are computed using Equations~\ref{eqn:pcnt}, \ref{eqn:cz}, \ref{eqn:cez}, 
\ref{eqn:clk}, \ref{eqn:cD}, and \ref{eqn:rn_rate}, respectively.

\subsection{Baseline Parameters and Input Data}

It is useful to define a baseline set of astrophysical, telescope, and instrument 
parameters, though later sections of this paper will explore sensitivity to key 
variables (e.g., distance, exozodiacal light levels, and coronagraph raw contrast).  
Baseline astrophysical parameters are given in Table~2, while telescope and 
instrument parameters are in Table~3.

Regarding astrophysical parameters, we adopt the standard, empirically-derived 
solar spectrum of \citet{wehrli1985}.  As mentioned in Section~\ref{subsec:zod}, 
we use a surface brightness of $M_{\rm{z},V}=23$~mag~arcsec$^{-2}$ for Solar 
System zodiacal light.  For exozodiacal light, we assume one zodi 
(i.e., $N_{\rm{ez}}=1$) with a surface brightness of 
$M_{\rm{ez},V}=22$~mag~arcsec$^{-2}$, which is a factor of $\sim\!\!2$ larger than 
Solar System zodiacal surface brightness, as exozodiacal dust both above and 
below the midplane contribute \citep{starketal2014}.

\begin{figure}
  \centering
  \includegraphics[trim = 1mm 1mm 1mm 1mm, clip, width=6.2in]{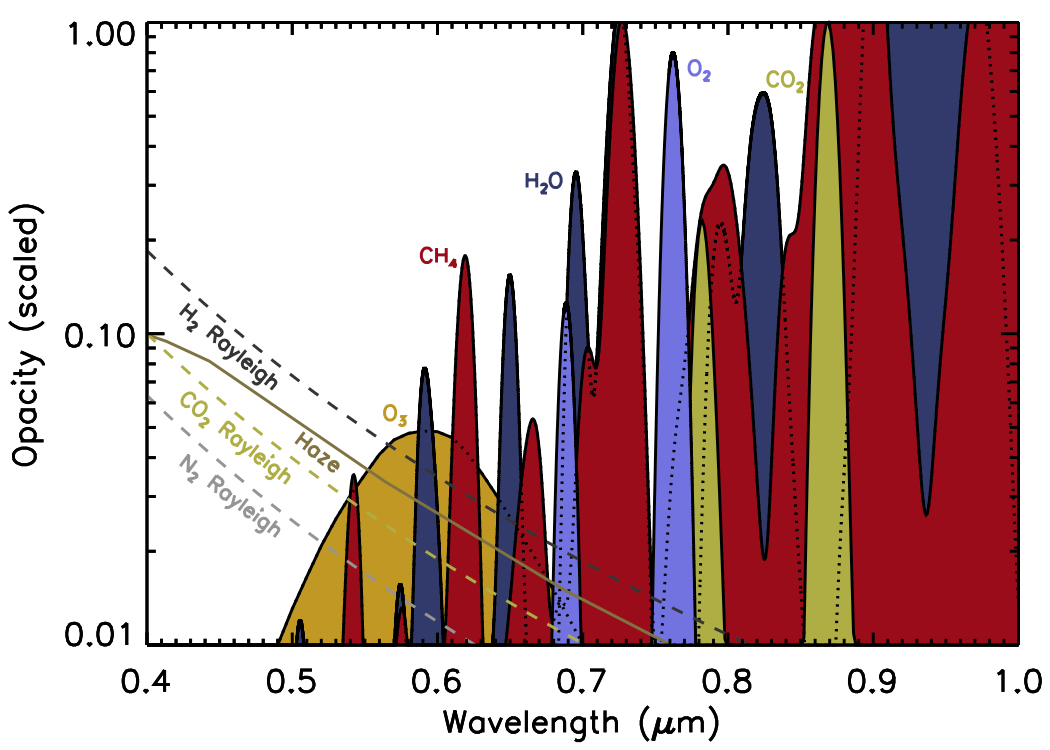}
  \caption{Scaled opacities for different species and aerosols between 0.4--1.0~$\mu$m at 
                $\mathcal{R}=70$, demonstrating the numerous bands and features present across 
                the visible wavelength range.  Scalings for gaseous absorption are constant with 
                wavelength for each species and arbitrary; these were chosen to highlight relevant 
                bands.  The haze opacity is for extinction by a Titan tholin-lke haze, and the Rayleigh 
                scattering opacities are scaled to that of CO$_{2}$ at 0.4~$\mu$m (and divided by 10).}
  \label{fig:bands}
\end{figure}

For baseline telescope and instrument parameters, we consider a 2-meter class 
telescope equipped with a coronagraph designed to achieve a raw contrast of 
$C=10^{-9}$ between $\theta_{\rm{IWA}}=2\lambda/D$ and 
$\theta_{\rm{OWA}}=10\lambda/D$.  The total system throughput is 
$\mathcal{T}=0.05$, and we mimic the decreasing sensitivity of CCDs at red 
wavelengths by parameterizing the quantum efficiency according to 
\begin{equation}
q = 
    \begin{cases}
     0.9, &  \lambda \leq 0.7~\mu\rm{m}  \\
     0.9\left(1 - \frac{\lambda-0.7}{0.3}\right), &  0.7~\mu\rm{m} < \lambda \leq 1~\mu\rm{m} \ .
  \end{cases}
\label{eqn:quanteff}
\end{equation}
We use a characteristic dark current of $5\times10^{-4}$~s$^{-1}$ and read noise 
of 0.1.  We assume a square aperture width of $1.5\lambda/D$ (i.e., $X=1.5$), 
which would be a sensible choice to help eliminate sensitivity to the sub-pixel 
(or sub-lenslet) location of the planet, whose PSF is sampled at $0.5\lambda/D$.  
Finally, we assume that the coronagraph is paired with a spectrometer that covers 
0.4--1.0~$\mu$m at  $\mathcal{R}=70$, which is sufficient to resolve many 
molecular absorption bands and features throughout the visible wavelength range 
(see Figure~\ref{fig:bands}).  Note that, while a resolution of 70 is small 
enough to resolve the A-band of molecular oxygen, this may not be the ideal resolution 
for detecting this band \citep{brandt&spiegel2014}.  Overall, our assumed parameters 
are meant to be representative of either the previously-mentioned 
WFIRST-AFTA mission or the Exo-C mission concept (see reports linked in the 
Introduction).

Our reflectivity data come from a number of different sources.  For Earth 
twins we use the extensively validated Virtual Planetary Laboratory 3-D 
spectral Earth model from \citet{robinsonetal2011}, which realistically 
simulates wavelength- and phase-dependent spectra of the Pale Blue 
Dot \citep{robinsonetal2010,robinsonetal2014b}.  In certain cases, we 
investigate super-Earths, which we take to have an Earth-like reflectivity 
and a radius of $1.5R_{\oplus}$, and Venus twins, with a characteristic 
reflectivity from the validated models of \citet{arneyetal2014}.  For Jupiter 
and Neptune twins, we use the observed geometric albedos from 
\citet{karkoschka1998} and a Lambertian phase function given by
\begin{equation}
  \Phi_{\rm{L}}(\alpha) = \frac{\sin\alpha + (\pi - \alpha)\cos \alpha}{\pi} \ ,
\label{eqn:lambert}
\end{equation}
although realistic planetary phase functions are not perfectly Lambertian 
\citep[e.g.,][their Figure~15]{cahoyetal2010}.  Recall that the phase function 
captures phase-dependent brightness changes as the planet moves through its 
orbit and that the brightness increase from quadrature to full phase is a factor of 
$\sim\!\! 3$.  Finally, we also consider ``warm'' and ``cool'' Jupiters and 
Neptunes, using the wavelength- and phase-dependent models of 
\citet{cahoyetal2010} for both a 0.8~AU and 2~AU (from a Sun-like star) Jupiter 
and Neptune (at $3\times$ and $30\times$ heavy element enhancement, 
respectively).  All planets are taken to be at a characteristic phase angle of 
$\alpha = \pi/2 = 90^{\circ}$ (i.e., quadrature).

\subsection{Noise}

The total number of counts, $C_{\rm{tot}}$, is computed from the total count rate 
using 
\begin{equation}
  C_{\rm{tot}} = \left( c_{\rm{p}} + c_{\rm{z}} + c_{\rm{ez}} + c_{\rm{lk}} + c_{\rm{D}} + c_{\rm{R}} \right) \Delta t_{\rm{exp}} = \left(c_{\rm{p}} + c_{\rm{b}} \right) \Delta t_{\rm{exp}} \ ,
\end{equation}
where $\Delta t_{\rm{exp}}$ is the exposure time, and we have defined the total 
background count rate, 
$c_{\rm{b}} =  c_{\rm{z}} + c_{\rm{ez}} + c_{\rm{lk}} + c_{\rm{D}} + c_{\rm{R}}$.
The total number of background counts would be 
$C_{\rm{b}} = c_{\rm{b}} \Delta t_{\rm{exp}}$.  Following \citet{brown2005}, the 
noise counts are taken as
\begin{equation}
  C_{\rm{noise}} = \sqrt{C_{\rm{p}} + 2C_{\rm{b}}} \ ,
\label{eqn:noise_counts}
\end{equation}
which assumes a background subtraction (e.g., via a spacecraft roll maneuver) is 
performed to detect the planet (implying that the single exposure photon counting 
noise limit is not achieved), and $C_{\rm{p}}$ is the total planet counts.  With the 
signal-to-noise ratio given by
\begin{equation}
  {\rm{SNR}} = \frac{ C_{\rm{p}} }{ C_{\rm{noise}} } \ ,
\end{equation}
we can compute the exposure time needed to achieve a given SNR by
\begin{equation}
  \Delta t_{\rm{exp}} = \frac{ c_{\rm{p}} + 2c_{\rm{b}} }{ c_{\rm{p}}^{2} }\rm{SNR}^{2} \ .
\label{eqn:t_exp}
\end{equation}

Note that the exposure time required to achieve a given SNR is distinct from 
the SNR required to detect a certain molecule \citep[e.g.,][]{brandt&spiegel2014}, 
or to distinguish between models with different heavy element enhancements 
\citep{maireetal2012}.  The former (given by Equation~\ref{eqn:t_exp}), when 
applied at the bottom of several molecular features (as is done in later sections), is 
representative of the requisite exposure time for determining molecular abundances 
and cloud properties for exoplanet atmosphere \citep{marleyetal2014}.  Such 
determinations require comparisons to cloudy atmospheric models, both at continuum 
wavelengths and across/in molecular bands, thereby simultaneously retrieving cloud 
and atmospheric parameters (including the abundances of key species).

We also explore the integration time required to simply detect a molecular feature 
by the deviation it causes from a flat continuum, which is sufficient for stating whether 
or not a certain species is present in an atmosphere.  Following \citet{misraetal2014}, 
we write the band SNR by defining the signal as the count difference between the 
spectral model and a model without absorption, determined by fitting the continuum 
on both sides of the absorption band.  The noise counts are summed over the band.   
Thus, if $c_{\rm{cont}}$ is the continuum count rate, then the band SNR is given by 
\begin{equation}
  {\rm{SNR_{band}}} = \frac{\sum\limits_{j} c_{{\rm{cont,}}j} -  c_{{\rm{p,}}j} }{ \sqrt{\sum\limits_{j} c_{{\rm{p,}}j} + 2c_{{\rm{b,}}j}} } \Delta t_{\rm{exp}}^{1/2} \ ,
\label{eqn:SNRband}
\end{equation}
where the sum is over all spectral elements (denoted by sub-script `$j$') within 
the molecular band.  Band detection times, in general, will be long for shallow 
and narrow features, whereas the previously discussed exposure times to reach 
a given SNR at the bottom of a band is shortest for shallow features.  These two 
times will be roughly equal for a feature that is one resolution element in width 
and that is moderately deep.

\section{Results}

The following subsections explore the detectability of, and integration time 
required to detect, key spectral features for various categories of planets.  We 
define a planet ``type'' as having fixed radius, reflectivity 
[$A\Phi\left(\alpha\right)$], and incident flux from its host star.  The latter 
requirement is sometimes referred to as the ``flux (or insolation) equivalent  
distance''.  Note that we focus on the visible wavelength range 
(0.4--1.0~$\mu$m), as exoplanet characterization in reflected light at 
near-infrared wavelengths is especially difficult for small telescopes due 
to IWA constraints.

To get a sense for the scales involved, we note that an Earth twin around a 
Sun-like star, using $A=0.2$, $\Phi\left(\pi/2\right) \sim 1/\pi$, 
$R_{\rm{p}}=R_{\oplus}$, and $r=1$~AU, would have the canonical planet-star 
flux ratio (from Equation~\ref{eqn:flxratio}) of $10^{-10}$.  Then, assuming our 
baseline parameters, and using $\lambda=0.55$~$\mu$m with a $8$~nm-wide 
bandpass, Equation~\ref{eqn:pcnt} gives $c_{\rm{p}} \sim 5\times10^{-4}$~s$^{-1}$ 
(or about 2~hr$^{-1}$) for an Earth twin at 5~pc.  By comparison, the leakage  
count rate is 20~hr$^{-1}$ for $C=10^{-9}$, while the zodiacal and exozodiacal 
light count rates are 2~hr$^{-1}$ and 4~hr$^{-1}$, respectively.  With a dark count 
rate of about 240~hr$^{-1}$ and a read noise count rate of 10~hr$^{-1}$, the exposure 
time needed to reach $\rm{SNR}=5$ is roughly $5\times10^{3}$~hr---an 
unfeasibly long integration time, but which could be reduced by decreasing spectral 
resolution.  For a Jupiter twin at 5~pc, with a geometric albedo of 0.5, the planet-star 
flux ratio increases to $\sim\!\!10^{-9}$, and the planet count rate increases to 
20~hr$^{-1}$ (although this twin, at 5.2 AU from its Sun-like host, would be outside 
$\theta_{\rm{OWA}}=10\lambda/D$ at quadrature) while the exozodiacal light count 
rate decreases to 0.2~hr$^{-1}$.  Here, then, the time needed to reach $\rm{SNR}=5$ 
is only 30 hours.

As demonstrated by these count rate example calculations, the dominant term in the 
noise counts (Equation~\ref{eqn:noise_counts})---for a wide range of planet types, 
distances, stellar effective temperatures, and wavelengths---is dark current.  This term 
is strongly influenced by two parameters: the dark current rate ($D_{\rm{e}^{-}}$) and the 
width of our assumed photometric aperture (controlled by $X$).  The latter term determines 
the area used to extract the planetary signal, which leads to a $X^{2}$ dependence in the 
dark noise count rate.  For our assumed baseline set of parameters and a characteristic 
distance of 10~pc, all of the planet types investigated in Section~\ref{sec:tint} and later 
have dark current as the dominant noise source, leading the exposure times required to 
achieve a given SNR to scale with the planet count rate as (roughly) 
$c_{\rm{D}}/c_{\rm{p}}^{2}$.  Note that this scaling leads to a strong dependence on 
planetary properties, implying that, for example, observing planets closer to full phase 
would drive down integration times by as much as an order of magnitude from the 
estimates provided below (which are given at quadrature phase).  Also, the overall 
importance of dark current in our results below is influenced by our baseline 
throughput assumption (selected to be consistent with, e.g., WFIRST-AFTA)---improved 
telescope and instrument throughputs will drive up the planetary signal, while also 
increasing the relative weights of noise due to stellar leakage and zodiacal and 
exozodiacal light.

\subsection{Planet-Star Flux Ratios}

\begin{figure}
  \centering
  \begin{tabular}{cc}
    \includegraphics[trim = 4mm 2mm 2mm 3mm, clip, width=3.1in]{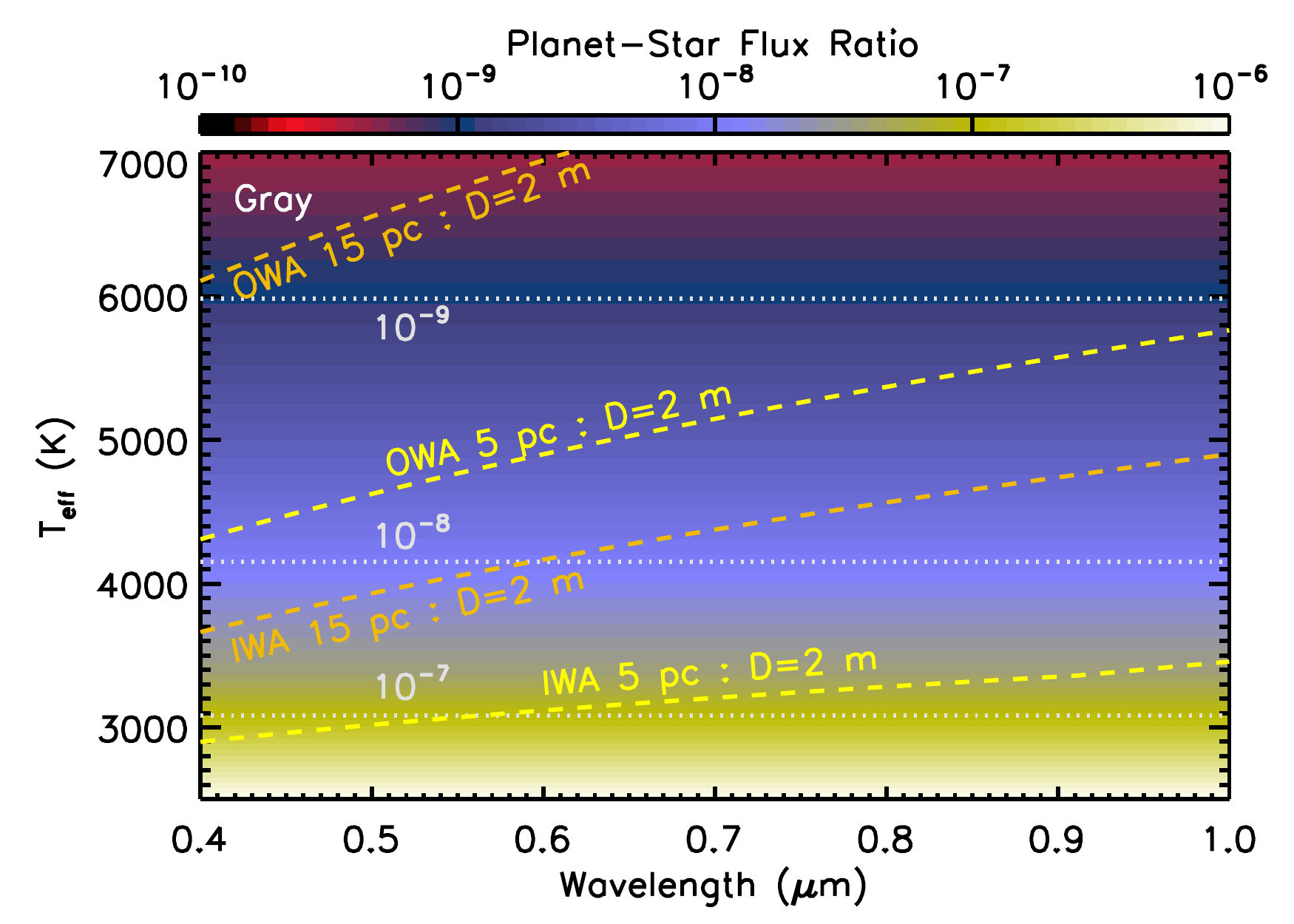} &
    \includegraphics[trim = 4mm 2mm 2mm 3mm, clip, width=3.1in]{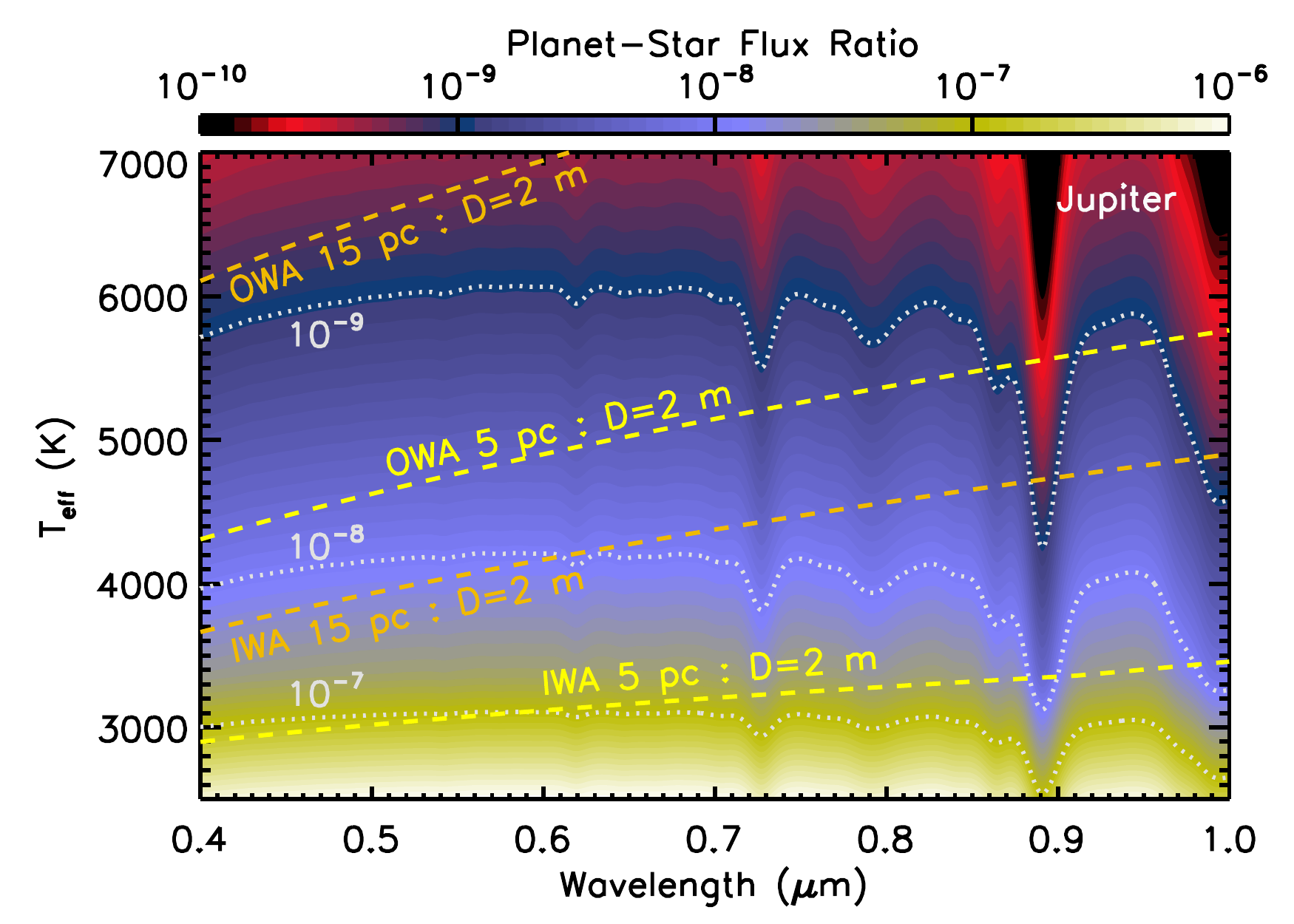} \\
    \includegraphics[trim = 4mm 2mm 2mm 3mm, clip, width=3.1in]{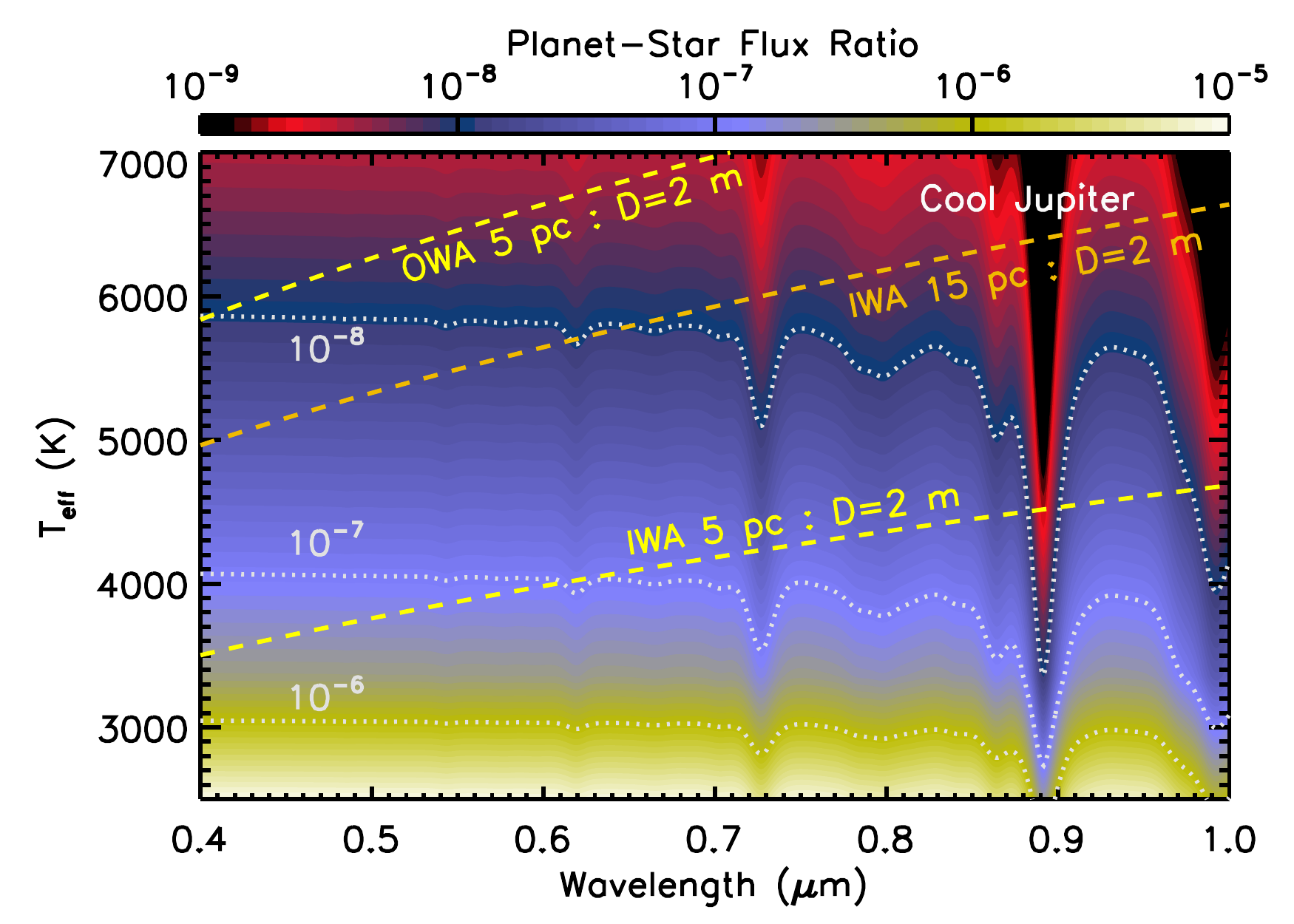} &
    \includegraphics[trim = 4mm 2mm 2mm 3mm, clip, width=3.1in]{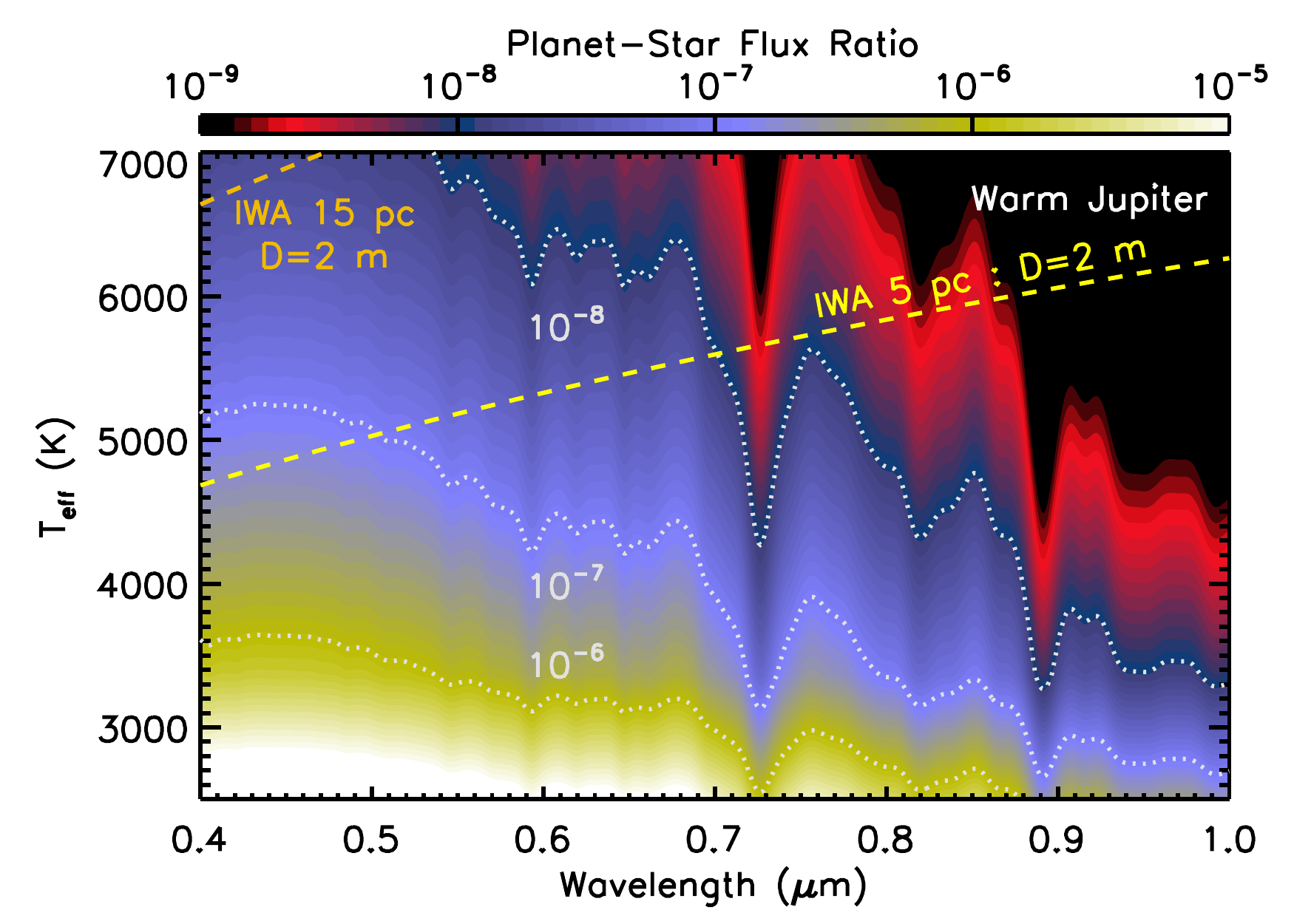} \\
  \end{tabular}
  \caption{Planet-star flux ratio color contours for a variety of Jupiter-sized worlds at  
                fixed flux equivalent distances around main sequence stars of different 
                effective temperatures.  Planet types are: (a) a gray planet with 
                $A\Phi(\alpha)=0.5/\pi$ and $F_{\rm{TOA}}=50.5$~W~m$^{-2}$ (top-left), (b) a 
                Jupiter twin with $F_{\rm{TOA}}=50.5$~W~m$^{-2}$ and reflectivity from 
                \citet{karkoschka1998} (top-right), (c) a cool Jupiter with 
                $F_{\rm{TOA}}=342$~W~m$^{-2}$ and reflectivity from \citet{cahoyetal2010}  
                (bottom-left), and (d) a warm Jupiter with 
                $F_{\rm{TOA}}=2140$~W~m$^{-2}$ and reflectivity from \citet{cahoyetal2010}  
                (bottom-right).  Inner and outer working angle constraints (at quadrature; dashed), 
                at $2\lambda/D$ and $10\lambda/D$, are shown for a 2-meter class telescope and 
                assuming the planet-star system are at 5~pc (yellow) and 15~pc (orange).}
  \label{fig:contrast_jupiters}
\end{figure}

Insight into the complexities associated with directly observing cool exoplanets comes 
from investigating planet-star flux ratios for our different types of worlds.  The first panel 
of Figure~\ref{fig:contrast_jupiters} shows the planet-star flux ratio as color contours  
for a Jupiter-sized world with $A\Phi\left(\alpha\right) = 0.5/\pi$ (independent of 
wavelength) and with a Jupiter-like top-of-atmosphere flux (fixed at 
$F_{\rm{TOA}}=50.5$~W~m$^{-2}$), as a function of main sequence host star effective 
temperature.  As we have assumed a gray reflectivity, the planet-star flux ratio only 
depends on $T_{\rm{eff}}$ (assuming a simple relationship between this and 
$R_{\rm{s}}$ on the main sequence), with
\begin{equation}
  \frac{F_{\rm{p}}}{F_{\rm{s}}} = \frac{A\Phi\left(\alpha \right)F_{\rm{TOA}}}{\sigma T_{\rm{eff}}^4}\left( \frac{R_{\rm{p}}}{R_{\rm{s}}} \right)^{2} \ ,
\end{equation}
where the dependence on planetary orbital distance is functionally removed.  Here, 
since the planet flux is fixed by $A\Phi\left(\alpha\right)$ and $F_{\rm{TOA}}$  
(regardless of stellar effective temperature), the planet-star flux \emph{ratio} 
increases for cooler stars as these stars are intrinsically fainter.

The planet-star flux ratio contours for the gray case in Figure~\ref{fig:contrast_jupiters} 
tell an incomplete story, however.  First, while the results indicate that detecting 
a Jupiter-sized planet around cool stars (with $T_{\rm{eff}}<4000$~K) would be 
straightforward with a coronagraph capable of achieving a raw contrast of only 
$10^{-8}$, it is important to remember that the probability that such cool stars host 
a gas giant is small \citep[of order a few percent,][]{johnsonetal2010,
montetetal2014}.  Additionally, the coronagraph inner and outer working angles will 
constrain the range of planet-star apparent separations (which are related to orbital 
separations) accessible to the telescope when observing a system at a given distance.  
These constraints are shown as dashed lines for distances of 5~pc and 15~pc in 
Figure~\ref{fig:contrast_jupiters}.  For fixed $F_{\rm{TOA}}$ and a system distance 
of 5~pc, the OWA would not allow observations of the gray Jupiter world at quadrature 
over the entire wavelength range for host stars with effective temperatures above 
5800~K.  When placed at 15~pc, the IWA can prevent observations at the longest 
wavelengths for stars with effective temperature below 4900~K.

The remaining panels in Figure~\ref{fig:contrast_jupiters} show non-gray cases, 
where realistic reflectivities are used.  Like the gray case, the Jupiter twins have 
$F_{\rm{TOA}}=50.5$~W~m$^{-2}$, while the cool and warm Jupiters have 
$F_{\rm{TOA}}=342$~W~m$^{-2}$ and $F_{\rm{TOA}}=2140$~W~m$^{-2}$, 
respectively.  With the planet-star flux ratio now a function of wavelength, these 
plots allow us to easily determine roughly what raw contrast must be achieved 
to detect different types of Jupiters at continuum wavelengths and within 
absorption features.  The Jupiter twins and the cool Jupiters are relatively 
reflective and have several methane bands covering a range of feature depths.  
Note that the Jupiter twins have low planet-star contrast ratios due to their 
small $F_{\rm{TOA}}$.  Additionally, despite their large $F_{\rm{TOA}}$, 
the warm Jupiters have low planet-star contrast ratios at longer wavelengths 
due to strong methane and water vapor bands as well as a lack of water clouds.

Figure~\ref{fig:contrast_neptunes} is similar to Figure~\ref{fig:contrast_jupiters} 
but for Neptune twins, cool Neptunes, and warm Neptunes.  Neptune twins have 
very low planet-star flux ratios due to a very small $F_{\rm{TOA}}$ 
(1.5~W~m$^{-2}$), and would also be difficult to detect for a wide range of 
stellar effective temperatures due to the telescope OWA restrictions.  The flux ratios 
improve for the cool and warm cases, but are lower than the equivalent 
Jupiter-sized cases as Neptune is 2.9 times smaller than Jupiter.  Also, the methane 
and water bands for these cases are deeper than those for the Jupiters, 
yielding flux ratios near $10^{-10}$--$10^{-11}$ over much of parameter space.

Continuing downwards in the size/mass regime, Figure~\ref{fig:contrast_terrests} 
shows planet-star contrast ratios for Venus twins 
($F_{\rm{TOA}}=2610$~W~m$^{-2}$), as well as for Earth twins and super-Earths 
(both with $F_{\rm{TOA}}=1370$~W~m$^{-2}$).  The Venus twins have 
comparatively high planet-star flux ratios, both due to their high reflectivity and 
large $F_{\rm{TOA}}$, and have nearly featureless spectra over the 
highlighted wavelength range.  Inner working angle constraints make Venus 
twins very difficult to detect for our chosen architecture, especially at longer 
wavelengths.  For Earths and super-Earths, the visible wavelength range 
contains key spectral features---the 0.76~$\mu$m O$_{2}$ A-band and the 
0.94~$\mu$m water band---but IWA constraints largely limit our ability to 
characterize such planets around cool stars.

The relationships between wavelength-dependent reflectivity, host star effective 
temperature, and planet-star flux ratio discussed above can be straightforwardly 
manipulated to show contours of stellar effective temperature where planet types 
achieve a given flux ratio.  Four such plots are shown in Figure~\ref{fig:teff_contours}, 
which illustrate the range of different stellar spectral types that can be investigated for a 
specified planet type and contrast performance.  These diagrams demonstrate how,  
for a fixed planet-star flux ratio, the depths of absorption bands can best be probed 
for cooler stars, where the low intrinsic stellar brightness compensates for the 
low planetary reflectivity.  As before, inner and outer working angle limitations 
are shown for planet-star systems at 5~pc and 15~pc.

\subsection{Integration Times: Full Spectral Range}
\label{sec:tint}

We now shift our attention to the integration times required to achieve a 
specified SNR for different planet types.  These exposure times determine 
whether or not it is realistically feasible to acquire the types of observations 
needed to characterize planets at a wide range of wavelengths, especially given 
constraints on mission duration and overall mission time allotted for exoplanet 
studies.  While we highlight $\rm{SNR}=5$ as a minimum value needed for 
characterization---\citet{marleyetal2014} were able to draw scientifically 
valuable conclusions from their simulated data at this SNR---the exposure 
times discussed here can be easily scaled to other SNRs since 
$\Delta t_{\rm{exp}} \propto \rm{SNR}^{2}$.

Wavelength-dependent integration times required to achieve 
$\rm{SNR}=5$ fixed $F_{\rm{TOA}}$ cases are shown for Jupiter twins and 
cool~Jupiters at a characteristic distance of 10~pc in 
Figure~\ref{fig:jupiter_Dt}.  Requisite integration times for the cool~Jupiters 
are promising, reaching $\rm{SNR}=5$ in 10--100~hr,  
even in the depths of some methane absorption bands.  For the Jupiter twins, 
detecting the bottom of these bands requires more than $10^{3}$ hours of 
observation, owing to these planets' larger orbital distances, although 
binning data to lower spectral resolution ($\mathcal{R}=$20--30) would 
drive down the requisite exposure times (by a factor of a few).
 
Figure~\ref{fig:neptune_Dt} shows integration times required to achieve 
$\rm{SNR}=5$ for cool~Neptunes at a distance of 10~pc.  Note that these 
integration times show a much larger dynamic range than the cool~Jupiter 
cases owing to the generally deeper methane bands in the geometric albedo 
spectrum of the Neptunes .  Overall, integration times shorter than 
$10^{3}$~hr are only achieved below 0.7~$\mu$m, owing to both strong 
absorption features and falling stellar spectra at longer wavelengths.

Finally, Figure~\ref{fig:earth_Dt} shows integration times to $\rm{SNR}=5$ 
for Earth twins and super-Earths, both taken to be at 5~pc distance.  To maintain 
integration times shorter than about $10^{3}$~hr, the Earth and super-Earth 
observations have been degraded to spectral resolutions of 10 and 20, 
respectively (which would not allow detection of the 0.76~$\mu$m A-band of 
molecular oxygen).  Outside of the 0.94~$\mu$m water band, the Earth twins 
have integration times of order $10^{3}$~hr, although achieving such 
observations for Sun-like stars and hotter would require a contrast noise floor 
better than $10^{-10}$.  By comparison, super-Earths avoid the contrast limitation 
over most of parameter space, and require integration times less than $500$~hr 
for most wavelengths (at $\mathcal{R}=20$).

\subsection{Integration Times: Molecular Features}
\label{sec:bands}

By highlighting key molecular bands in certain planet type spectra, we can 
investigate the integration times required to detect these features.  We explore 
both the integration time required to achieve a given SNR at the bottom of a 
molecular band (Equation~\ref{eqn:t_exp}), and the integration time required to 
detect a band (or bands) of certain species (Equation~\ref{eqn:SNRband}).  
Recall that the former is indicative of the integration time required to achieve 
observations suitable for abundance determinations (where detections of the 
bottoms of multiple bands are required), while the latter indicates the time 
required to determine whether or not a species is present.  As before, we set 
the detection SNR to be 5, which can easily be scaled to other SNR values.  
Also, for all the bands discussed below, our baseline spectral resolution 
($\mathcal{R}=70$) is sufficient to resolve these bands 
\citep{desmaraisetal2002}.

Figure~\ref{fig:jupiter_det} shows integration times for detection of species 
and bottoms of bands for cool Jupiters as a function of distance to the planetary 
system.  We focus on methane, which is the dominant absorber for the cool 
Jupiters (and Neptunes) at visible wavelengths.  For detecting the bottoms of 
bands, we highlight the 0.73~$\mu$m and 0.89~$\mu$m methane bands, which 
are, respectively, moderate and strong bands.  For detecting methane, we use 
the 0.73~$\mu$m band, although required integration times for detecting the 
0.89~$\mu$m band are similar, owing to a trade-off between detector quantum 
efficiency (lower at these wavelengths) and band strength (higher).  We also 
show that the integration time required to detect methane is decreased by about 
50\% if both bands are used.  Including the 0.62~$\mu$m methane band does 
not further decrease integration times, as the shallowness of this band leads 
to long required integration times for detection.

Figure~\ref{fig:neptune_det} is similar to Figure~\ref{fig:jupiter_det}, except for 
cool Neptunes.  For detecting band bottoms, we use the 0.54~$\mu$m and 
0.62~$\mu$m methane bands, as detecting the bottom of longer wavelength 
bands would require contrast ratios better than $10^{-10}$.  We show integration 
times for methane detection using the 0.62~$\mu$m band, and including both 
this band and the 0.73~$\mu$m decreases requisite times by about 50\%.

Finally, Figures~\ref{fig:earth_det} and \ref{fig:superearth_det} show integration 
times for band bottom and species detections for Earth twins and super-Earths, 
respectively.  Here, we emphasize the 0.76~$\mu$m A-band of molecular oxygen 
and the 0.94~$\mu$m water vapor band.  The A-band is a key potential biosignature, 
while the water vapor band is essential for discerning planetary habitability.  Note that 
the time required to detect the bottom of the A-band is much shorter than the time 
to detect the presence of the band.  This is due to the convolution of the 
high-resolution Earth reflectance spectrum with the spectrometer line-shape 
function, which serves to decrease the depth of this band at $\mathcal{R}=70$ 
and, thus, makes detection of the band difficult.  Thus, the integration time required 
to reach the bottom of the A-band at a SNR of 5 is not much different from the 
time to reach the same SNR in the surrounding continuum.  We note that requisite 
integration times for Earths or super-Earths placed at the outer edge of the habitable 
zone are typically 5--10 times longer due to the lower stellar flux incident on these 
planets.

\subsection{Sensitivity to Key Parameters}

Our baseline parameter set assumed that every planet-star system has one 
exozodi and that the coronagraph could deliver a raw contrast of $10^{-9}$.  
However, exozodi levels can vary by orders of magnitude, as can the designed 
raw contrast (depending on which mission architecture is adopted).  Thus, we 
explore sensitivity to these two parameters by examining their influence 
on our ability to characterize a planet for key atmospheric constituents.

Figure~\ref{fig:jupiter_Dt_sens} shows the integration time required to 
achieve $\rm{SNR}=5$ in the bottom of the 0.73~$\mu$m methane band as well 
as the integration time required to detect this band  $\rm{SNR_{band}}=5$ for our 
cool~Jupiter model.  Panels demonstrate sensitivity to exozodi levels and raw 
contrast performance at a fixed distance of 10~pc.  For these studies, the 
coronagraph IWA prevents observations of this planet type around stars cooler 
than 5,300~K.  As the 0.73~$\mu$m methane band is not particularly deep or wide 
in the cool Jupiter reflectance spectrum, integration times for band detection are 
longer than for detecting the bottom of the band.  Nevertheless, even at poor raw 
contrast performance and/or larger ($N_{\rm{ez}}\gtrsim10$) levels of exozodi, 
integration times for either band bottom or species detection are below 100~hr.

Figure~\ref{fig:neptune_Dt_sens} is similar to Figure~\ref{fig:jupiter_Dt_sens}, 
but for the cool~Neptune model.  Here we use the 0.62~$\mu$m methane band, 
because, as mentioned earlier, the bottom of the 0.73~$\mu$m band is at a 
contrast ratio of less than $10^{-10}$.  We place the planet-star systems at 
a distance of 5~pc.  Integration times for band detection are now shorter than 
for detecting the bottom of the band, as the 0.62~$\mu$m methane band is 
deeper in the cool Neptune reflectance spectrum than the 0.73~$\mu$m band 
in the cool Jupiter spectrum.

Finally, Figure~\ref{fig:superearth_Dt_sens} shows sensitivity tests for the 
integration times required to detect the 0.76~$\mu$m oxygen A-band for 
1.5$R_{\oplus}$ super-Earths.  We place the planet-star systems at a distance 
of 3.7~pc, which is roughly the distance to $\tau$ Ceti.  We do not consider the 
0.94~$\mu$m water vapor band as Figure~\ref{fig:superearth_det} shows that 
detecting this band at 3.7~pc would require nearly a $10^{4}$~hr integration 
time.  Also, we do not consider the integration time required to detect the bottom 
of the band as this is shorter than the time to detect the band.  

\subsection{Towards Large UV/Optical/Infrared Telescopes}

Large UV-Optical-Infrared telescopes have been proposed as important 
tools for continuing NASA's vision for exploring and characterizing exoplanets 
beyond the James Webb Space Telescope mission \citep{kouveliotouetal2014}.  
The capabilities of such a 10-meter class space-based telescope, if equipped with 
a coronagraph, can be investigated using the formalism outlined above.  As an 
example, Figure~\ref{fig:contours_luvoir} shows integration times to $\rm{SNR}=5$ 
across a broad wavelength range, and sensitivity to exozodi levels for the 
0.76~$\mu$m oxygen A-band for Earth twins, as would be the primary target of a 
LUVOIR mission.  We have adopted an extended wavelength range (0.4--3~$\mu$m), 
and have assumed the distance to these systems is 10~pc.  We have extended 
the OWA to a more generous $20\lambda/D$ and use $C=10^{-10}$, as could 
hopefully be achieved by such a distant-future mission.  Also, we extend the IWA to 
$3\lambda/D$, as telescope stability for a 10-meter class observatory would make 
achieving smaller IWAs very challenging.  (Note that this larger IWA is outside the 
$2\lambda/D$ ``practical limit to small IWA coronagraphy'' determined by 
\citet{mawetetal2014}.)  Finally, we assume that the visible and near-infrared wavelength 
ranges would be handled by two separate detectors, with respective pixel/lenslet sizes 
determined at 0.4~$\mu$m and 1.0~$\mu$m.  For the near-infrared detector, we adopt 
a dark current, read noise, and quantum efficiency that are representative of HgCdTe 
detectors \citep{morgan&siegler2015}.  Specifically, we use 
$D_{\rm{e}^{-}}=10^{-3}$~s$^{-1}$, $R_{\rm{e}^{-}}=1$, and $q=0.85$.  The visible 
wavelength detector is the same as our baseline model, except we adopt a more 
optimistic dark current of $D_{\rm{e}^{-}}=10^{-4}$~s$^{-1}$.

For the adopted LUVOIR architecture and for Earth twins at 10~pc, most of the 
wavelength-$T_{\rm{eff}}$ parameter space is dominated by noise from dark current, 
which is not influenced by telescope diameter.  If not for the larger dark current rates 
of the near-infrared HgCdTe detector, exozodiacal light would be the dominant noise 
source at longer wavelengths, owing to its strong wavelength dependence 
($\propto \lambda^{4}F_{\rm{s},\lambda}$).  Note also that the wavelength-dependent 
IWA at $3\lambda/D$ limits our ability to obtain a full spectrum through the near-infrared 
for all but the hottest stars.  For Earth twins at 5~pc, the IWA restrictions are less strict 
(although the OWA can restrict observations at quadrature for stars hotter than 5,500~K), 
and leakage from the star can dominate over other noise sources at visible wavelengths 
for stars hotter than about 6,000~K.  This all assumes that the telescope system can be 
cooled to a sufficiently low temperature to minimize thermal noise contributions to 
observations in the near-infrared (i.e., below about 80~K).  As an example of the 
capabilities of a LUVOIR mission, Figure~\ref{fig:spec_luvoir} shows a simulated 
observation of an Earth twin orbiting a solar twin, assuming the system is at 10~pc and 
an integration time of 200~hr.  Note that IWA constraints cut off the observations at the 
longest wavelengths.

\section{Discussion}

Direct characterization of exoplanets in reflected light will require moving beyond 
V-band detections, and will utilize a wide range of wavelengths.  This added 
spectral dimension influences our understanding of spectrographic observations with 
coronagraph-equipped telescopes in a variety of ways.  First, planet-star flux ratios 
are a strong function of wavelength, owing to the presence of gaseous and aerosol 
absorption/extinction features \citep[e.g.,][]{marleyetal1999,sudarskyetal03,burrowsetal04,
cahoyetal2010}.  Thus, while a coronagraph may provide a raw contrast, or achieve a 
contrast floor, that can detect certain planet types at continuum wavelengths, this 
performance may not permit adequate SNRs in molecular absorption bands, thereby 
impeding characterization attempts.

\begin{figure}
  \centering
  \begin{tabular}{cc}
    \includegraphics[trim = 4mm 2mm 2mm 3mm, clip, width=3.1in]{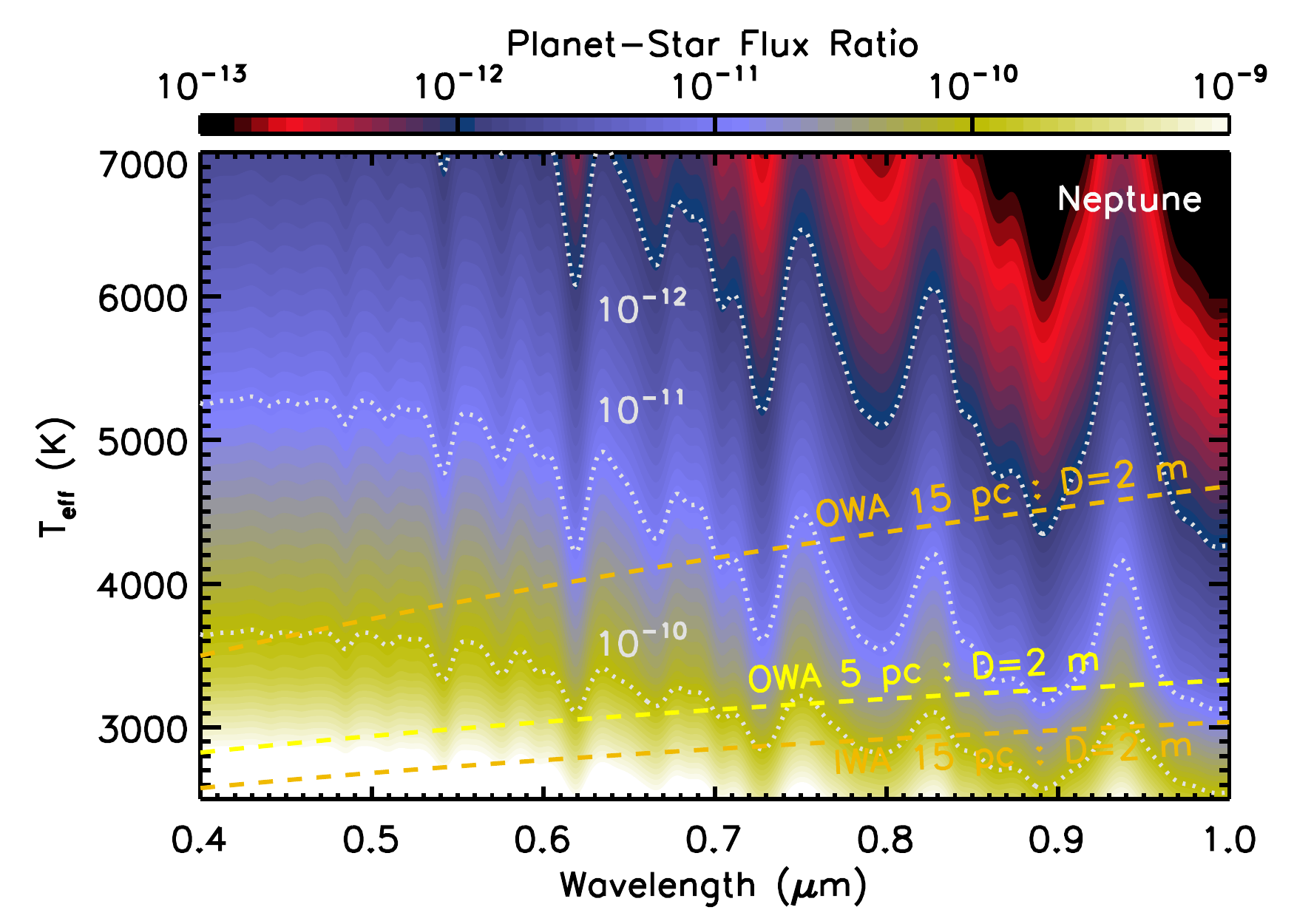} &
    \includegraphics[trim = 4mm 2mm 2mm 3mm, clip, width=3.1in]{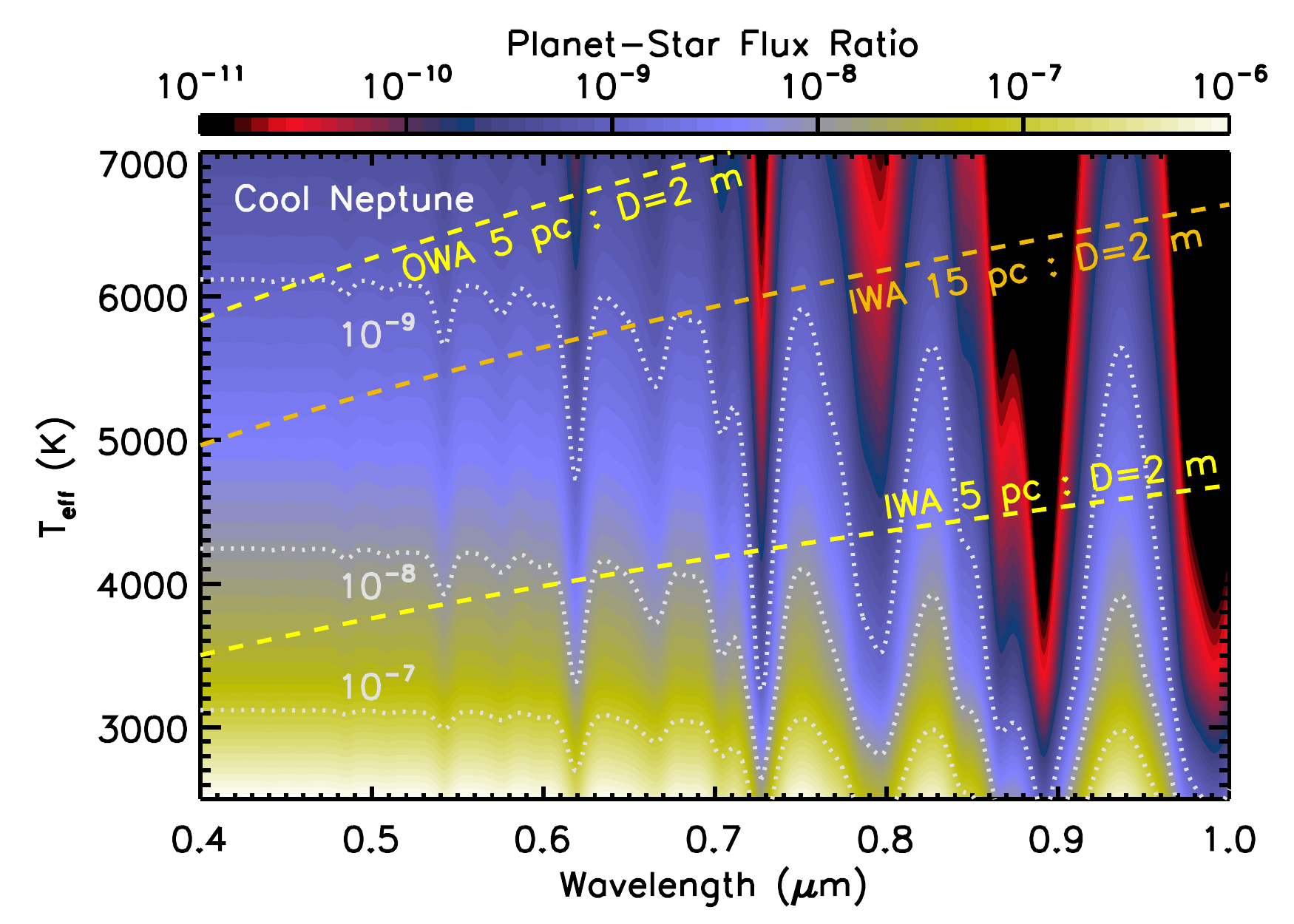} \\
    \includegraphics[trim = 4mm 2mm 2mm 3mm, clip, width=3.1in]{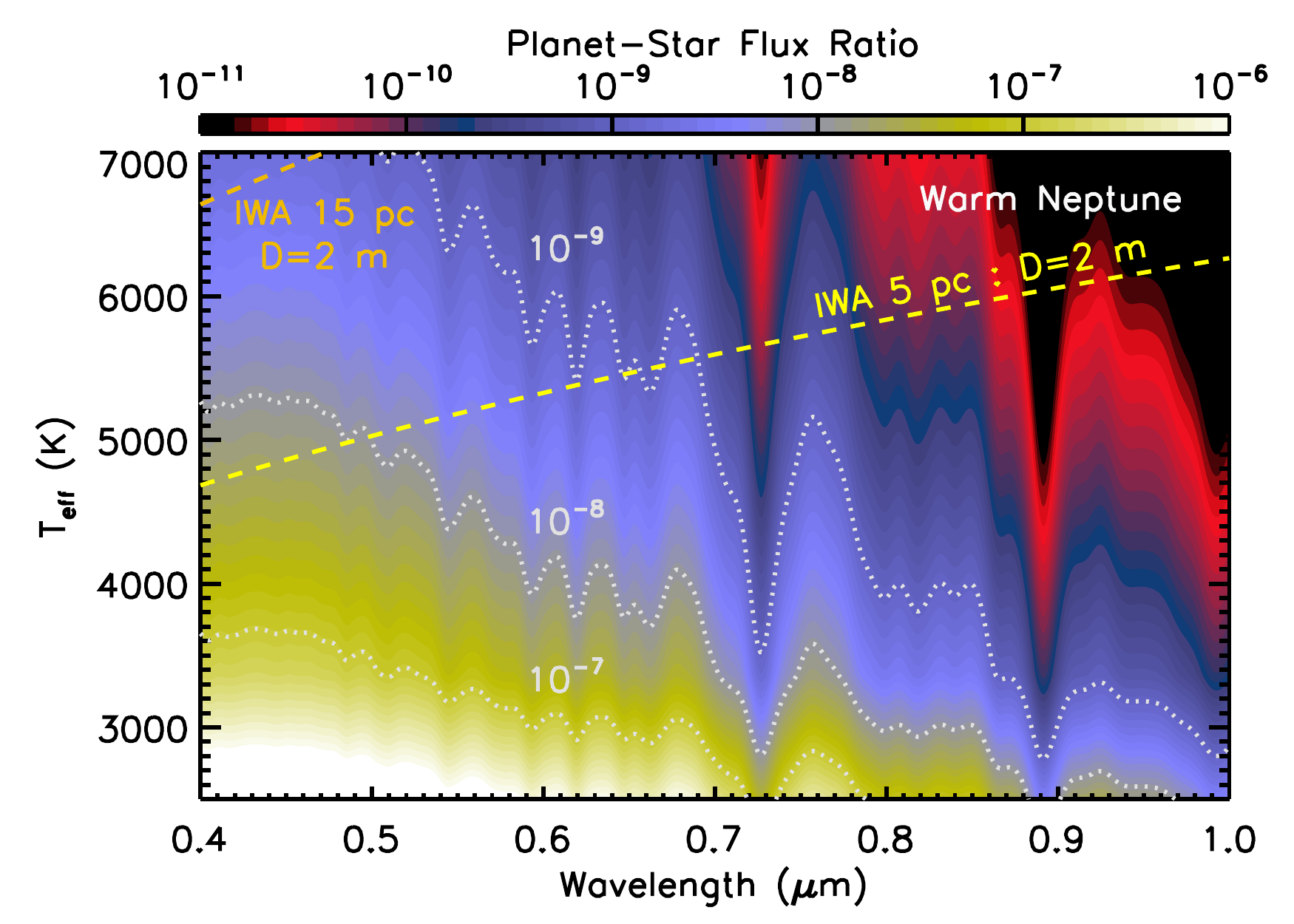} \\
  \end{tabular}
  \caption{Same as Figure~\ref{fig:contrast_jupiters}, but for Neptune-sized worlds.  
                Planet types are: (a) a Neptune twin with $F_{\rm{TOA}}=1.5$~W~m$^{-2}$ 
                and reflectivity from \citet{karkoschka1998} (top-left), (b) a cool Neptune with 
                $F_{\rm{TOA}}=342$~W~m$^{-2}$ and reflectivity from \citet{cahoyetal2010}  
                (top-right), and (c) a warm Neptune with $F_{\rm{TOA}}=2140$~W~m$^{-2}$ 
                and reflectivity from \citet{cahoyetal2010}  (bottom-left).  Inner and outer working 
                angle constraints are also as in Figure~\ref{fig:contrast_jupiters}.}
  \label{fig:contrast_neptunes}
\end{figure}

\begin{figure}
  \centering
  \begin{tabular}{cc}
    \includegraphics[trim = 4mm 2mm 2mm 3mm, clip, width=3.1in]{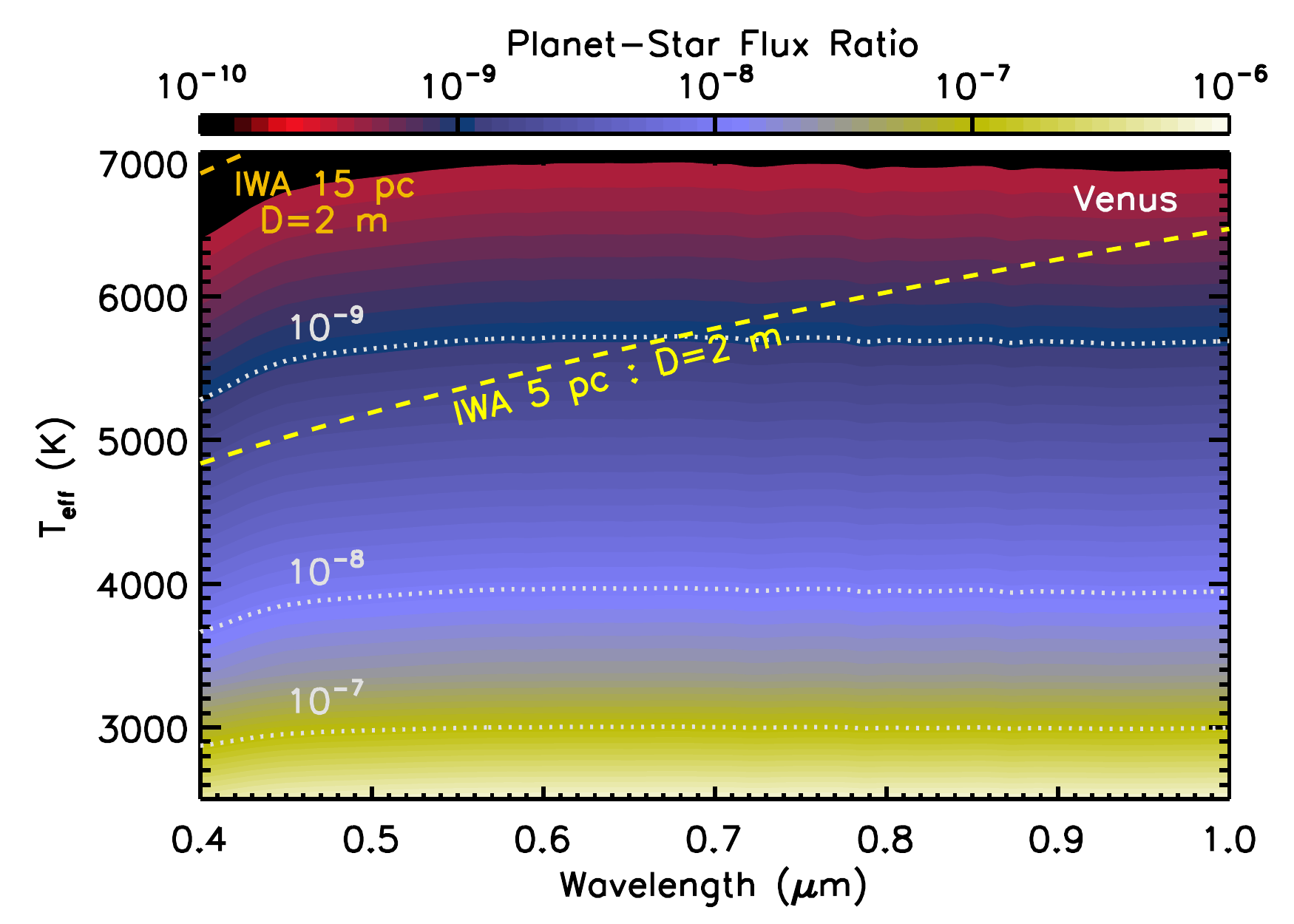} &
    \includegraphics[trim = 4mm 2mm 2mm 3mm, clip, width=3.1in]{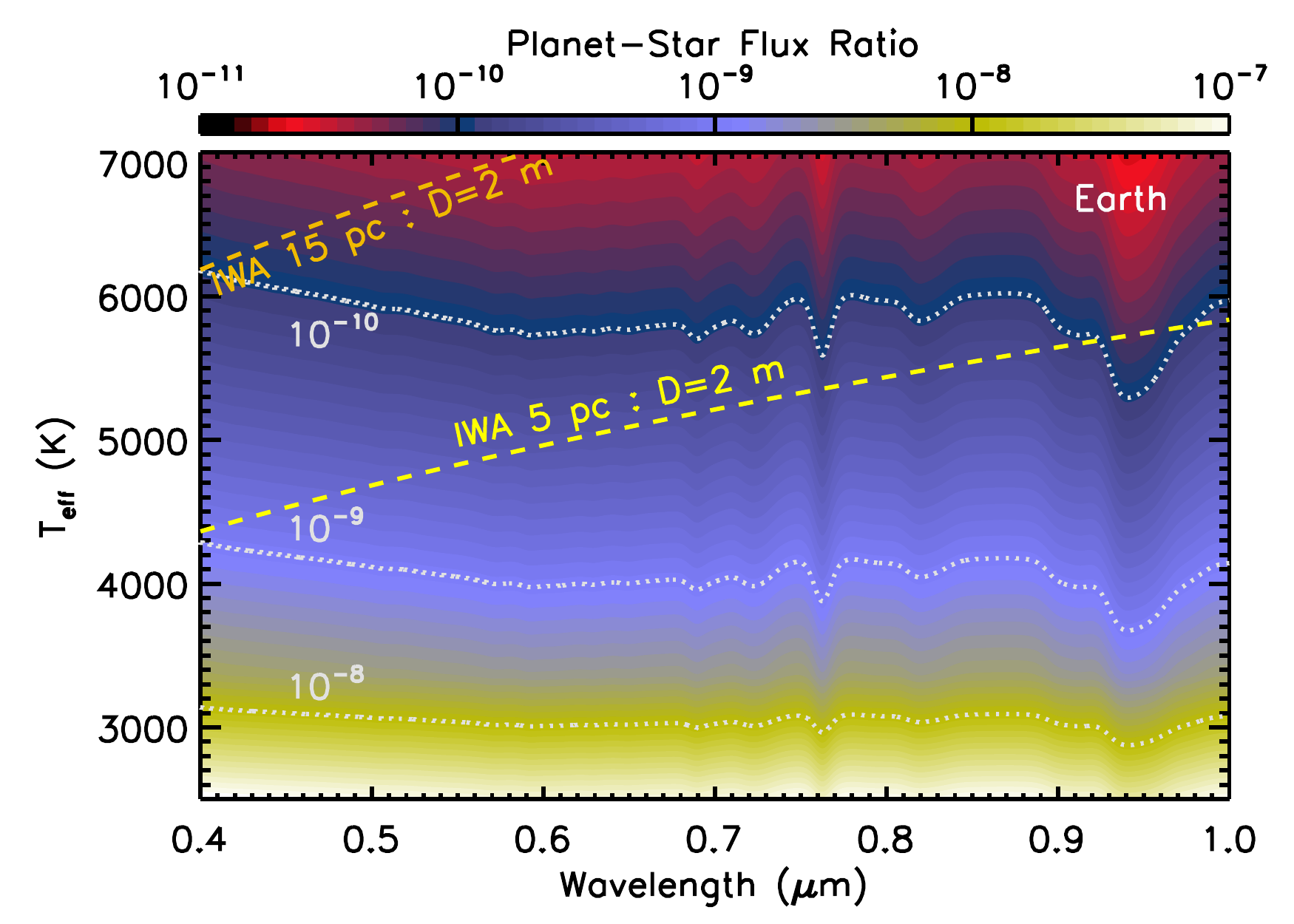} \\
    \includegraphics[trim = 4mm 2mm 2mm 3mm, clip, width=3.1in]{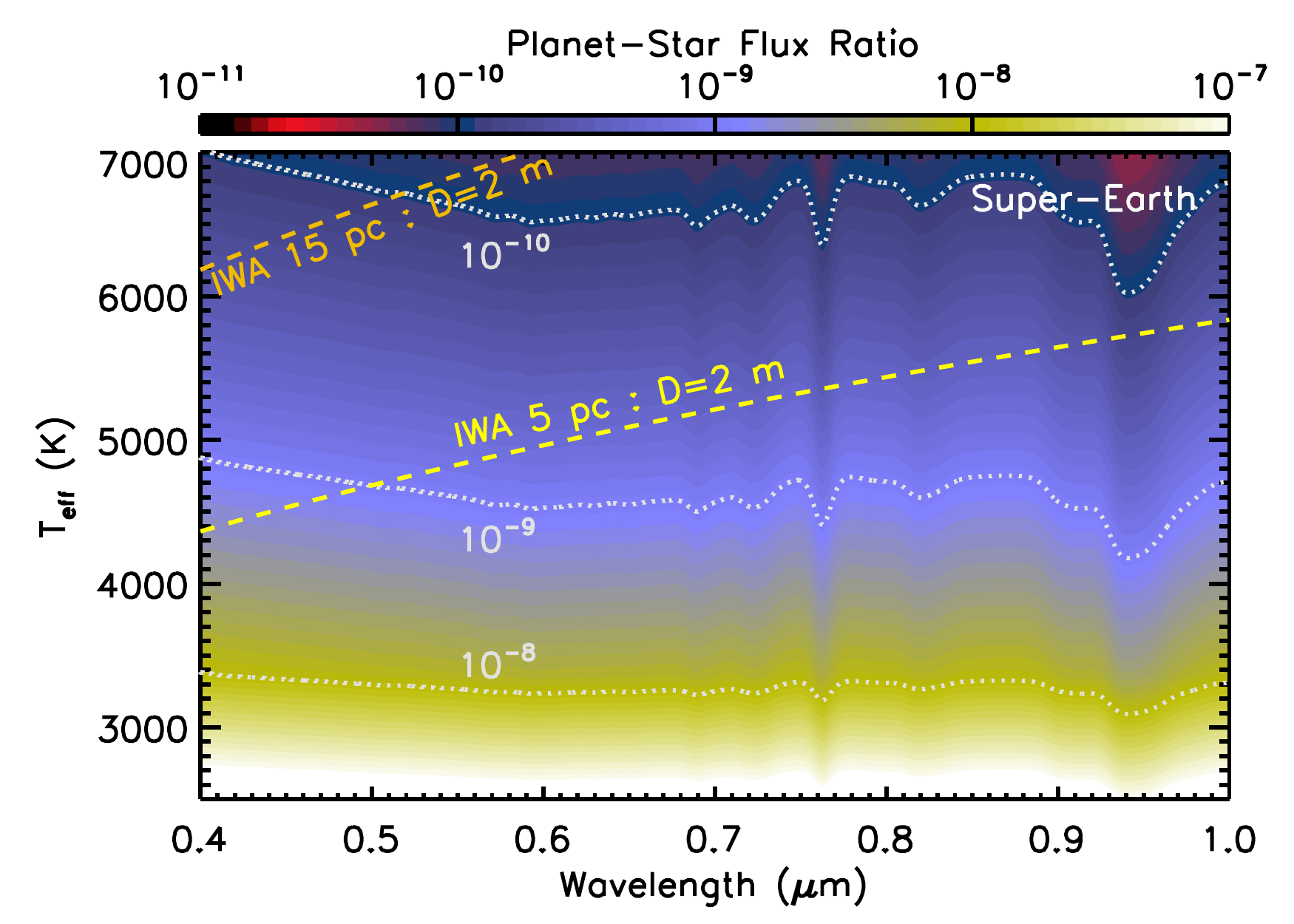} \\
  \end{tabular}
  \caption{Same as Figure~\ref{fig:contrast_jupiters}, but for terrestrial worlds.  
                Planet types are: (a) a Venus twin with $F_{\rm{TOA}}=2610$~W~m$^{-2}$ 
                and reflectivity from an application of a line-by-line radiative transfer model 
                \citep{meadows&crisp1996} to Venus \citep{arneyetal2014} (top-left) (b) an 
                Earth twin with $F_{\rm{TOA}}=1370$~W~m$^{-2}$ and reflectivity from 
                \citet{robinsonetal2010} (top-right), and (c) a super-Earth with 
                $1.5R_{\oplus}$, $F_{\rm{TOA}}=1370$~W~m$^{-2}$ and Earth-like 
                reflectivity from \citet{robinsonetal2010} (bottom-left).  Inner and outer 
                working angle  constraints are also as in Figure~\ref{fig:contrast_jupiters}.}
  \label{fig:contrast_terrests}
\end{figure}

\begin{figure}
  \centering
  \begin{tabular}{cc}
    \includegraphics[trim = 4mm 2mm 2mm 3mm, clip, width=3.1in]{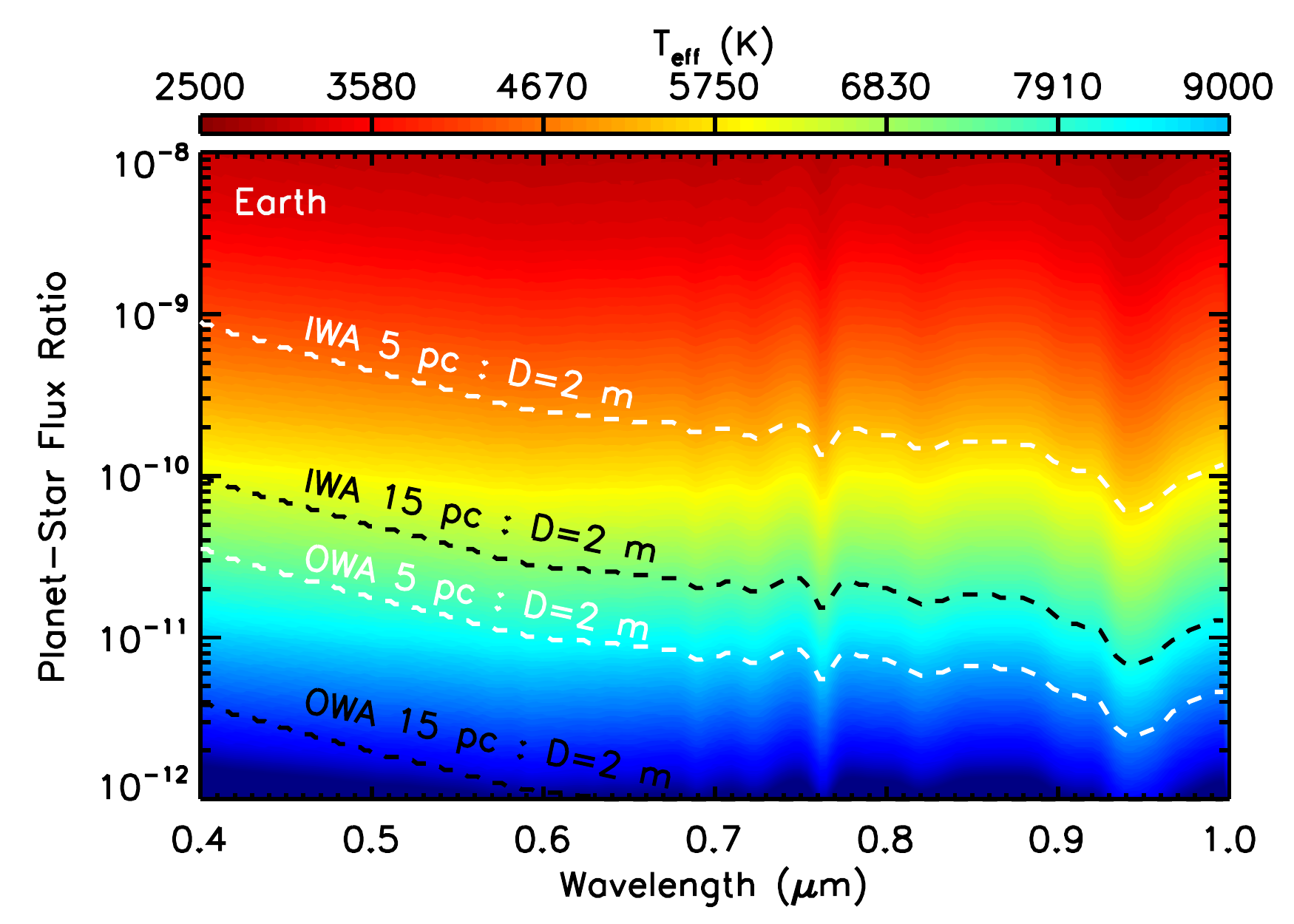} &
    \includegraphics[trim = 4mm 2mm 2mm 3mm, clip, width=3.1in]{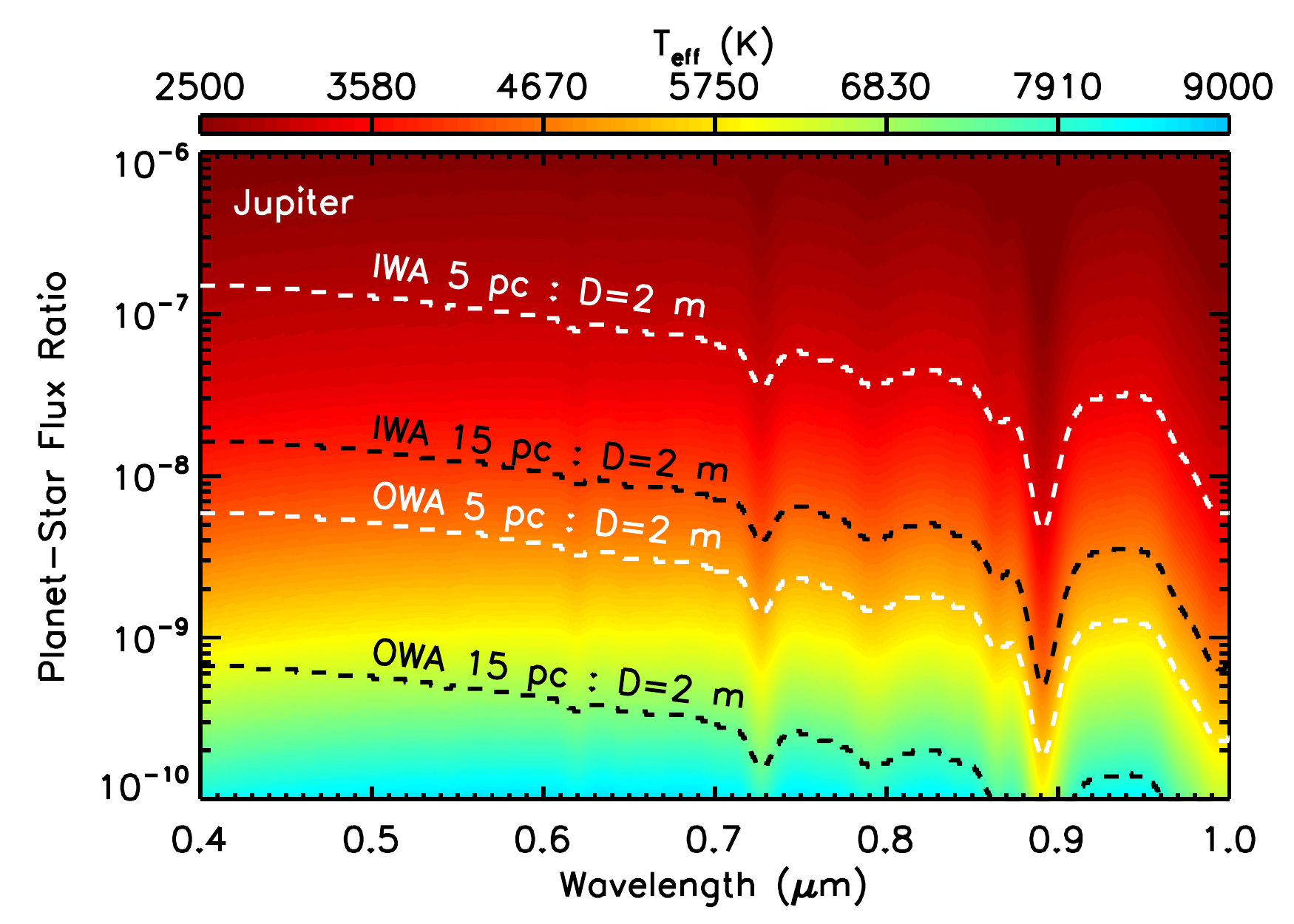} \\
    \includegraphics[trim = 4mm 2mm 2mm 3mm, clip, width=3.1in]{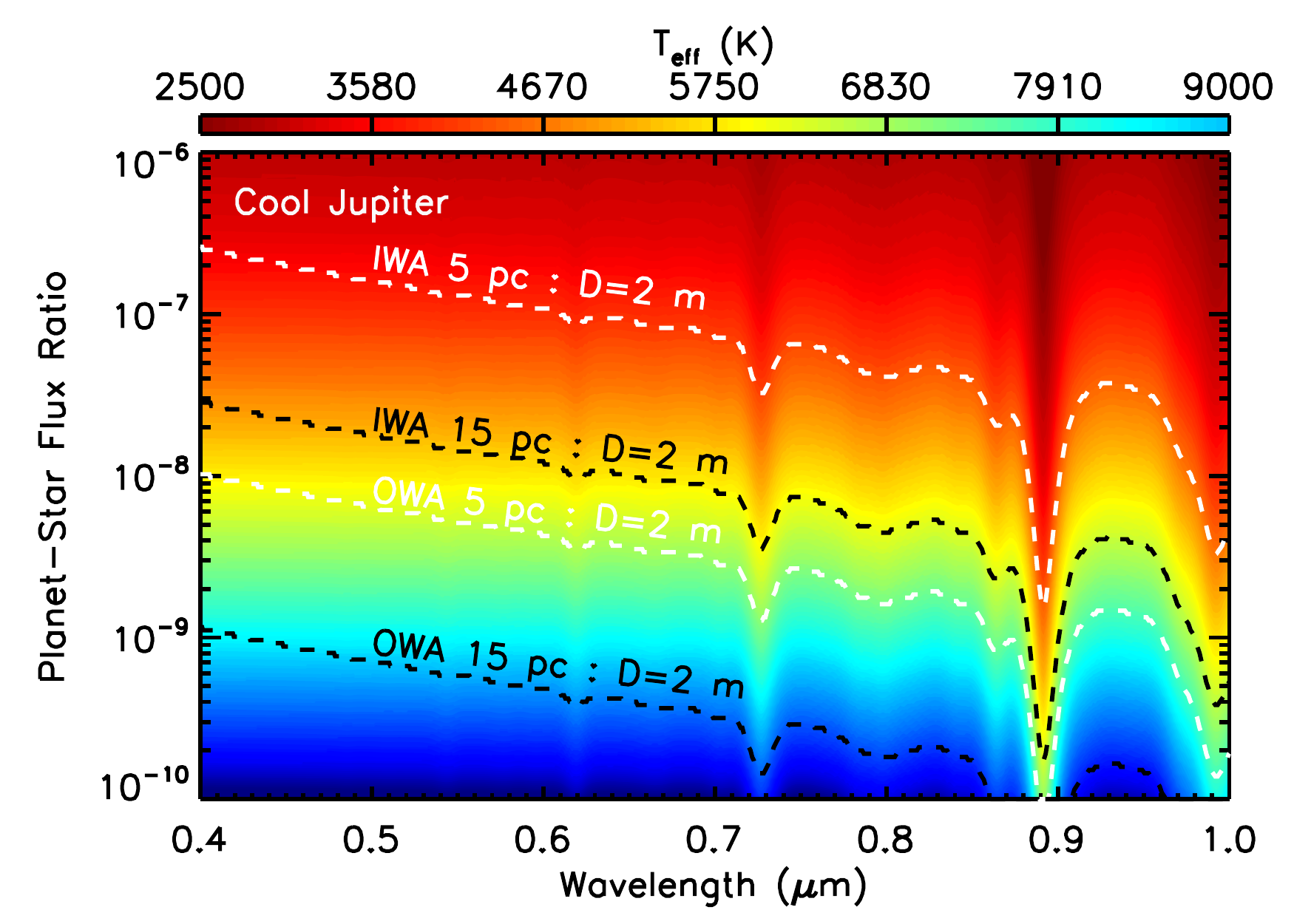} &
    \includegraphics[trim = 4mm 2mm 2mm 3mm, clip, width=3.1in]{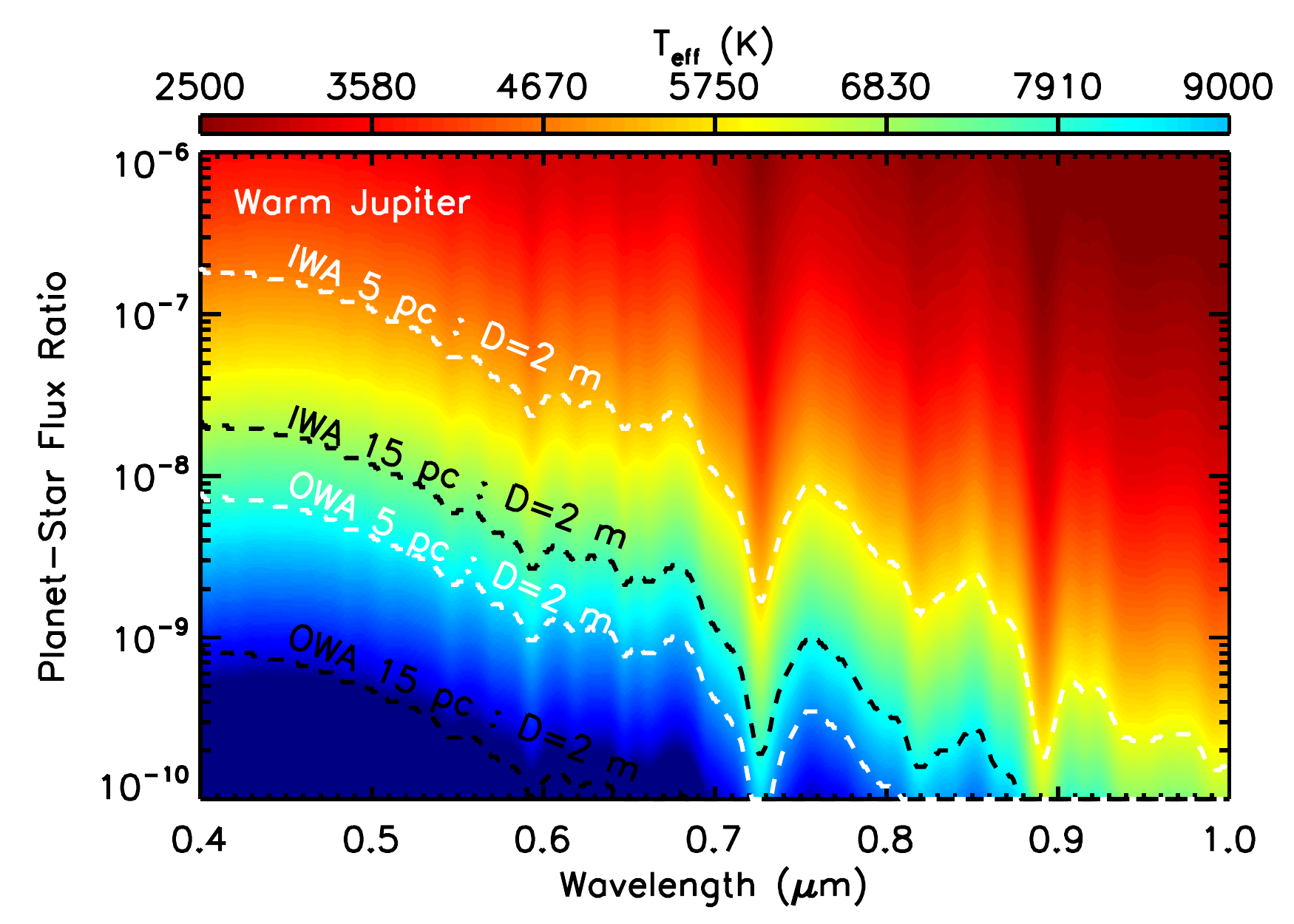} \\
  \end{tabular}
  \caption{Contours of stellar effective temperature around which a certain planet 
                type achieves a given planet-star flux ratio.  Jupiter-sized cases are the 
                same as Figure~\ref{fig:contrast_jupiters}, while the Earth case is as in 
                Figure~\ref{fig:contrast_terrests}.  Inner and outer working angle 
                constraints are also shown, and follow those in Figure~\ref{fig:contrast_jupiters}.}
  \label{fig:teff_contours}
\end{figure}

Second, as coronagraphs have inner and outer working angles that depend on 
wavelength, the variety of planets accessible for observation will depend on 
distance from the Solar System.  The interplay between planet-star flux ratio, 
wavelength, host star effective temperature, and inner and outer working angle 
constraints are depicted for different planet types in 
Figures~\ref{fig:contrast_jupiters}--\ref{fig:contrast_terrests}.  These contour 
diagrams are extremely useful for understanding the types of observations that 
could be made by a given mission architecture and, to our knowledge, are the 
first of their kind.  Additionally, these diagrams depict an important reality of 
exoplanet spectral characterization: wavelength-dependent instrument inner 
and/or outer working angles can cause portions of a exoplanetary spectrum 
to be unobservable, depending on the apparent planet-star separation at the 
time of observation.

Recall that we have defined a planet ``type'' based, in part, on insolation and 
size.  Thus, Figures~\ref{fig:contrast_jupiters}--\ref{fig:contrast_terrests} show 
that a $2\lambda/D$ IWA for a 2-meter class telescope will strongly limit our 
ability to study warm Jupiters and Neptunes (at distances equivalent to 0.8~AU 
from the Sun) and Venus-like planets.  Cool Jupiters and Neptunes (at distances 
equivalent to 2~AU from the Sun) are much more accessible.  Interestingly, the 
presence of thick water clouds in the atmospheres of the cool Jupiters and 
Neptunes make these worlds very reflective compared to their warm 
counterparts \citep{sudarskyetal2000,cahoyetal2010}.  Hence, the planet-star flux 
ratios in Figures~\ref{fig:contrast_jupiters} and \ref{fig:contrast_neptunes} are 
broadly similar for these planet types, even though the insolation levels of the 
warm planet types is over six times larger.

\begin{figure}
  \centering
  \begin{tabular}{cc}
    \includegraphics[trim = 4mm 2mm 2mm 3mm, clip, width=3.1in]{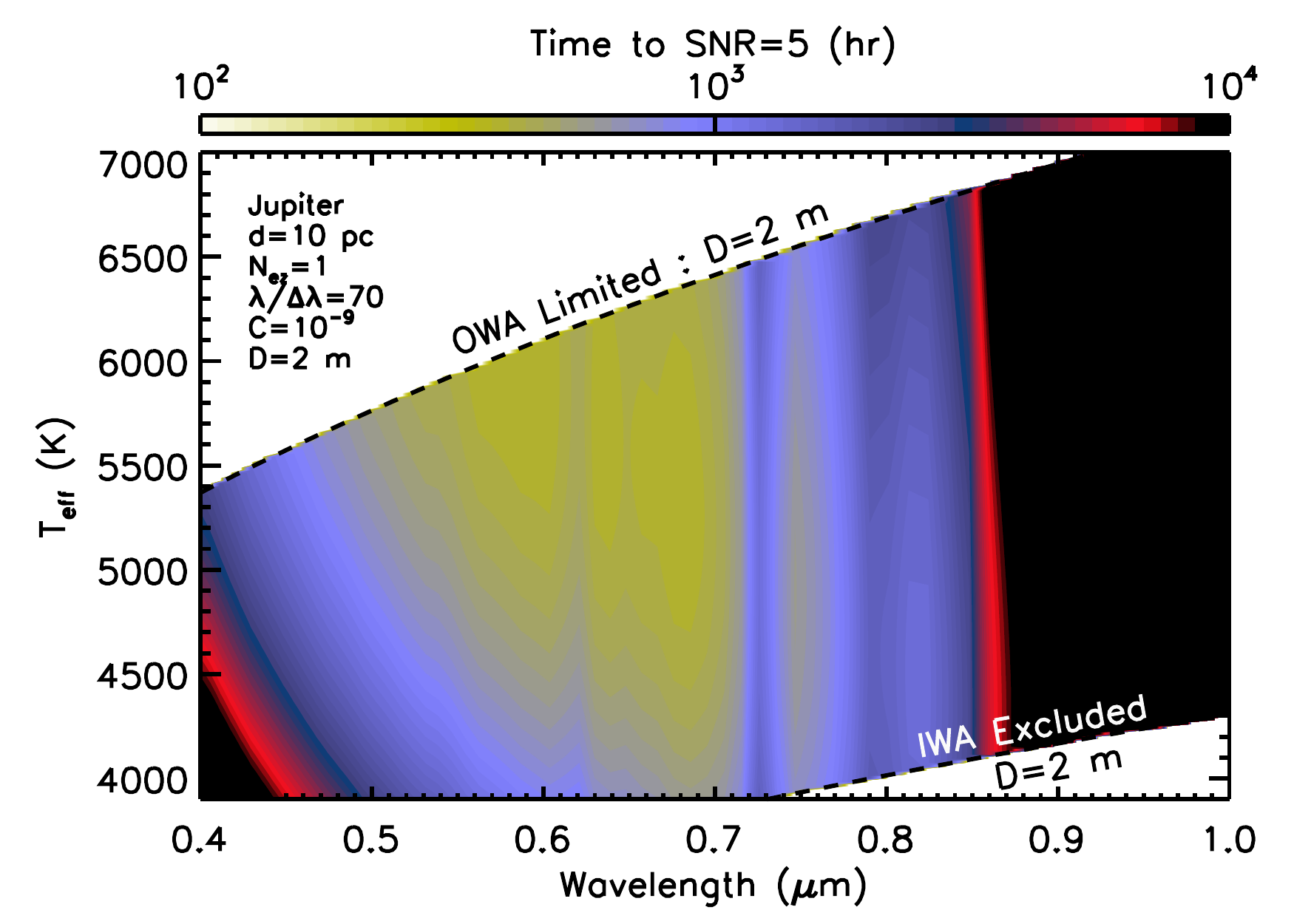} &
    \includegraphics[trim = 4mm 2mm 2mm 3mm, clip, width=3.1in]{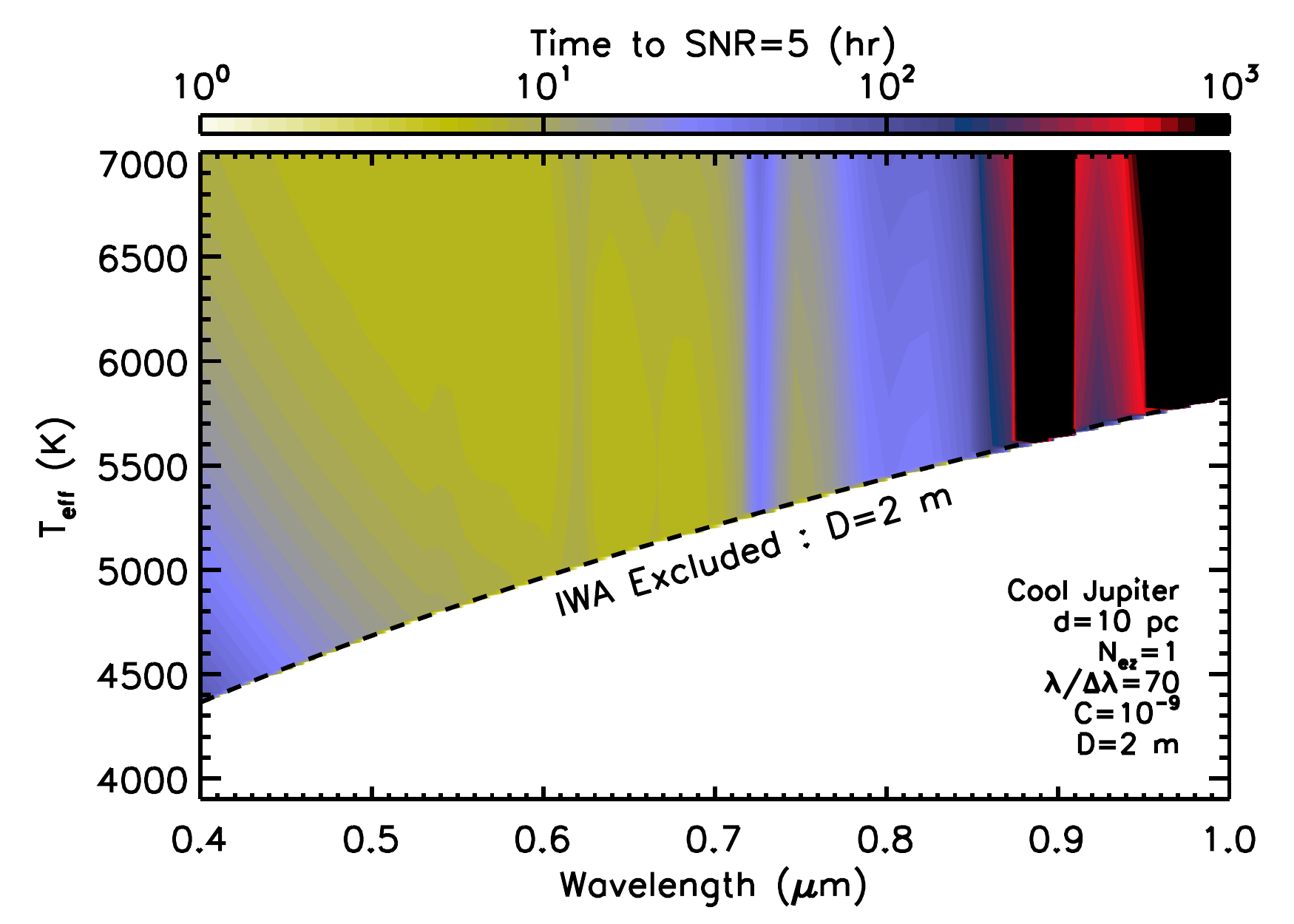} \\
  \end{tabular}
  \caption{Contours of integration time required to achieve $\rm{SNR}=5$ for a Jupiter twin 
                (left) and a cool Jupiter (right) at 10~pc.  These planet types are placed at the  
                flux equivalent distance of main sequence stars of different effective  
                temperatures, as in Figure~\ref{fig:contrast_jupiters}.  Note the two different color-bar 
                scales.  Regions excluded/limited by  a $2\lambda/D$ and $10\lambda/D$ inner and 
                outer working angle (respectively) are demarcated by a dashed line.}
  \label{fig:jupiter_Dt}
\end{figure}

\begin{figure}
  \centering
  \includegraphics[trim = 4mm 2mm 2mm 3mm, clip, width=6.2in]{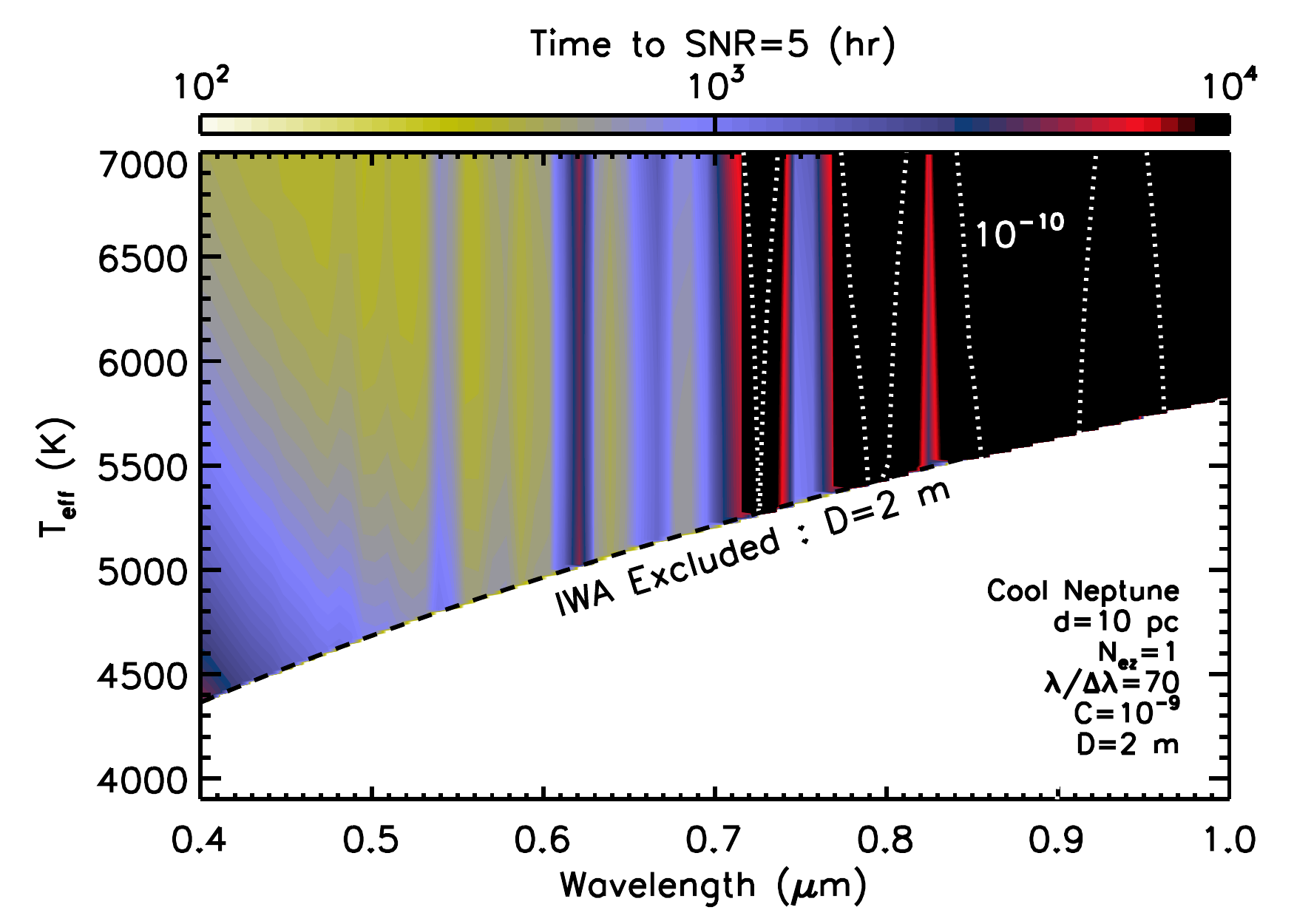}
  \caption{Same as Figure~\ref{fig:jupiter_Dt}, but for a cool Neptune placed at the 
                flux equivalent distance, as in Figure~\ref{fig:contrast_neptunes}.  Regions 
                with planet-star flux ratios smaller than $10^{-10}$ are above the labeled dotted 
                line.}
  \label{fig:neptune_Dt}
\end{figure}

\begin{figure}
  \centering
  \begin{tabular}{cc}
    \includegraphics[trim = 4mm 2mm 2mm 3mm, clip, width=3.1in]{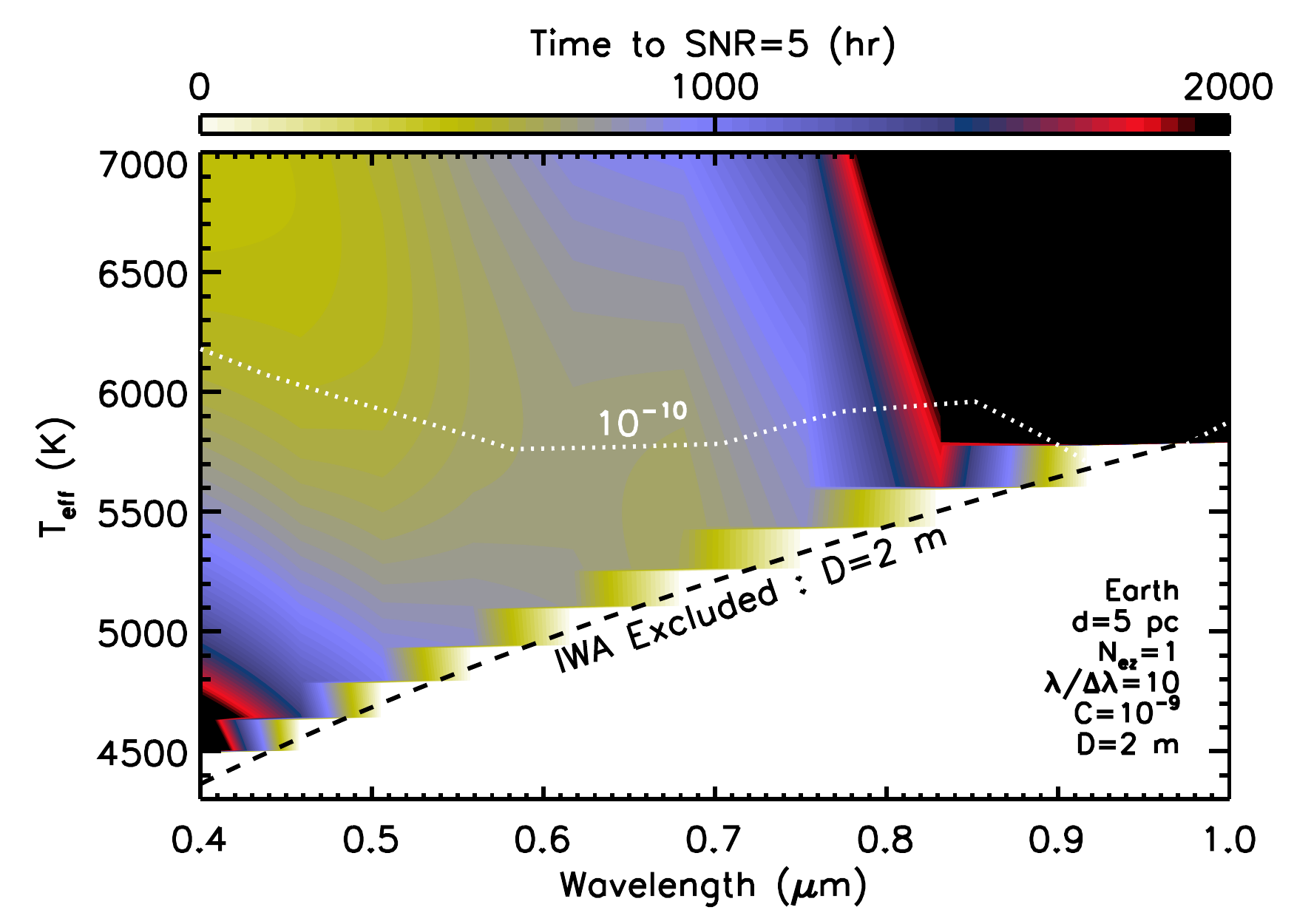} &
    \includegraphics[trim = 4mm 2mm 2mm 3mm, clip, width=3.1in]{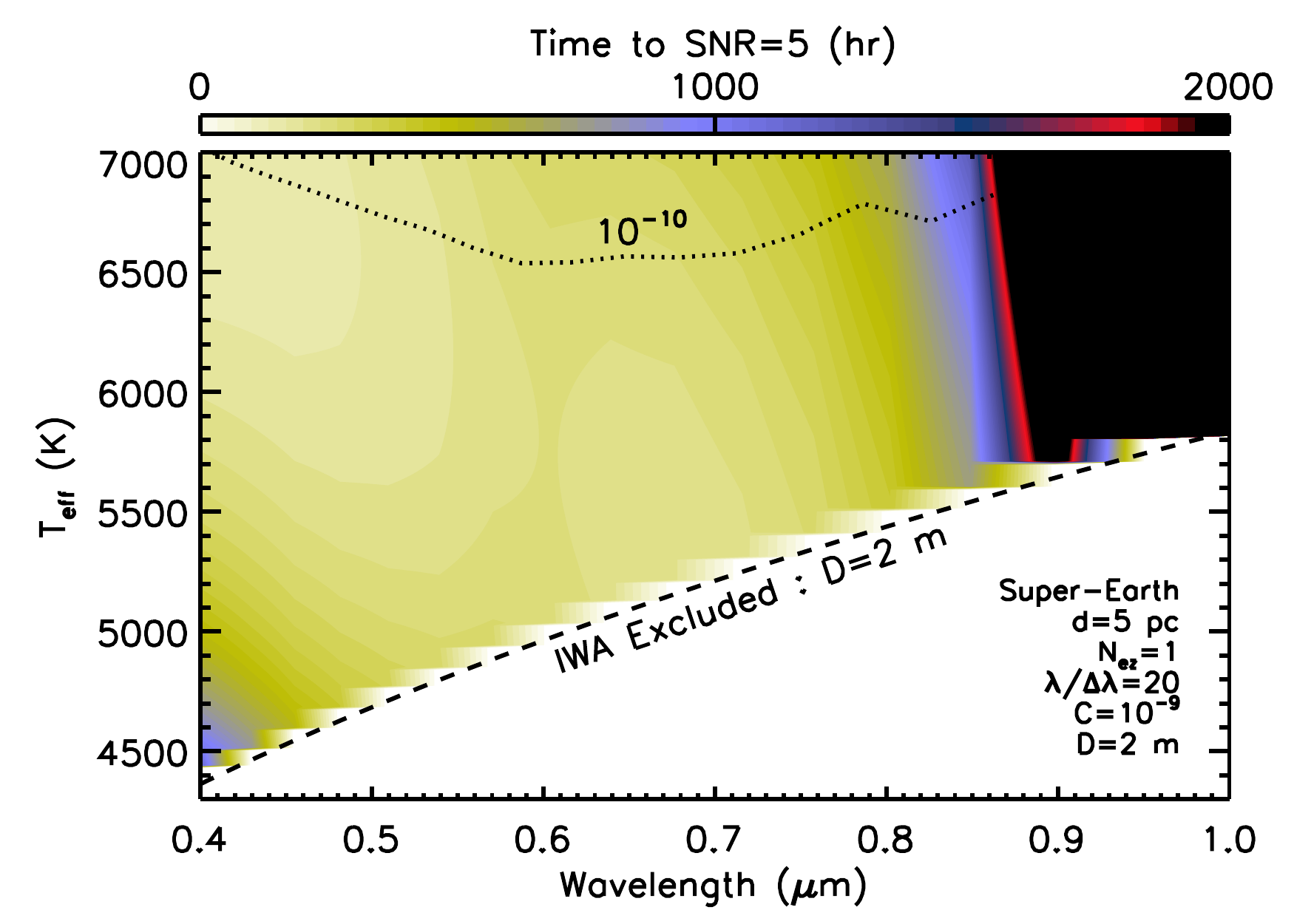} \\
  \end{tabular}
  \caption{Same as Figure~\ref{fig:jupiter_Dt}, but for an Earth twin (left) and a 
                $1.5R_{\oplus}$ super-Earth (right) at 5~pc.  Both are  
                placed at Earth's flux equivalent distance (as in 
                Figure~\ref{fig:contrast_terrests}).  To maintain reasonable integration times, 
                higher-resolution spectra ($\mathcal{R}=70$) have been degraded to $\mathcal{R}=10$ 
                and $\mathcal{R}=20$ for the Earth twins and super-Earths, respectively. Regions with 
                planet-star flux ratios smaller than $10^{-10}$ are above the labeled dotted line.}
  \label{fig:earth_Dt}
\end{figure}

\begin{figure}
  \centering
  \begin{tabular}{cc}
    \includegraphics[trim = 4mm 2mm 2mm 3mm, clip, width=3.1in]{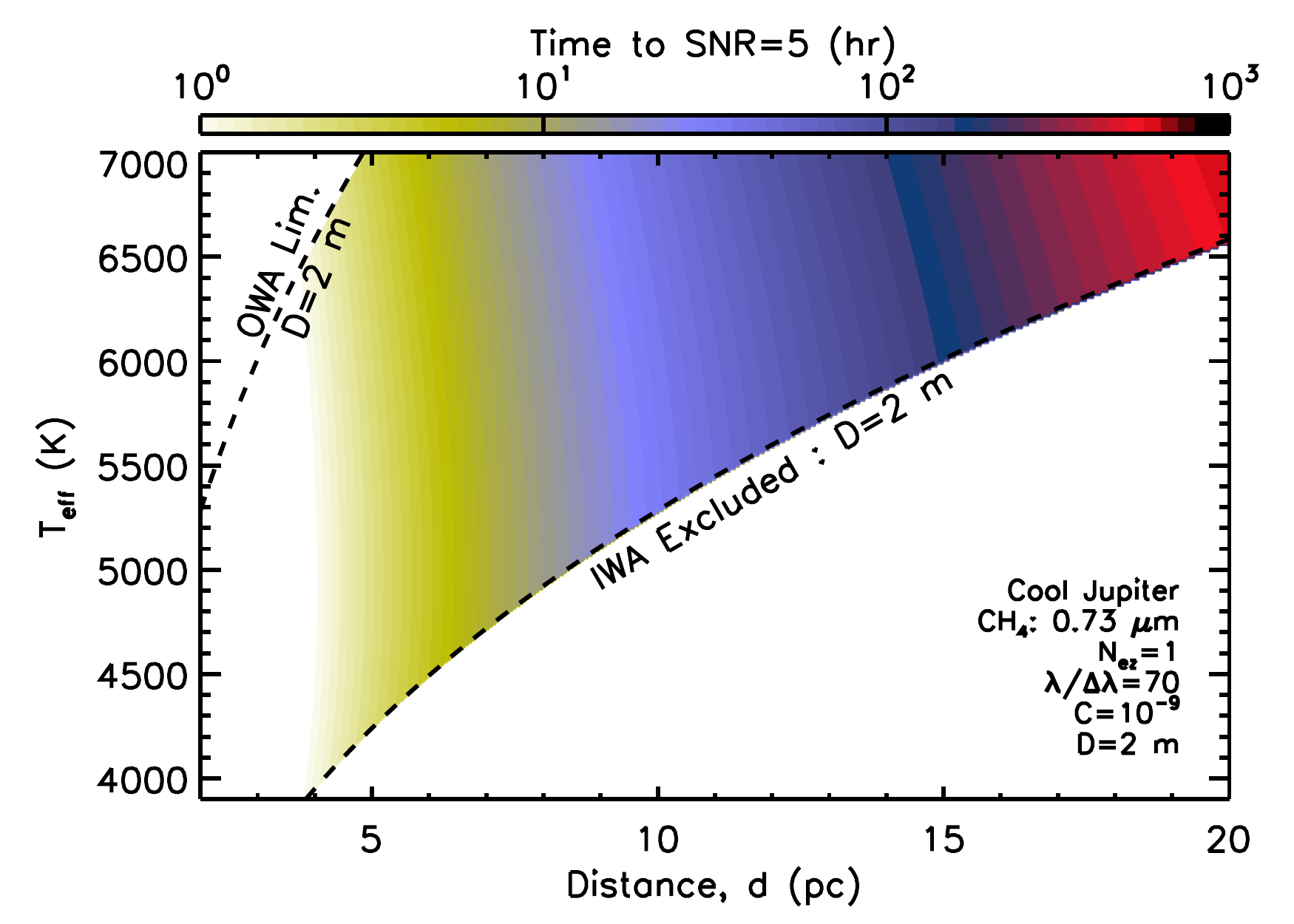} &
    \includegraphics[trim = 4mm 2mm 2mm 3mm, clip, width=3.1in]{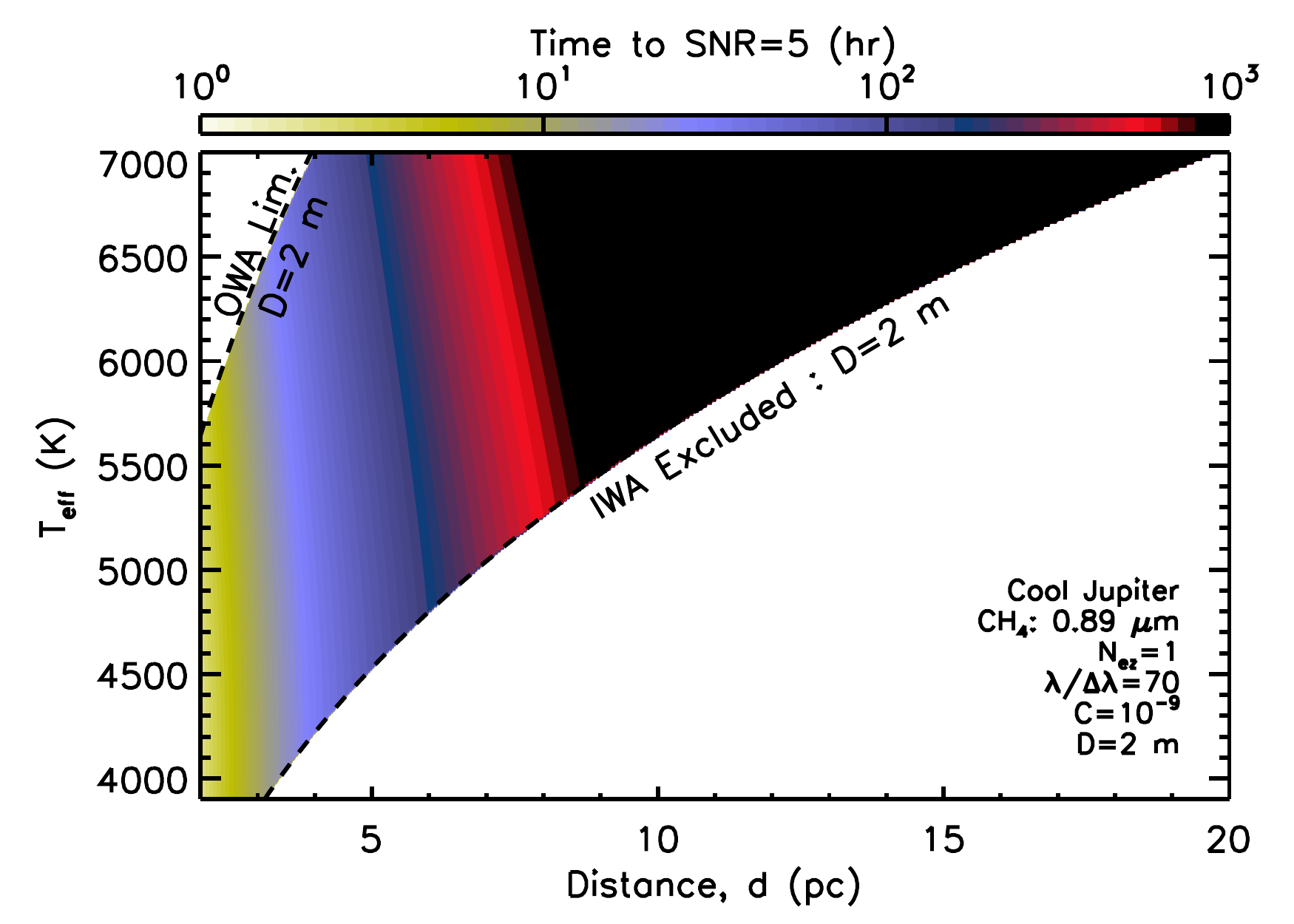} \\
    \includegraphics[trim = 4mm 2mm 2mm 3mm, clip, width=3.1in]{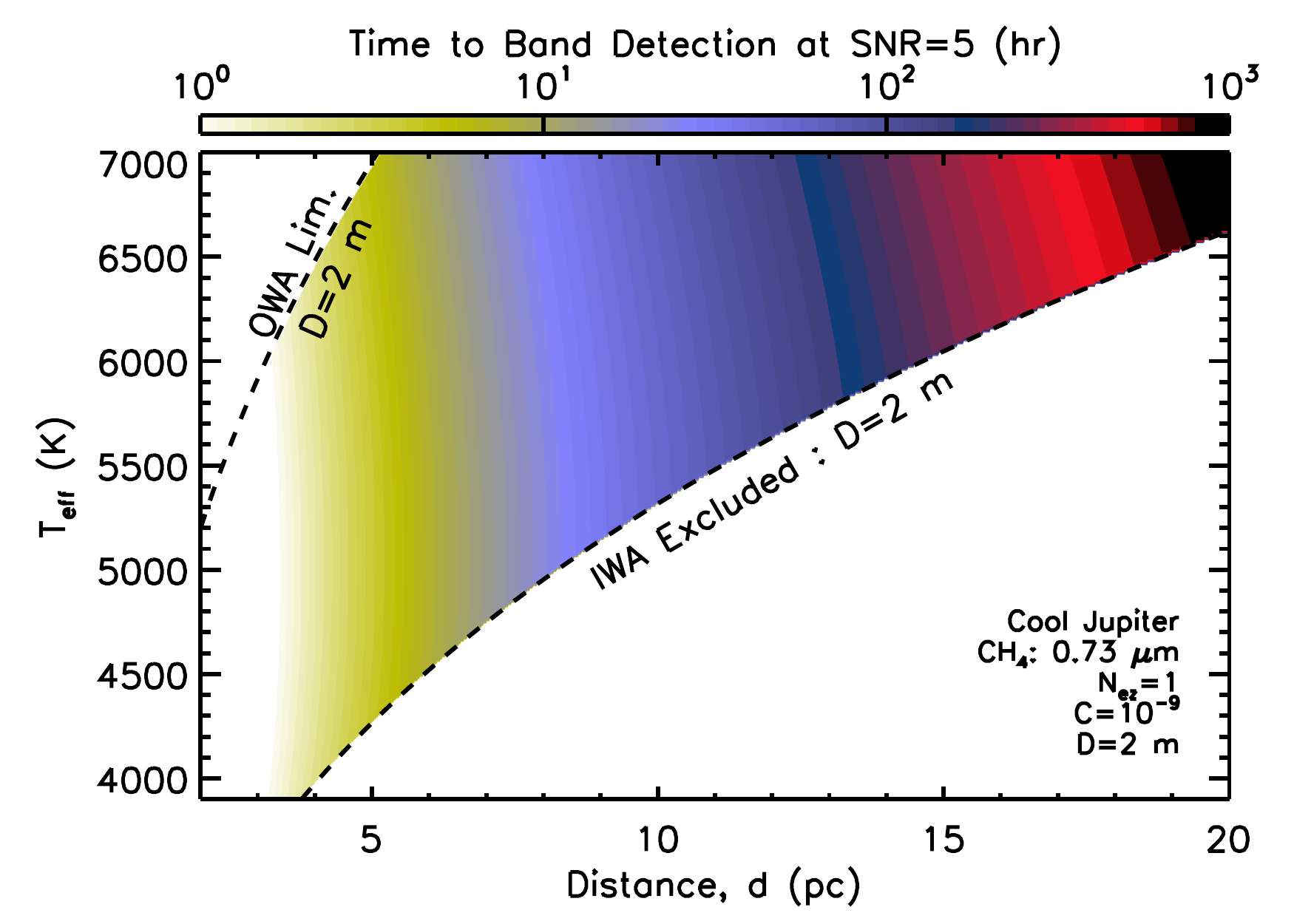} &
    \includegraphics[trim = 4mm 2mm 2mm 3mm, clip, width=3.1in]{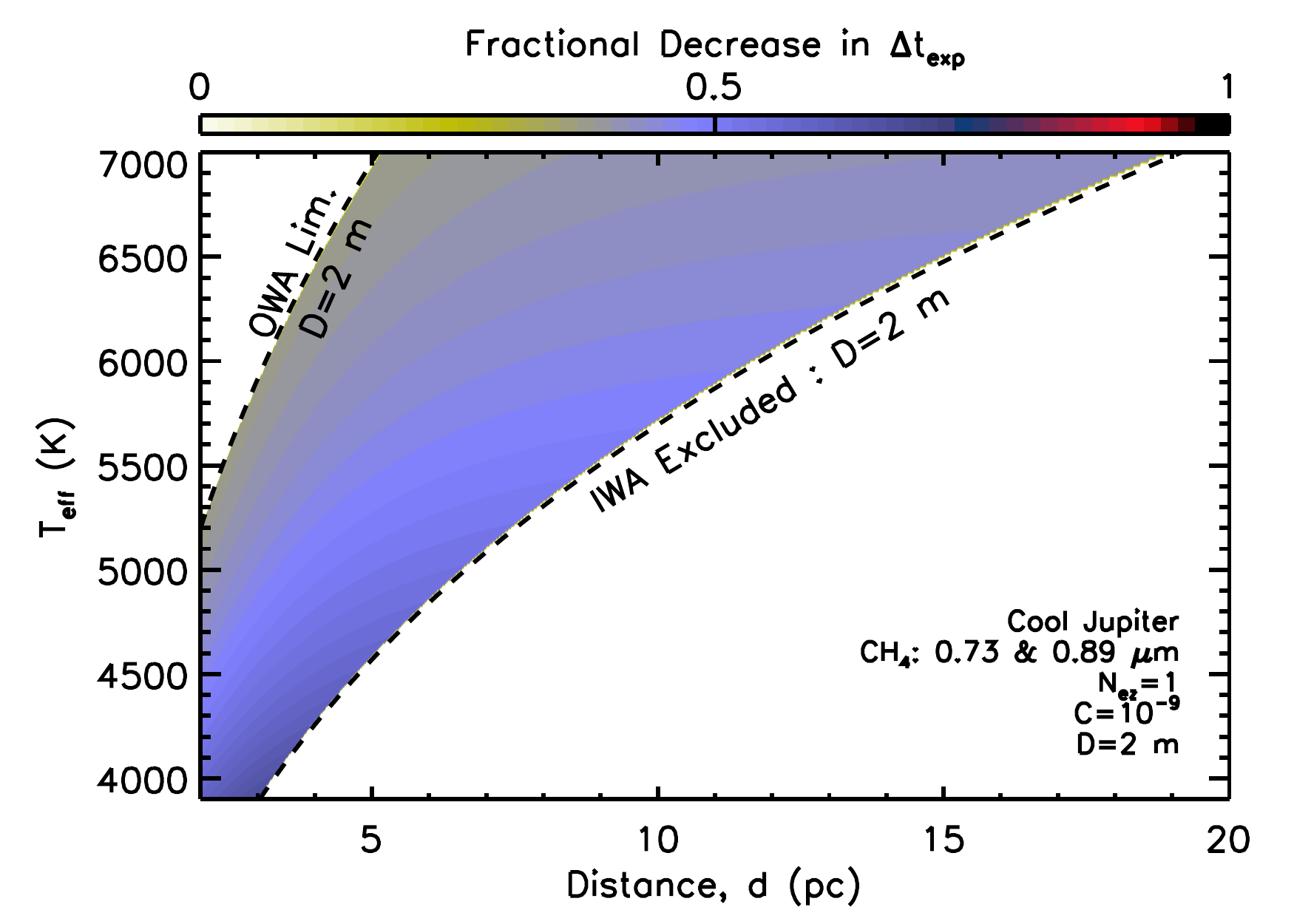} \\
  \end{tabular}
  \caption{Contours of integration time required to achieve $\rm{SNR}=5$ in the 
                bottom of key absorption bands as well as contours of required integration 
                time to detect a band at $\rm{SNR_{band}}=5$ for cool Jupiters.  We 
                highlight the 0.73~$\mu$m and 0.89~$\mu$m methane bands.  The lower 
                right plot shows how the band detection time decreases when both methane 
                features are used.  Regions excluded/limited by a $2\lambda/D$ and 
                $10\lambda/D$ inner and outer working angle (respectively) are demarcated 
                by a dashed line.}
  \label{fig:jupiter_det}
\end{figure}

\begin{figure}
  \centering
  \begin{tabular}{cc}
    \includegraphics[trim = 4mm 2mm 2mm 3mm, clip, width=3.1in]{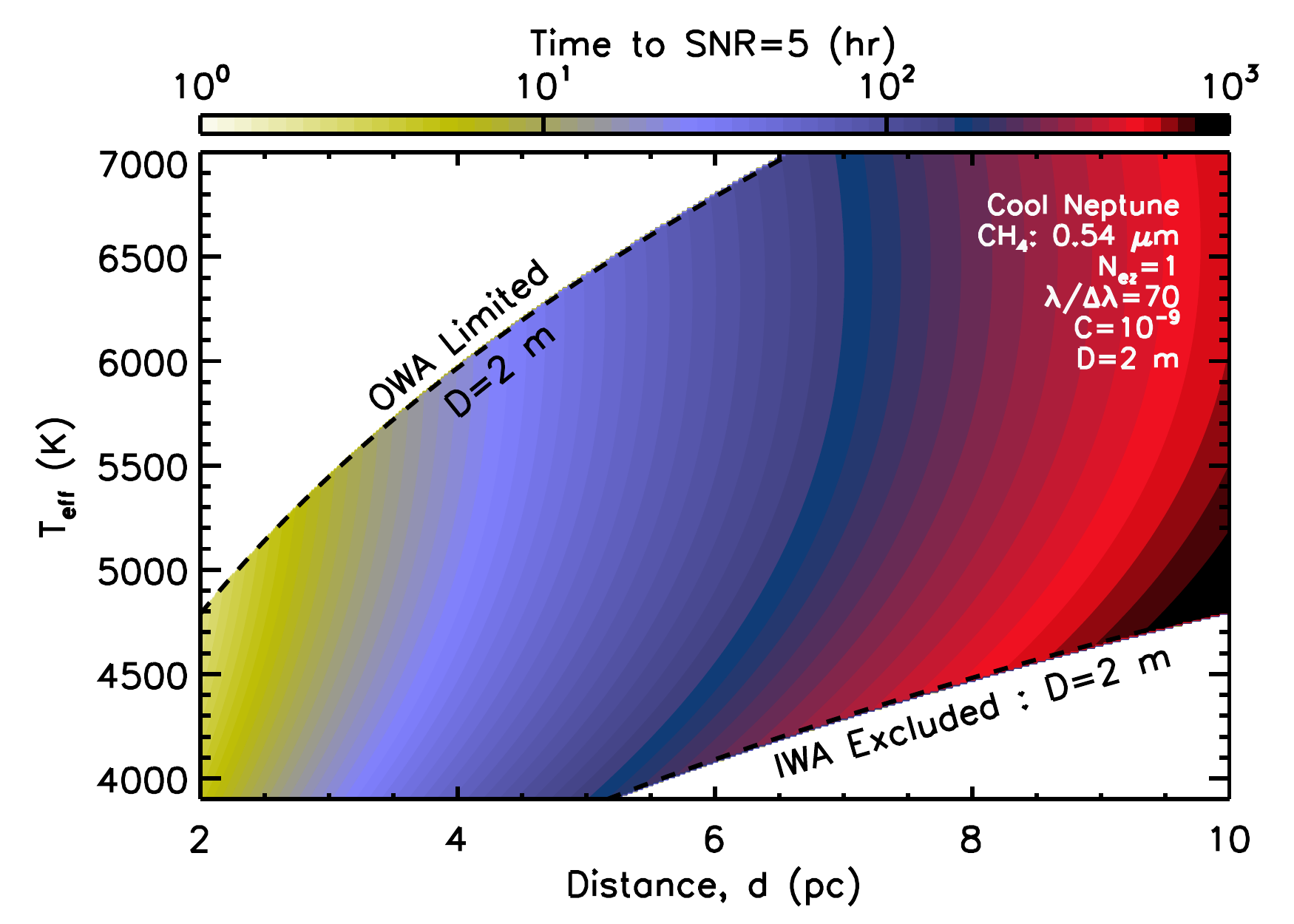} &
    \includegraphics[trim = 4mm 2mm 2mm 3mm, clip, width=3.1in]{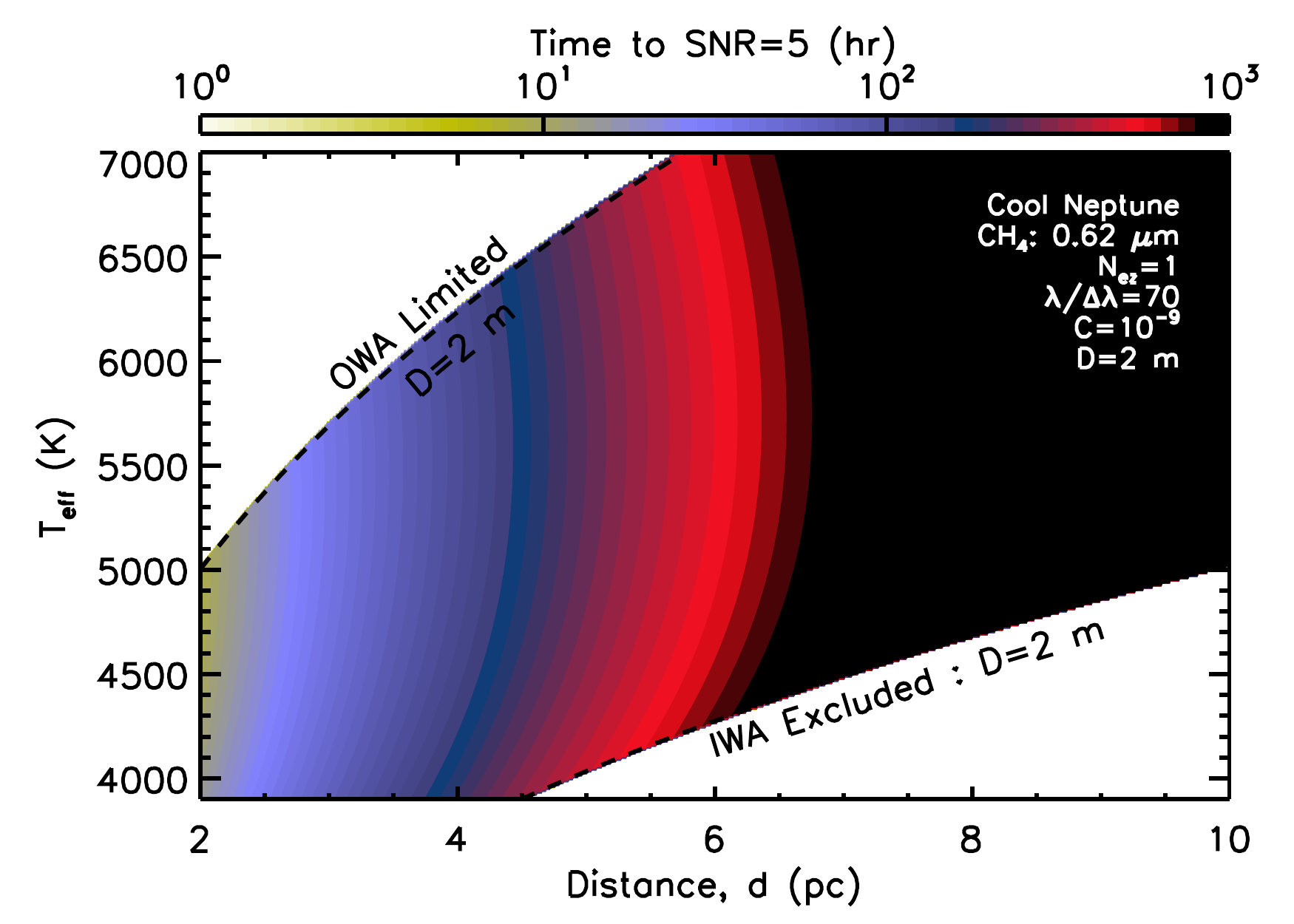} \\
    \includegraphics[trim = 4mm 2mm 2mm 3mm, clip, width=3.1in]{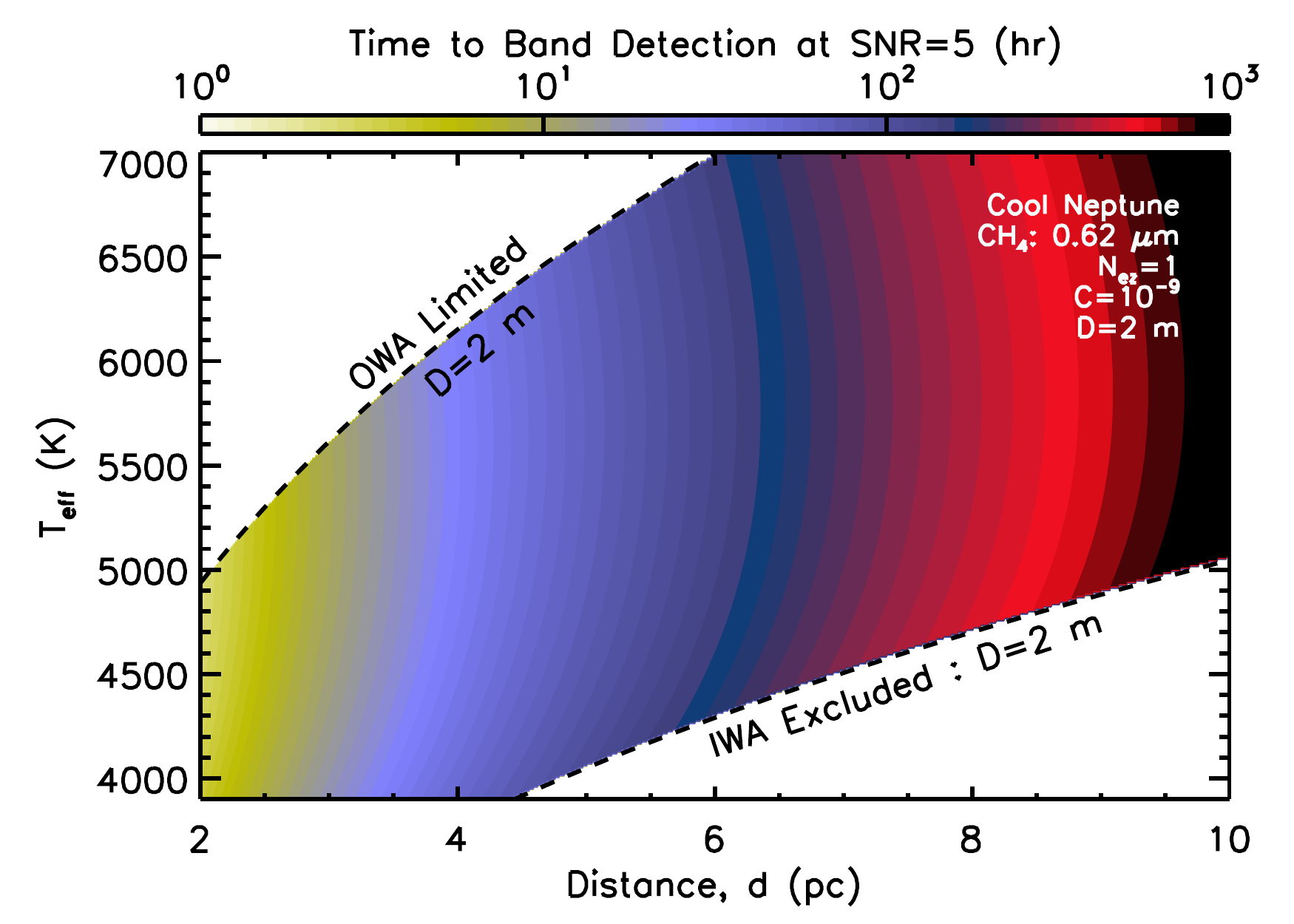} &
    \includegraphics[trim = 4mm 2mm 2mm 3mm, clip, width=3.1in]{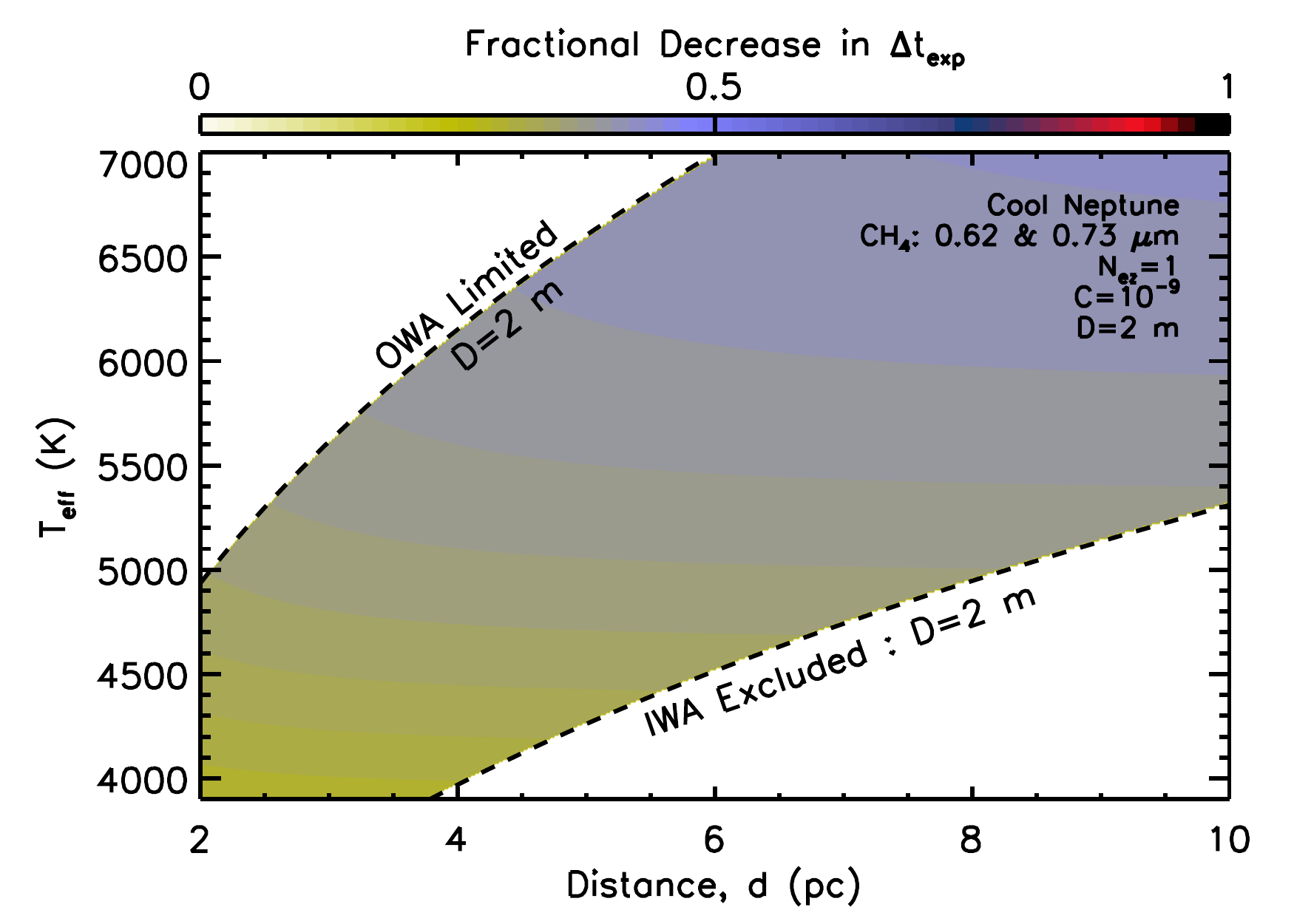} \\
  \end{tabular}
  \caption{Similar to Figure~\ref{fig:jupiter_det}, except now for cool Neptunes.  Here, 
                for detecting the bottom of features, we highlight the 0.54~$\mu$m and 
                0.62~$\mu$m methane bands, as the longer wavelength bands have 
                planet-star flux ratios of $10^{-10}$ or smaller in their bases.  For band 
                detections, we use the 0.73~$\mu$m methane band, as its depth leads to 
                shorter required integration times than bands at shorter wavelengths.  The 
                lower right plot shows how the band detection time decreases when both 
                the 0.62~$\mu$m and 0.73~$\mu$m methane features are used.}
  \label{fig:neptune_det}
\end{figure}

\begin{figure}
  \centering
  \begin{tabular}{cc}
    \includegraphics[trim = 4mm 2mm 2mm 3mm, clip, width=3.1in]{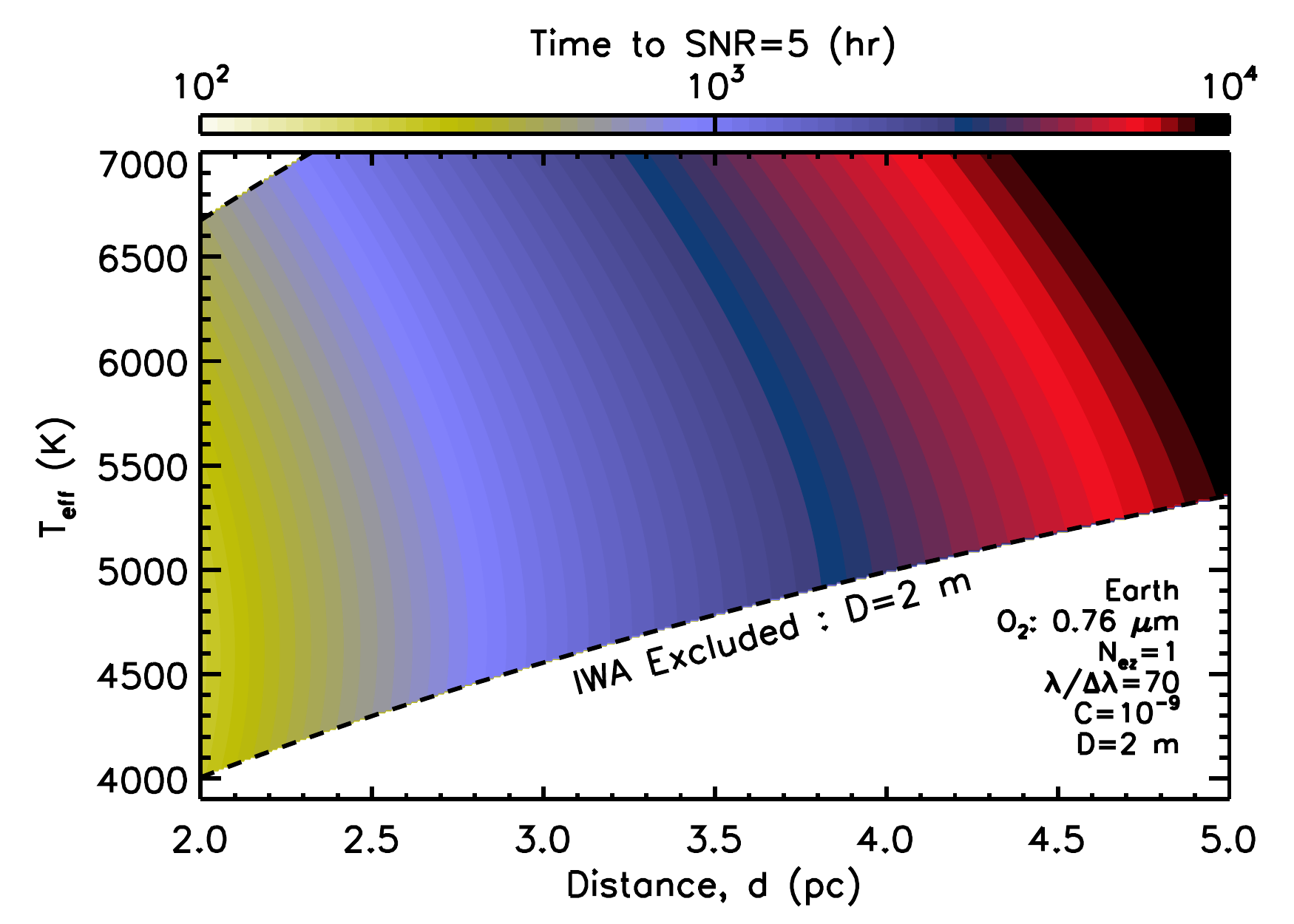} &
    \includegraphics[trim = 4mm 2mm 2mm 3mm, clip, width=3.1in]{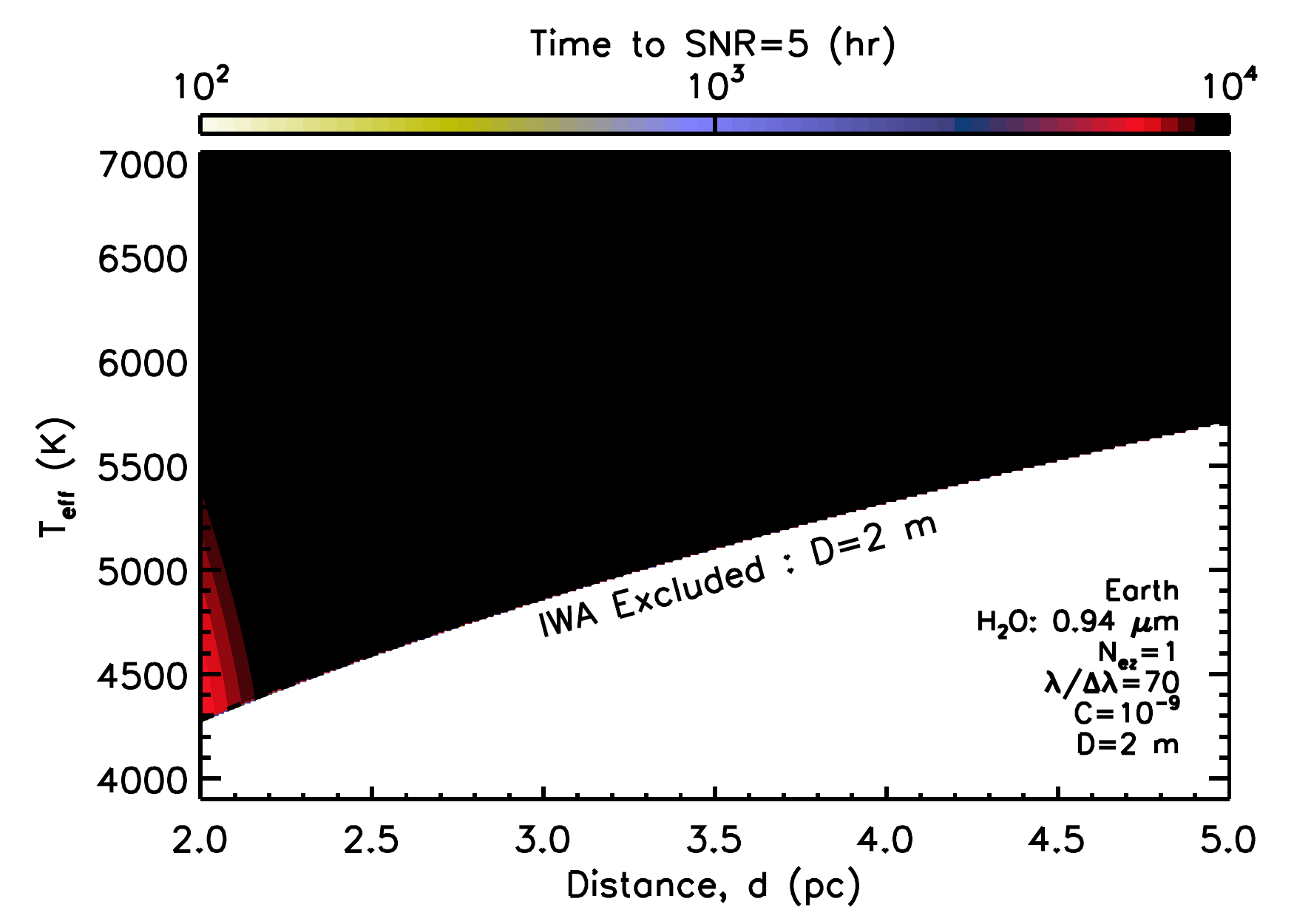} \\
    \includegraphics[trim = 4mm 2mm 2mm 3mm, clip, width=3.1in]{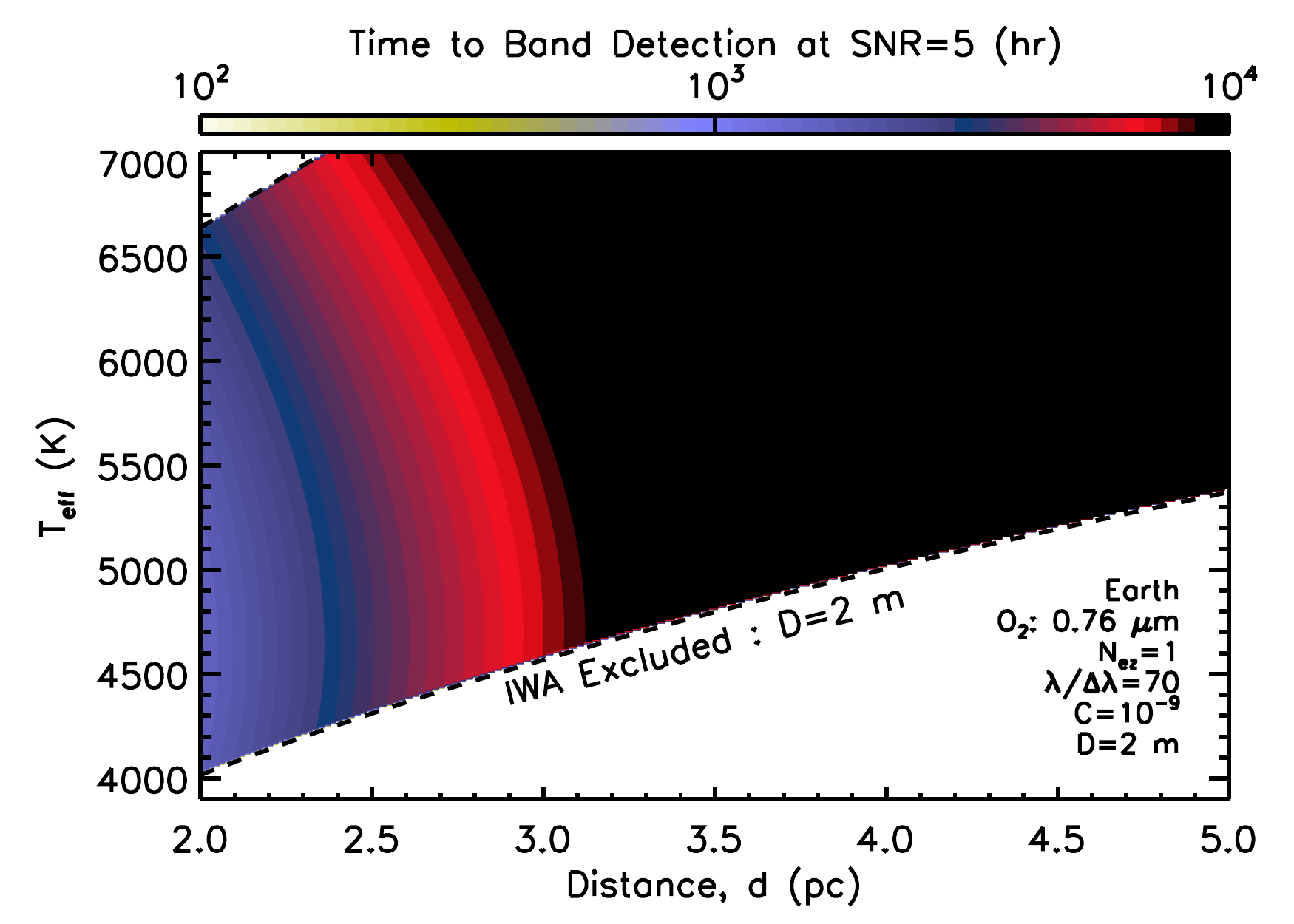} &
    \includegraphics[trim = 4mm 2mm 2mm 3mm, clip, width=3.1in]{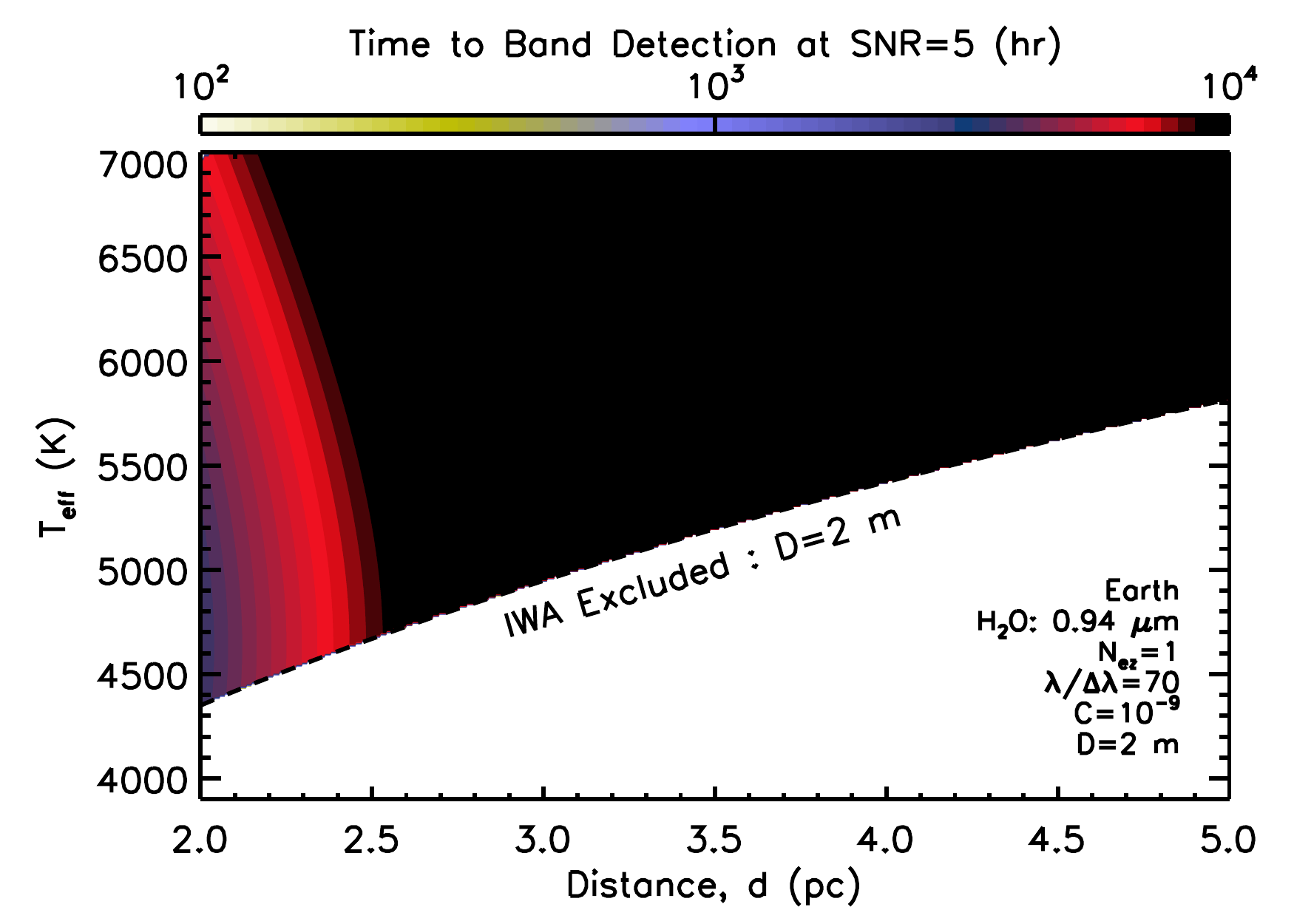} \\
  \end{tabular}
  \caption{Similar to Figure~\ref{fig:jupiter_det}, except now for Earth twins.  Here, 
                we focus on the 0.76~$\mu$m A-band of molecular oxygen and the 
                0.94~$\mu$m water vapor band.}
  \label{fig:earth_det}
\end{figure}

\begin{figure}
  \centering
  \begin{tabular}{cc}
    \includegraphics[trim = 4mm 2mm 2mm 3mm, clip, width=3.1in]{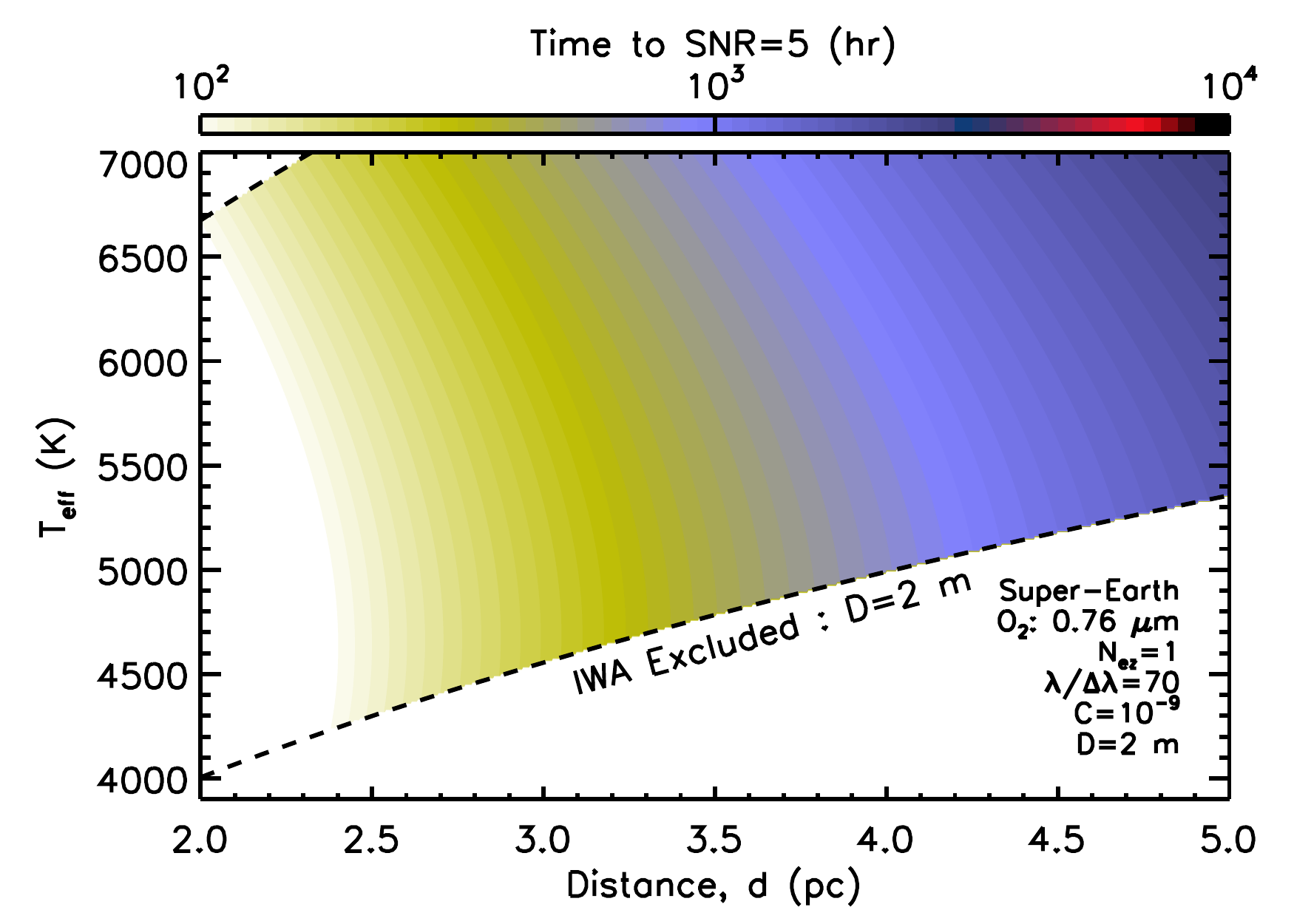} &
    \includegraphics[trim = 4mm 2mm 2mm 3mm, clip, width=3.1in]{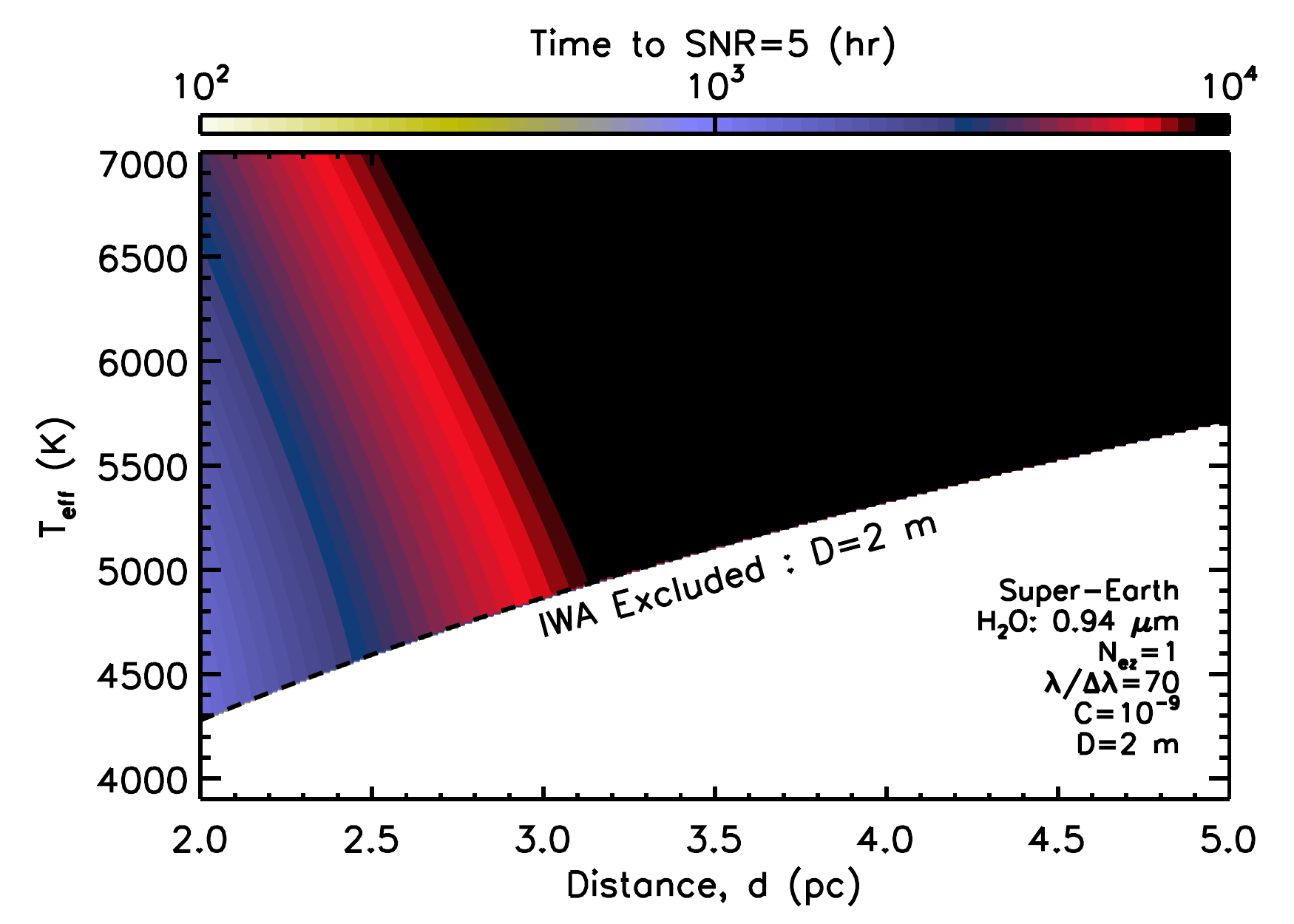} \\
    \includegraphics[trim = 4mm 2mm 2mm 3mm, clip, width=3.1in]{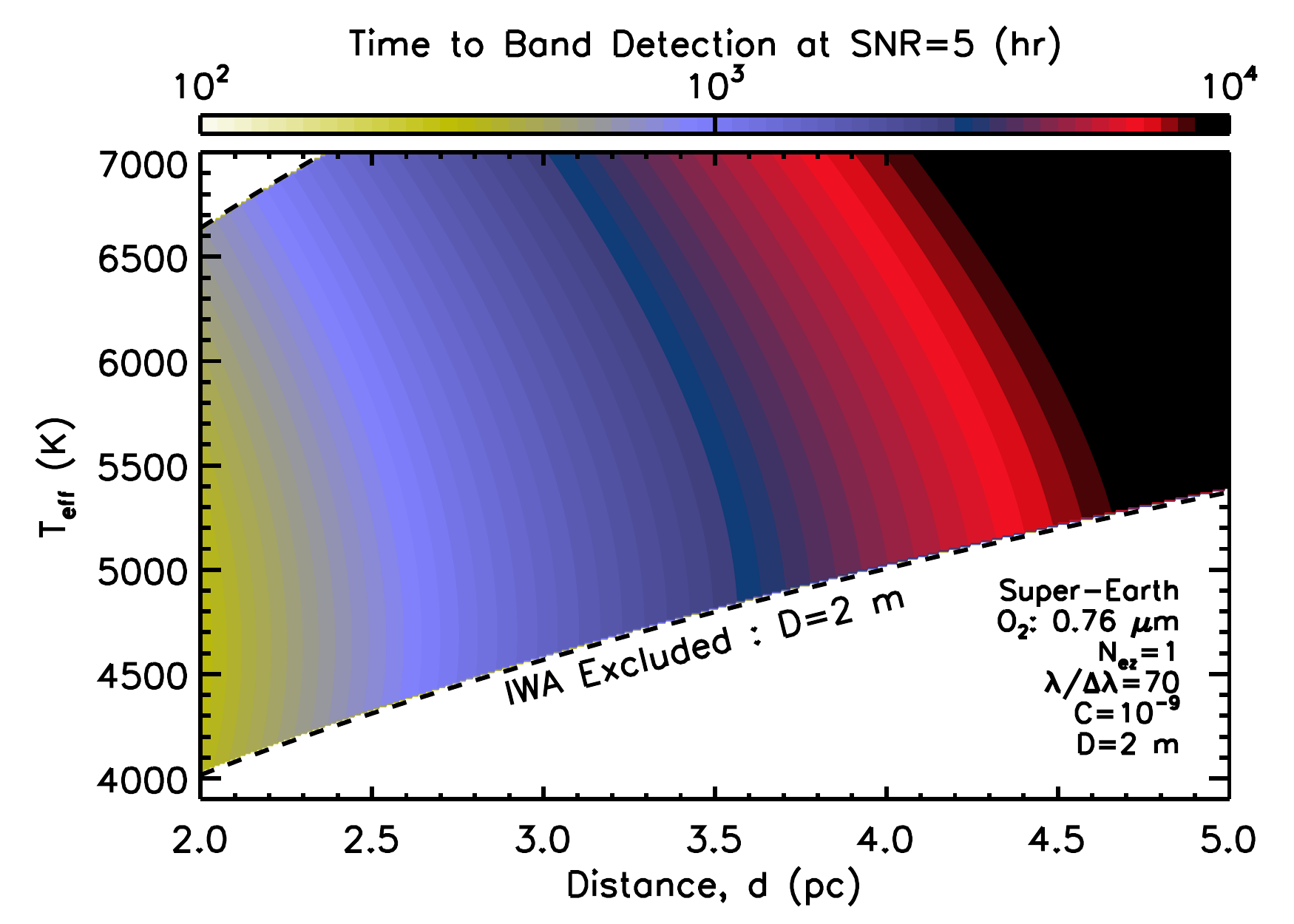} &
    \includegraphics[trim = 4mm 2mm 2mm 3mm, clip, width=3.1in]{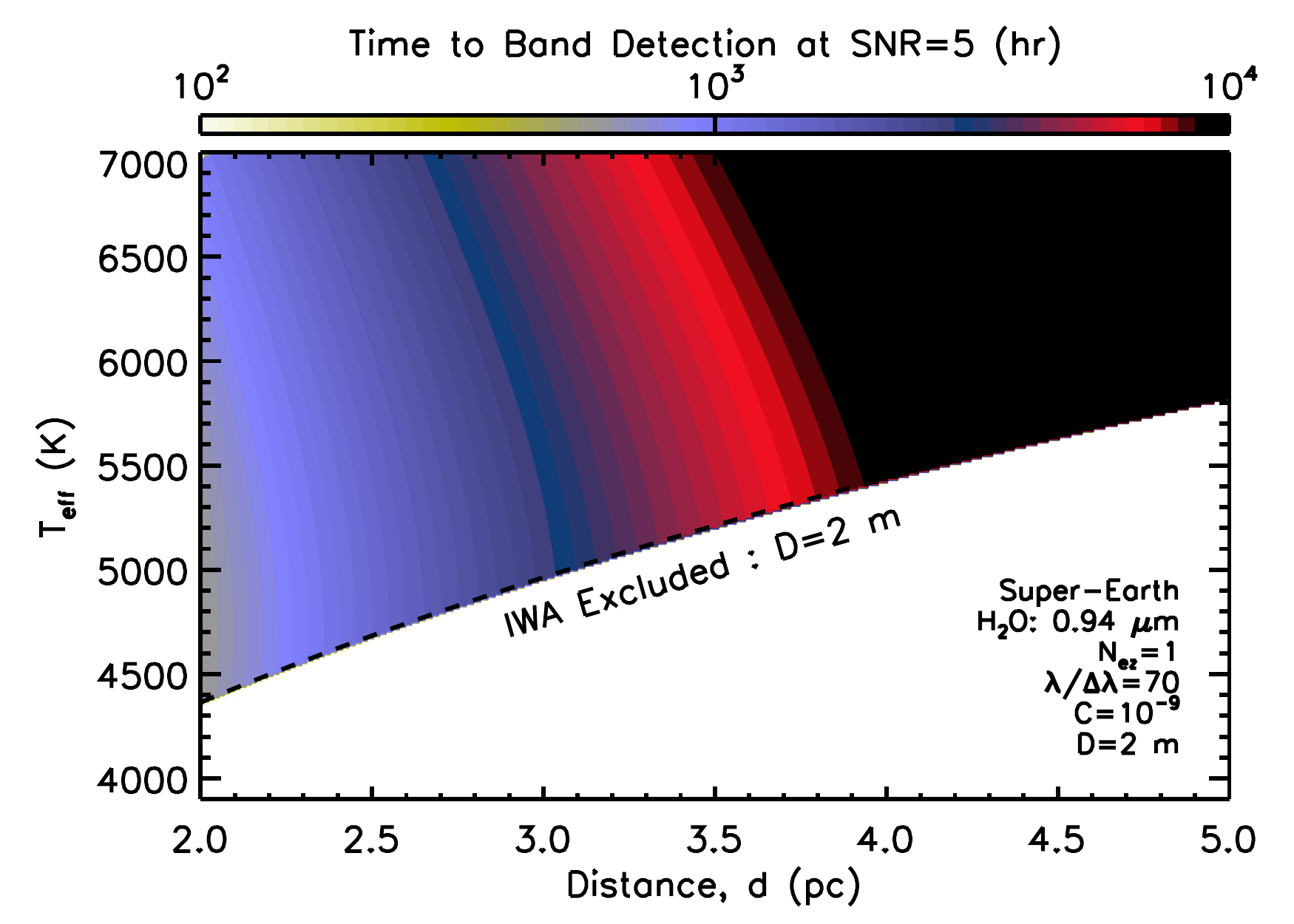} \\
  \end{tabular}
  \caption{The same as Figure~\ref{fig:earth_det}, except for 1.5$R_{\oplus}$ 
                super-Earths.}
  \label{fig:superearth_det}
\end{figure}

\begin{figure}
  \centering
  \begin{tabular}{cc}
    \includegraphics[trim = 4mm 2mm 2mm 3mm, clip, width=3.1in]{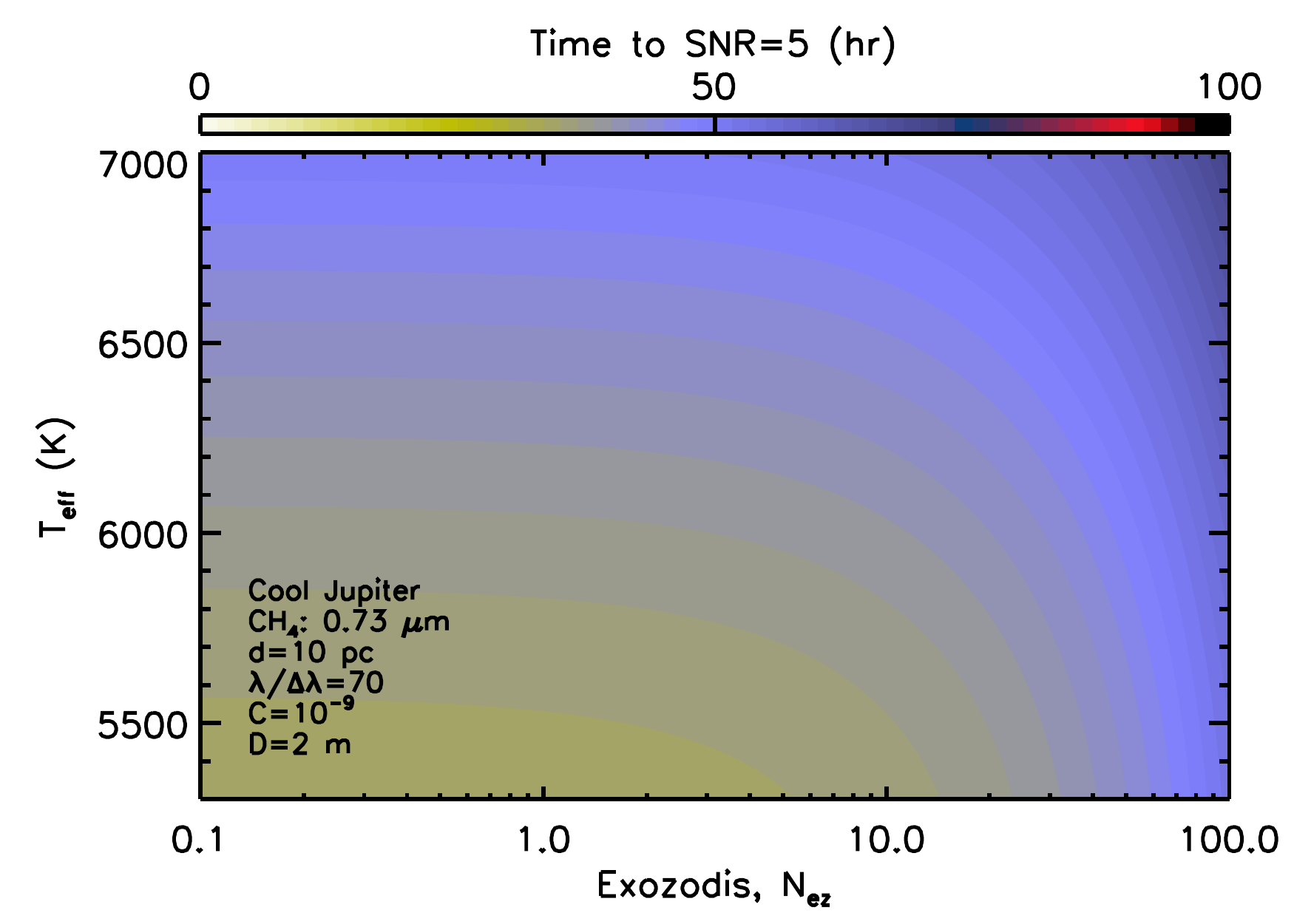} &
    \includegraphics[trim = 4mm 2mm 2mm 3mm, clip, width=3.1in]{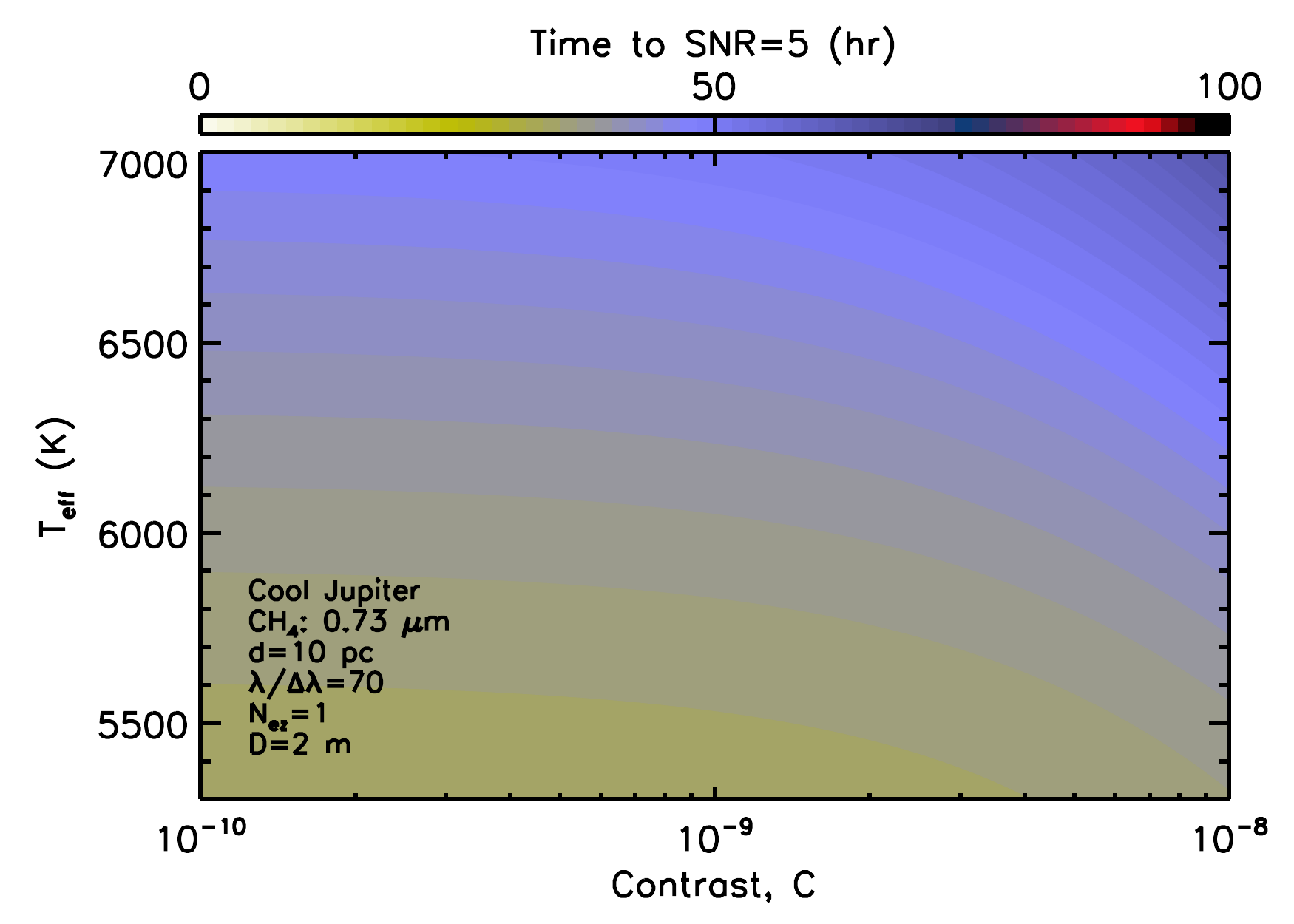} \\
    \includegraphics[trim = 4mm 2mm 2mm 3mm, clip, width=3.1in]{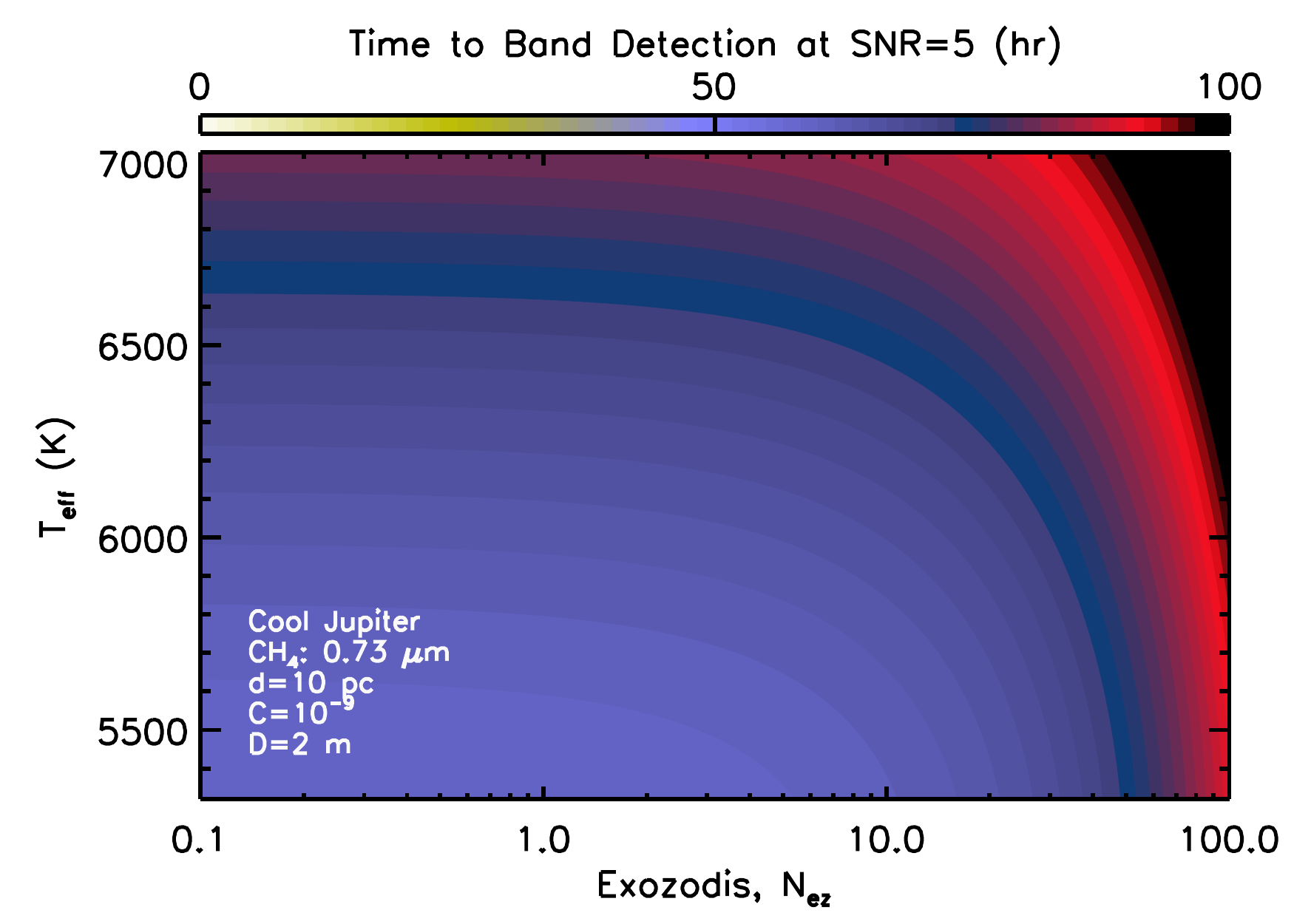} &
    \includegraphics[trim = 4mm 2mm 2mm 3mm, clip, width=3.1in]{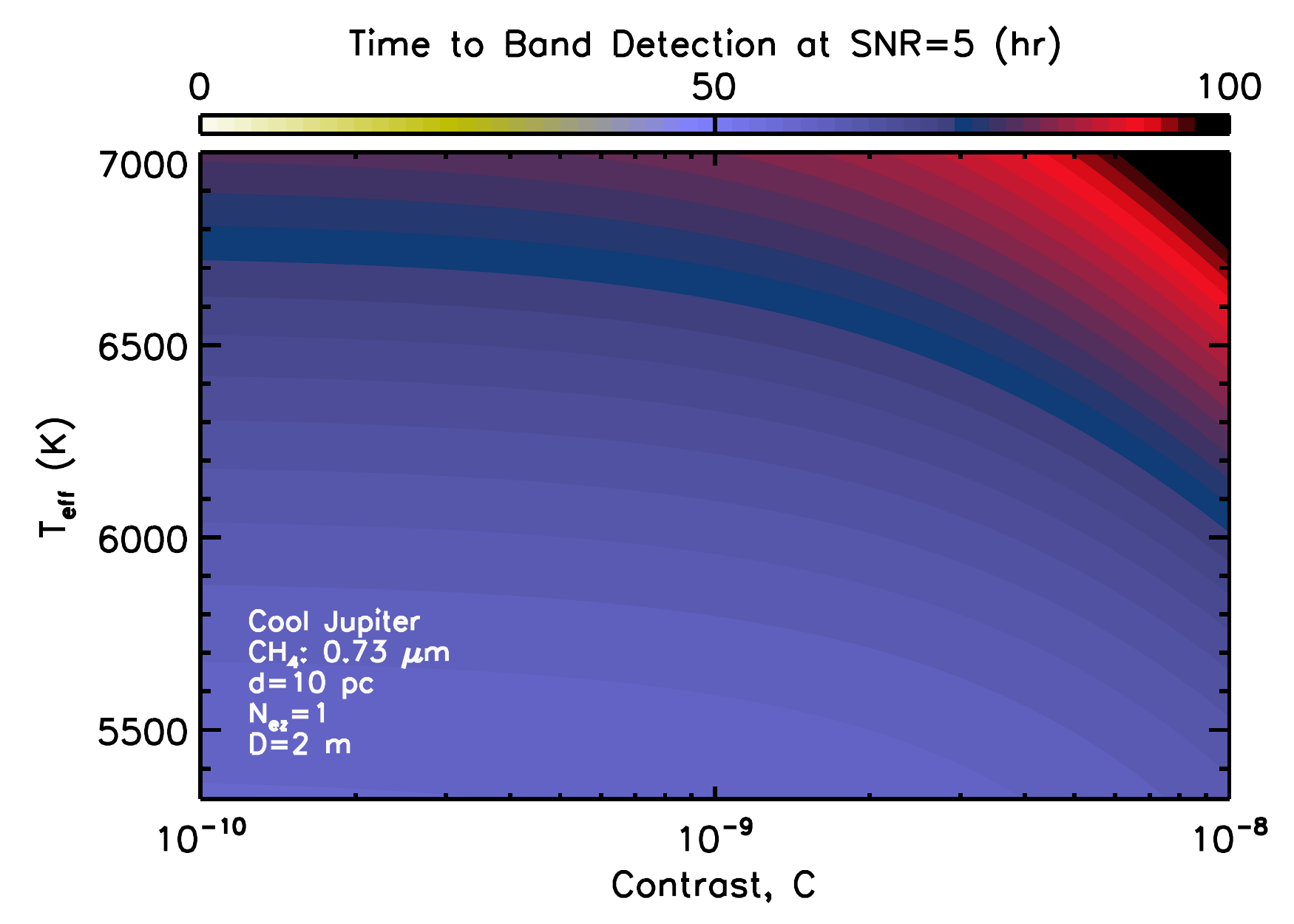} \\
  \end{tabular}
  \caption{Contours of integration time required to achieve $\rm{SNR}=5$ in the bottom 
                of the 0.73~$\mu$m methane band (top) or to detect this band at 
                $\rm{SNR_{band}}=5$ (bottom) for a cool Jupiter at a distance of 10~pc for 
                various levels of exozodis and raw contrast.  Note that the IWA limits the range of 
                $T_{\rm{eff}}$ that can be investigated (as the wavelength and distance are 
                fixed).}
  \label{fig:jupiter_Dt_sens}
\end{figure}

\begin{figure}
  \centering
  \begin{tabular}{cc}
    \includegraphics[trim = 4mm 2mm 2mm 3mm, clip, width=3.1in]{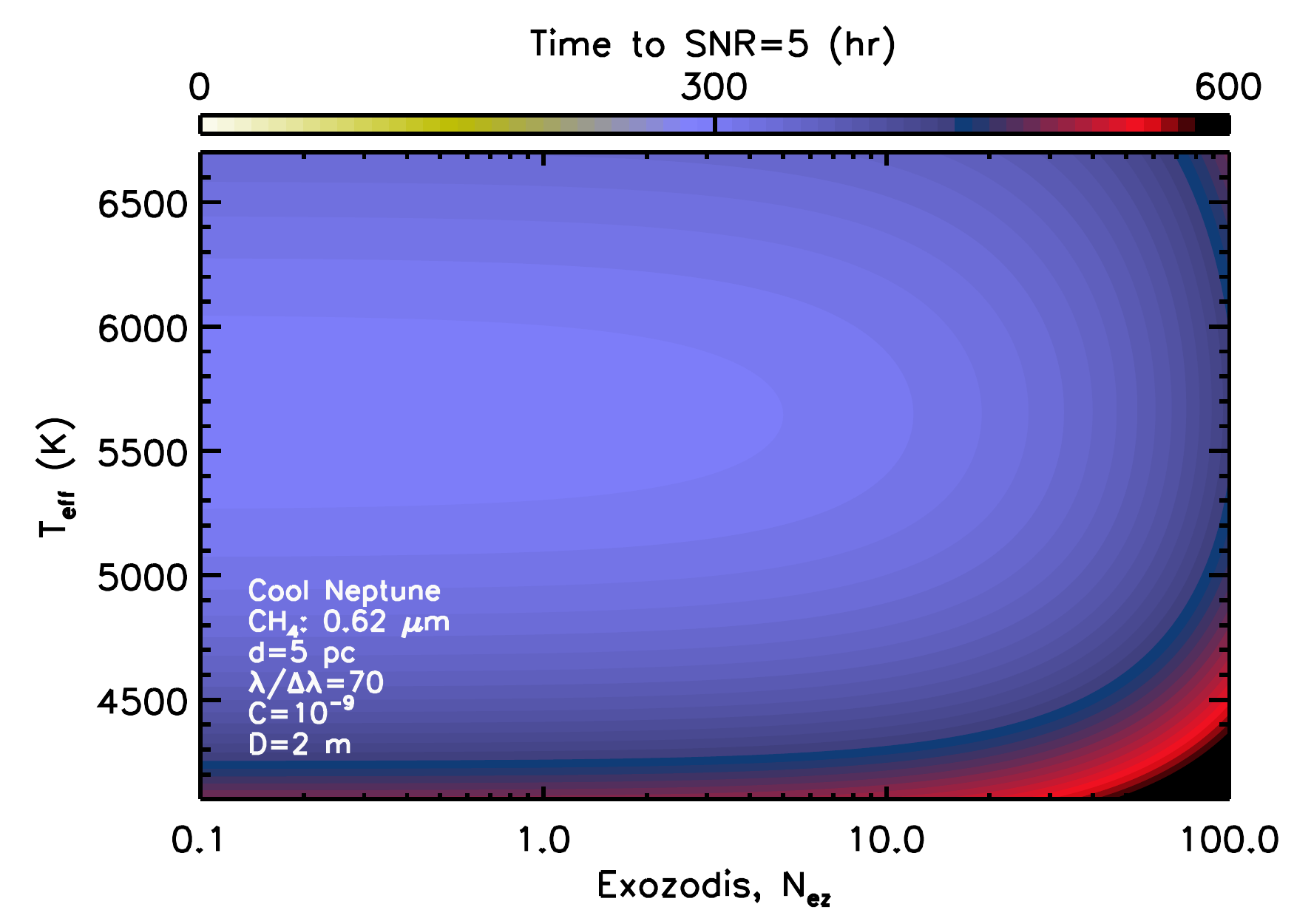} &
    \includegraphics[trim = 4mm 2mm 2mm 3mm, clip, width=3.1in]{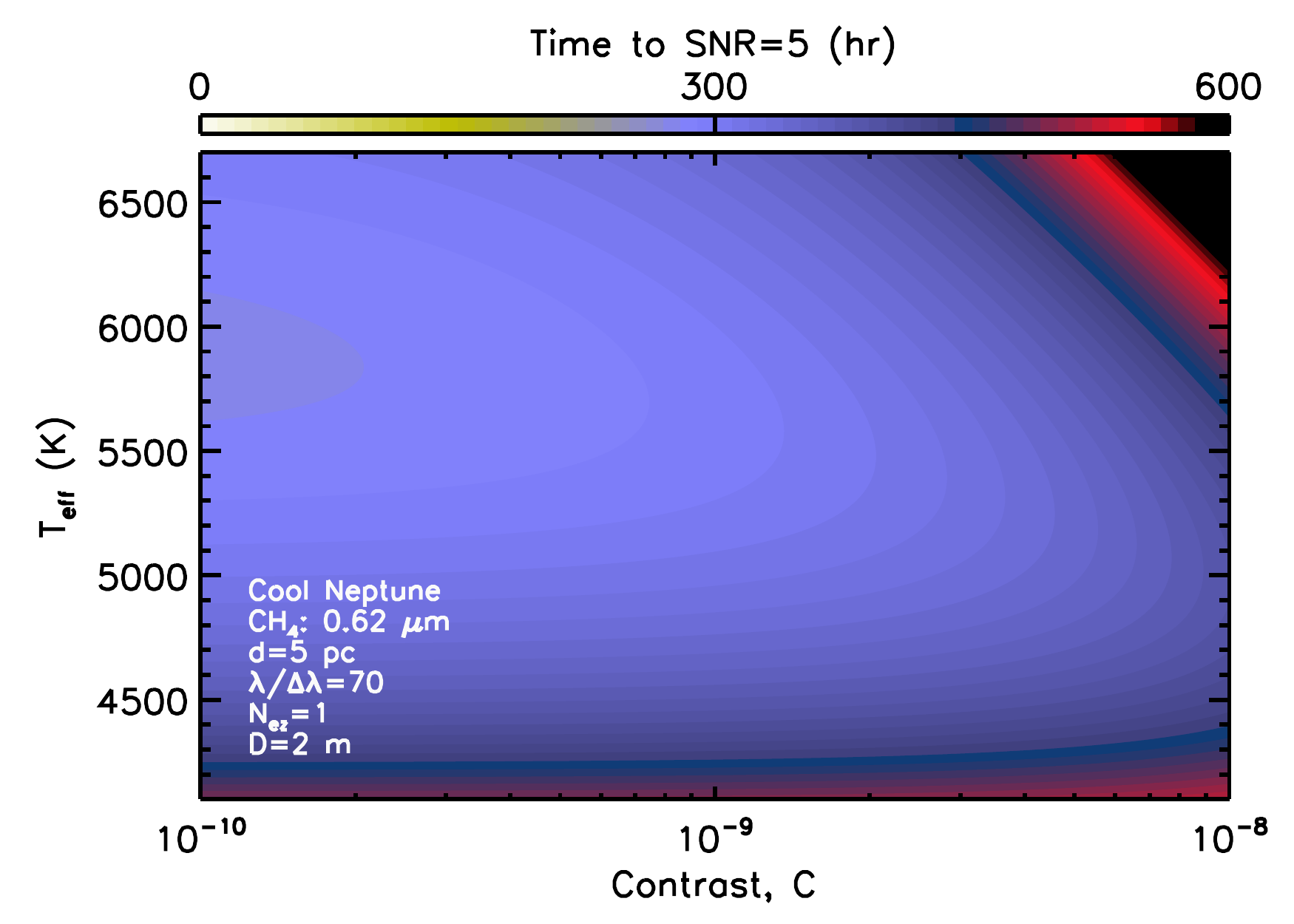} \\
    \includegraphics[trim = 4mm 2mm 2mm 3mm, clip, width=3.1in]{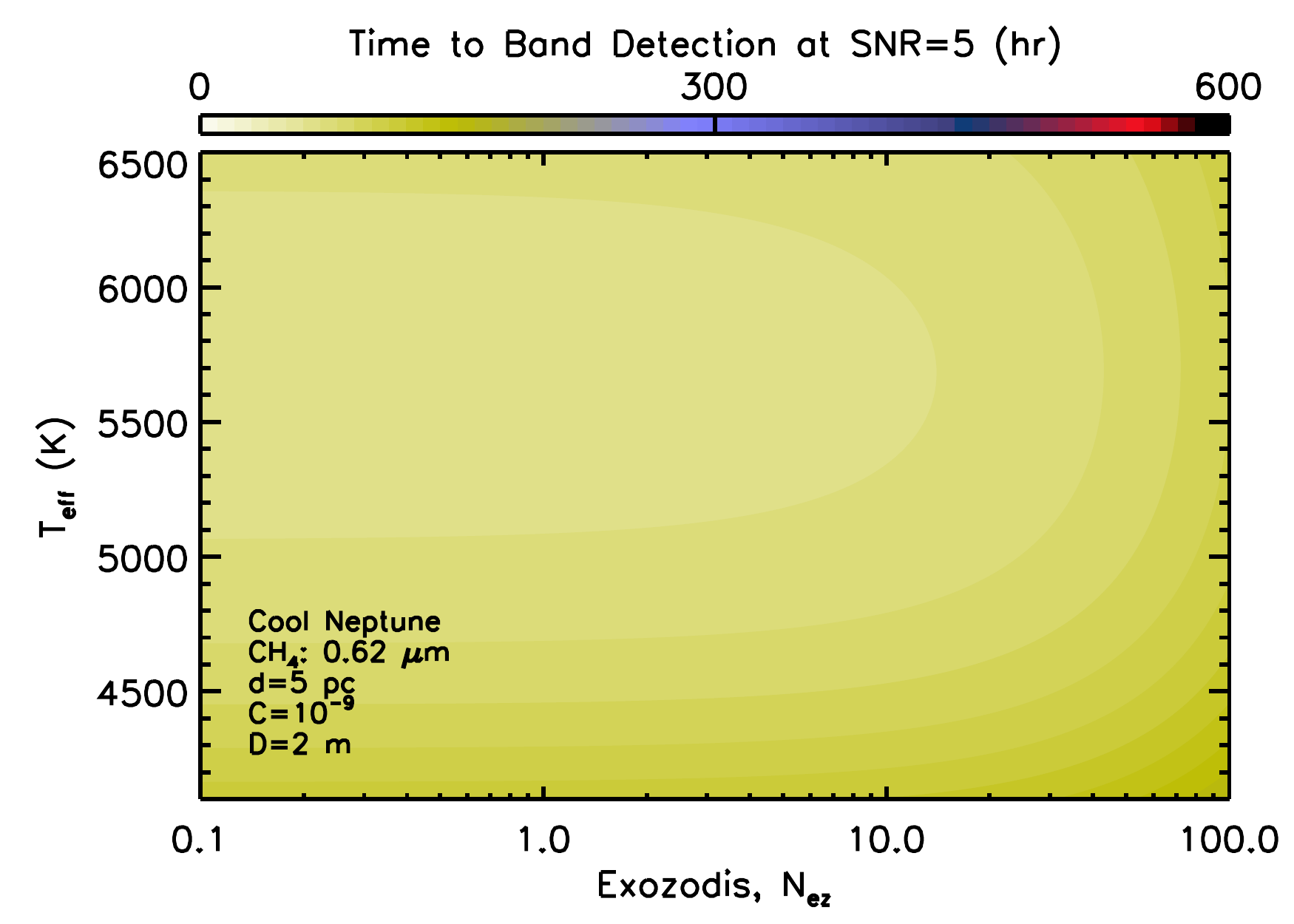} &
    \includegraphics[trim = 4mm 2mm 2mm 3mm, clip, width=3.1in]{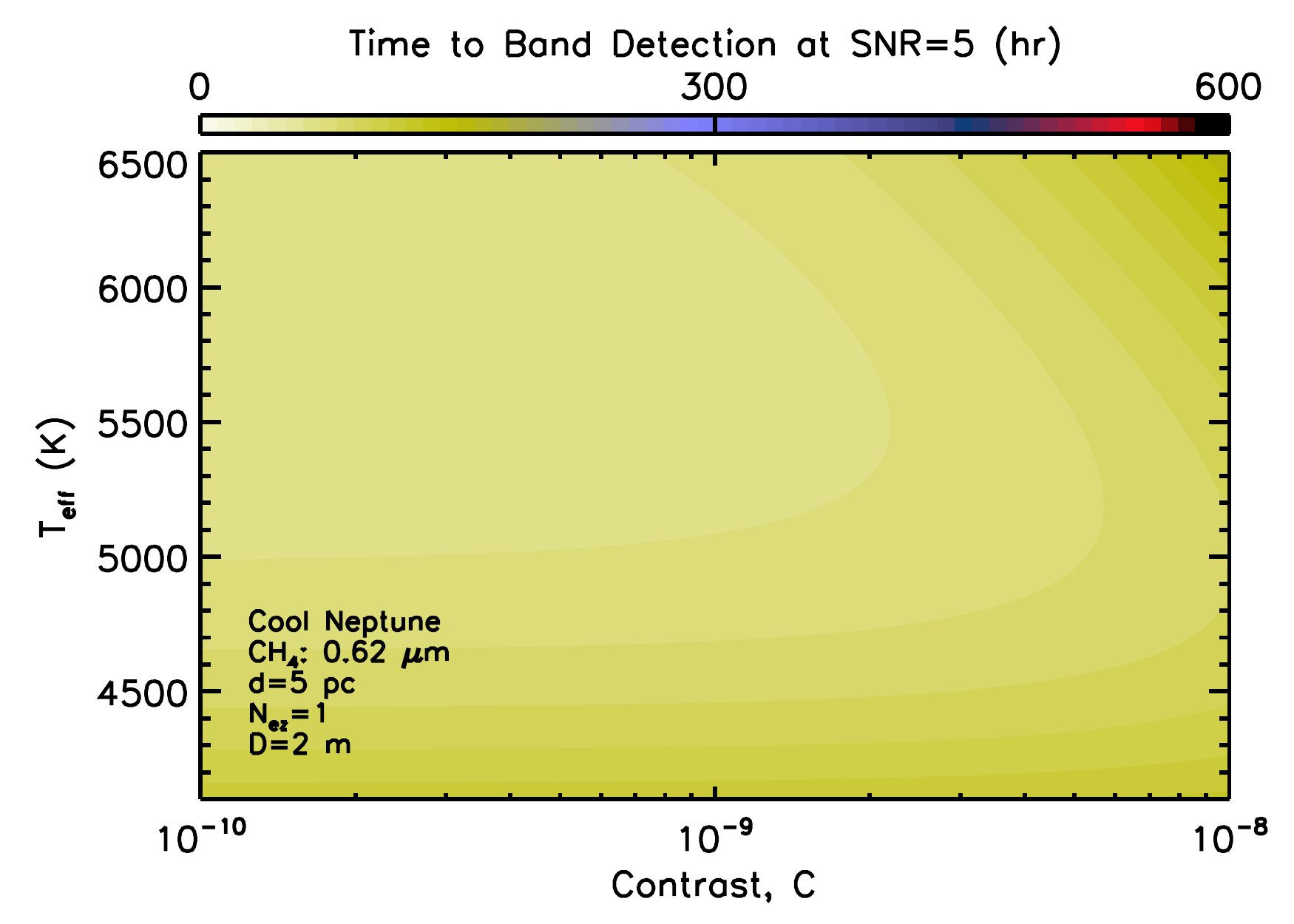} \\
  \end{tabular}
  \caption{Similar to Figure~\ref{fig:jupiter_Dt_sens} except for cool Neptunes at a 
                distance of 5~pc.  Here the 0.62~$\mu$m methane band is used for the 
                feature bottom values, while the 0.73~$\mu$m methane band is used for 
                the band detection values.}
  \label{fig:neptune_Dt_sens}
\end{figure}

\begin{figure}
  \centering
  \begin{tabular}{cc}
    \includegraphics[trim = 4mm 2mm 2mm 3mm, clip, width=3.1in]{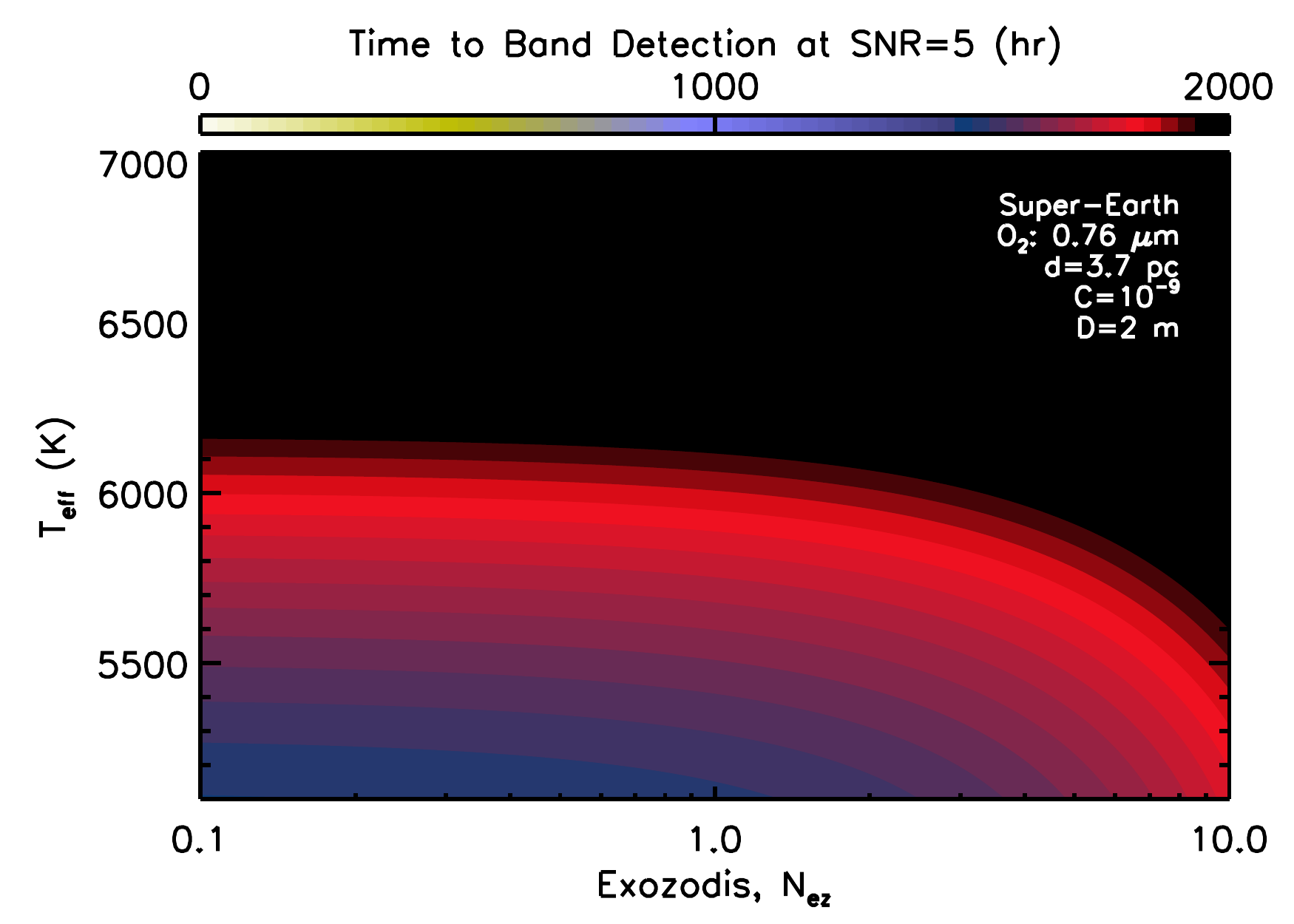} &
    \includegraphics[trim = 4mm 2mm 2mm 3mm, clip, width=3.1in]{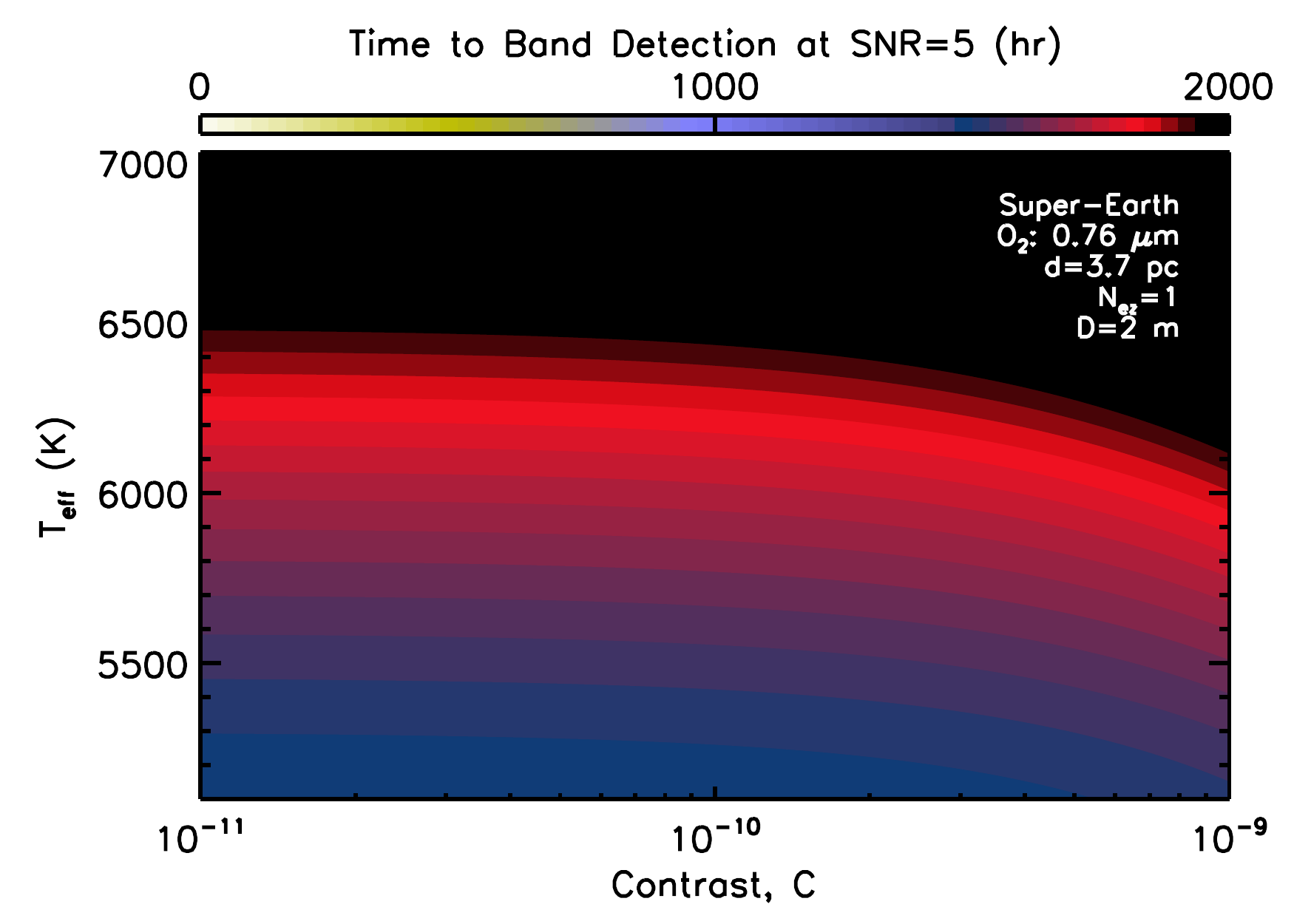} \\
  \end{tabular}
  \caption{Similar to Figure~\ref{fig:jupiter_Dt_sens} except for the required band 
                detection time for the 0.76~$\mu$m A-band of molecular oxygen for 
                super-Earths at 3.7~pc.}
  \label{fig:superearth_Dt_sens}
\end{figure}

\begin{figure}
  \centering
  \begin{tabular}{cc}
    \includegraphics[trim = 4mm 2mm 2mm 3mm, clip, width=3.1in]{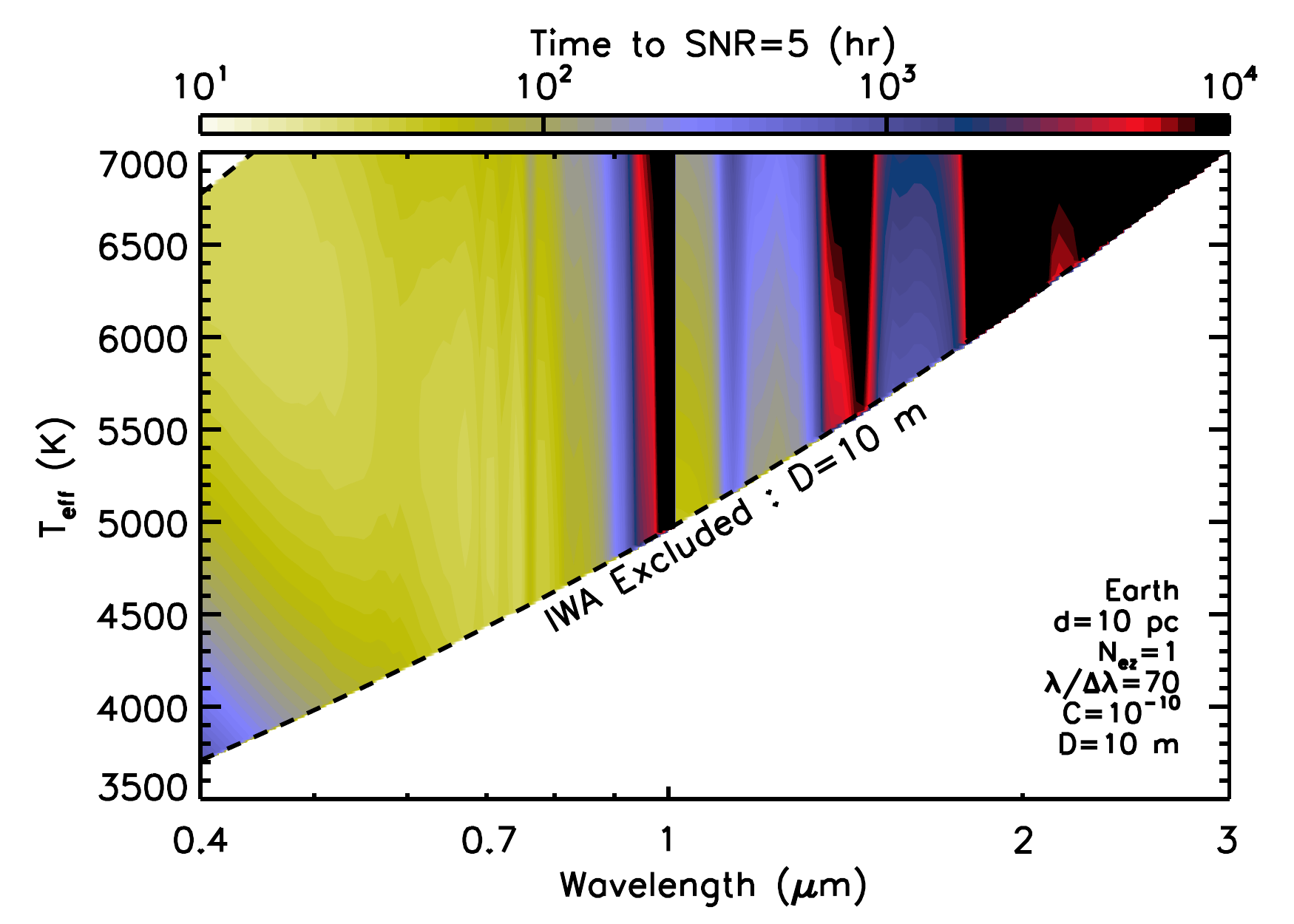} &
    \includegraphics[trim = 4mm 2mm 2mm 3mm, clip, width=3.1in]{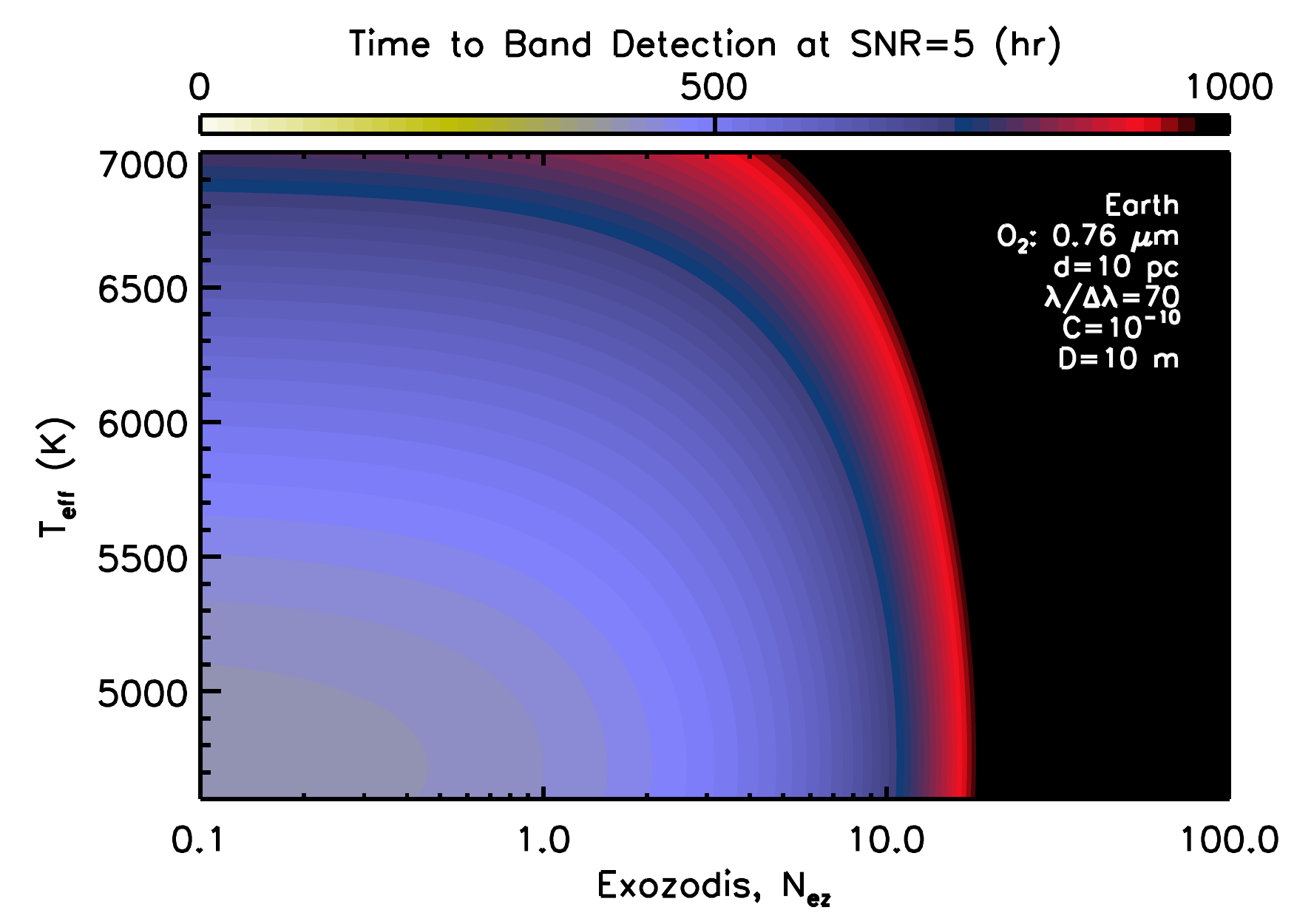} \\
  \end{tabular}
  \caption{Results for a 10-meter class space-based telescope and coronagraph, capable of 
                $10^{-10}$ raw contrast between an IWA of $3\lambda/D$ and an OWA of  
                $20\lambda/D$.  For Earth twins around stars of different effective temperatures, 
                panels show the integration times to achieve $\rm{SNR}=5$ and sensitivity to 
                exozodi levels for integration times in the A-band of molecular oxygen at 
                0.76~$\mu$m.}
  \label{fig:contours_luvoir}
\end{figure}

\begin{figure}
  \centering
  \includegraphics[trim = 4mm 2mm 2mm 3mm, clip, width=6.2in]{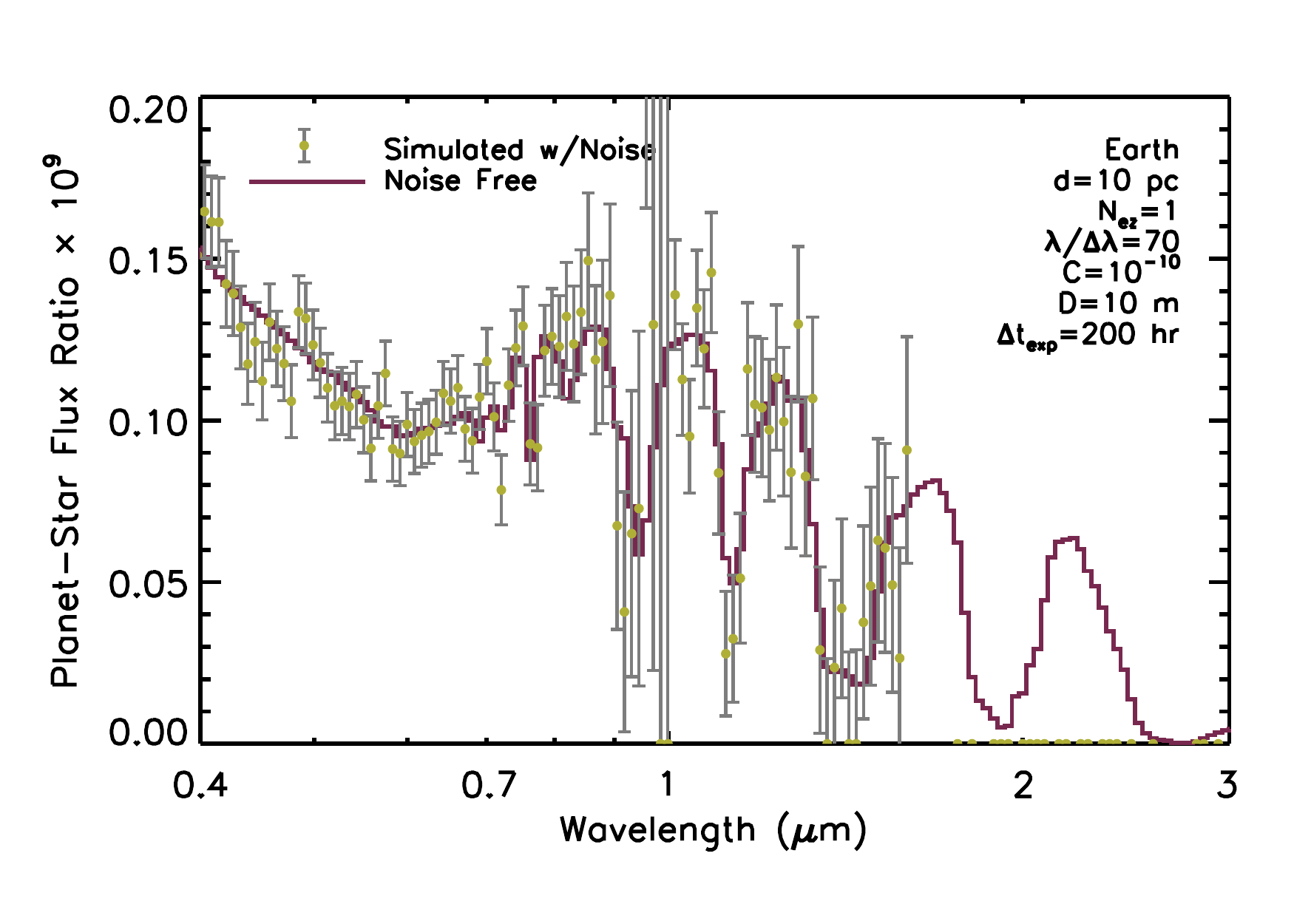}
  \caption{Simulated spectral observation of an Earth twin around a solar twin at 10~pc for 
                a LUVOIR mission assuming a 200~hr integration.  The dark-red line is a noise 
                free spectrum, while the yellow points are a simulated spectrum (with 1-$\sigma$
                error bars in gray).}
  \label{fig:spec_luvoir}
\end{figure}

For our baseline parameters, integration times for cool~Jupiters are relatively 
short, even out to 10~pc (Figure~\ref{fig:jupiter_Dt}).  Detecting the bottom of 
the 0.73~$\mu$m methane band, which could indicate different planet 
formation scenarios as it can distinguish between different amounts of heavy 
element enhancement \citep{cahoyetal2010}, can be accomplished with a 
10~hr integration time (or less) for distances as large as 7~pc and with 100~hr 
integration times out to 13~pc (Figure~\ref{fig:jupiter_det}).  Integration times 
required to detect the 0.73~$\mu$m methane band are similar, and could be 
decreased by about 50\% by observing both the 0.73~$\mu$m and 
0.89~$\mu$m methane bands.  Detecting the bottom of the 0.89~$\mu$m 
band in less than 100~hr is limited to cool Jupiters at distances of 6~pc or 
less, where the overall longer integration times for detecting the bottom of 
this band are driven up by falling detector quantum efficiencies at these 
wavelengths.  Finally, these integration times are largely insensitive to raw 
contrast performance and exozodi levels (Figure~\ref{fig:jupiter_Dt_sens}).  
Regarding the latter, dust clearing by giant planets will likely prevent large 
exozodi levels in the vicinity of a giant \citep[e.g.,][]{papaloizouetal2007}, 
limiting the range of realistic values for $N_{\rm{ez}}$ in 
Figure~\ref{fig:jupiter_Dt_sens}.

Resolution 70 spectra at constant SNR for Jupiter twins are limited by long 
integration times in methane bands at long wavelengths, which can exceed 
$10^{3}$~hr (Figures~\ref{fig:jupiter_Dt}).  Such long integration times are 
unrealistic (especially on a shared resource like WFIRST-AFTA---the 
\textit{Hubble} Deep Field used less than 200~hr of exposure time).  Only for 
continuum wavelengths between about 0.5--0.8~$\mu$m do integration times 
for $\rm{SNR}=5$ spectra fall below $10^{3}$~hr for a Jupiter twin at 10~pc.  
While integration times for Jupiter twins around closer (i.e., less than about 
5~pc) stars would be much shorter, telescope OWA constraints would limit 
study of twins in these nearby systems to cool stars (i.e., roughly 4,300~K or 
less for the orbit to be entirely within the OWA).  Note, however, that the OWA 
constraint is less strict than the IWA constraint, as some fraction of the orbit, 
both towards gibbous and crescent phases, can occur inside the OWA, 
depending on orbit parameters and orientation 
\citep[see, e.g.,][]{greco&burrows2015}.
 
For the cool~Neptunes, a coronagraph contrast noise floor of $10^{-10}$ would 
strongly limit observations at longer wavelengths (Figures~\ref{fig:neptune_Dt}), 
although detecting the 0.62~$\mu$m methane band with integration times below 
100~hr can be done out to distances of about 5~pc (Figure~\ref{fig:neptune_det}).
As this band is relatively deep in the reflectance spectrum of cool Neptunes, 
integration times for detecting the bottom of the 0.62~$\mu$m band are longer, 
remaining below 100~hr only out to 3--4~pc.  The 0.54~$\mu$m methane band 
is shallower, so detecting its bottom can be done with integration times comparable 
to the band detection times for the 0.62~$\mu$m band.

Inner working angle constraints strongly limit the ability of a 2-meter class 
telescope to study Earths and super-Earths at 1~AU flux equivalent distances from 
their host stars.  Covering the entire 0.4--1~$\mu$m spectral range at a distance of 
5~pc (or more) requires solar twins (or hotter).  For Earth twins, this implies 
planet-star flux ratios of $10^{-10}$--$10^{-11}$, which could be improved by a factor 
of $\left(R_{\rm{p}}/R_{\oplus}\right)^{2}$ for super-Earths (assuming similar reflectivity).  
Additionally, maintaining integration times below even $10^{3}$~hr at continuum 
wavelengths for these worlds at 5~pc would require degrading spectra to 
$\mathcal{R}=10$--20 to achieve $\rm{SNR}=5$ (Figure~\ref{fig:earth_Dt}), notably 
ruling out detection of the 0.76~$\mu$m oxygen A-band \citep{brandt&spiegel2014}.  
Such long integration times may be an unrealistic price to pay for characterizing 
potentially habitable worlds.  Additionally, a $2\lambda/D$ coronagraph IWA could 
interfere with our ability to detect the 0.94~$\mu$m water vapor band, a key indicator 
of habitability, for 5~pc Earths and super-Earths around Sun-like stars and cooler.  
These IWA constraints could be alleviated by investigating worlds at the outer edge of 
the habitable zone \citep{kastingetal93,kopparapuetal2013}, but the low incident flux 
on these worlds pushes integration to times $\gtrsim\!10^{3}$~hr (assuming an 
Earth-like size and reflectivity), even at $\mathcal{R}=10$.

Shorter integration times and more favorable contrast ratios argue for super-Earths 
as being more attractive targets than Earth-sized worlds.  Even for these worlds, 
though, detecting the oxygen A-band with integration times less than $10^{3}$~hr 
requires targets to be nearer than about 3~pc (Figure~\ref{fig:jupiter_Dt_sens}).  Only 
a few stars meet the distance and effective temperature requirements 
($\alpha$~Centauri A and B, $\epsilon$~Eridani), and most of these stars are in 
multi-star systems.  Detecting the 0.94~$\mu$m water vapor band in super-Earth 
spectra with integration times below $10^{3}$~hr will be extremely challenging, if 
not impossible, owing in large part to extremely low detector quantum efficiencies 
at these wavelengths.  This outlook could be improved by using detectors with better 
quantum efficiency at red visible wavelengths, although these detectors tend to have 
larger dark currents, which might diminish any possible improvements.


While our results show that characterizing Earth twins with a 2-meter class, 
coronagraph-equipped telescope will be unlikely, 10-meter class LUVOIR telescopes 
show much more promising results (Figure~\ref{fig:contours_luvoir}).  Such a mission 
could achieve $\mathcal{R}=70$ visible wavelength spectra at $\rm{SNR}=5$ for Earth 
twins at a characteristic distance of 10~pc for integration times of order 100~hr.  Falling 
stellar spectra and strong water vapor absorption drive up integration times at near-infrared 
wavelengths.  Additionally, falling CCD quantum efficiency will drive up required integration 
times for species and band bottom detection at red visible wavelengths, especially for the 
0.94~$\mu$m water vapor band.  Using the near-infrared HgCdTe detector at these 
wavelengths can improve required integration times by an order of magnitude, even when 
accounting for the larger dark current and read noise in these detectors.  Finally, note that 
telescope IWA constraints can prevent observations of large parts of the near-infrared 
wavelength range for Earth twins orbiting Sun-like stars at 10~pc, and exozodi levels 
larger than $N_{\rm{ez}}=10$ begin to rapidly drive up integration times.

Moving beyond requisite integration times and sensitivities, several 
of our assumptions about our instrumental, astrophysical, and planetary parameters 
warrant further discussion.  Most notably, current experimental coronagraphs designed 
to achieve raw contrasts suitable for exoplanet characterization typically operate only 
over a $\sim\!\!20$\% bandpass \citep[e.g.,][]{traugeretal2013}, which is limited by 
chromatic effects within the wavefront control system \citep[although progress is being 
made on some aspects of this problem, e.g.,][]{newmanetal2015}.  Thus, if only a single 
coronagraph is used, observing over the full 0.4--1.0~$\mu$m range used in this work 
would require 4--5 separate integrations, thereby driving up overall characterization time.  
This feature of coronagraphs is not entirely negative, as integration times in different 
bandpasses could be tailored to prevent achieving unnecessarily large SNRs in bright 
spectral regions.  Also, while we include stellar leakage from the coronagraph, we did 
not consider how systematic errors from speckle subtraction might impact performance 
and increase required integration times, all of which would depend on telescope stability, 
the capabilities of the coronagraph wave-front control system, and how well speckles 
can be modeled \citep{kristetal2008}.

Regarding astrophysical assumptions, recall that our noise models assumed that 
the star-planet systems were isolated in the instrument field-of-view.  In reality, stellar 
companions \cite[which are common, see][]{duchene&kraus2013} and background 
objects will contribute light to the observations, possibly within the planetary PSF.  Stray 
light from a companion could impact the systematic contrast noise floor and/or the raw 
contrast performance.  While simplistic for companion star scenarios, our sensitivity 
studies to raw contrast performance ($C$) provide an estimate for how integration 
times would change if a companion were to increase the contrast background.  Also, 
we assume smoothly varying exozodi structure that only depends on distance from 
the host star.  Real exozodiacal disks will have more complex structure, such as 
clumps and bands, and studies of such structure would be a major science driver of 
a space-based coronagraph mission.

While we have presented a range of planet models and types, the true 
diversity of exoplanets will be much greater.  Additionally, some of our planet types 
(the warm and cool Jupiters and Neptunes) are based on models that do not include 
atmospheric photochemistry.  For the Solar System gas and ice giants, this chemistry 
leads to haze formation, which causes lower albedos at blue wavelengths 
\citep{karkoschka94}.  Thus, we are likely underestimating integration times for these 
worlds at the shortest wavelengths.  Also, our placement of planet types at flux equivalent 
distances attempts, to first order, to maintain self-consistent atmospheric and cloud 
structures for these worlds.  This treatment does not account for the shift of cooler stellar 
spectral energy distributions to longer wavelengths, where planets tend to be more 
absorbing \citep{marleyetal1999}.  So, our planet types around cooler stars would be 
absorbing more stellar flux, implying that an ``absorbed flux'' equivalent distance is 
further from the host star than our incident bolometric flux equivalent distance.

\section{Conclusions}

Space-based 2-meter class exoplanet characterization missions have the potential to 
provide spectra at moderate to high SNRs (i.e., 5 or better) for a variety of planets, some 
of which have no analog in our Solar System.  We find that:
\begin{itemize}
  \item for cool Jupiters (i.e., Jupiters at 2~AU flux equivalent distance from the Sun), methane 
           can be detected with integration times shorter than 100~hr out to a distance of 13~pc, 
           and with integration times less than 10~hr at 5~pc.
  \item spectra of cool Neptunes at 10~pc will be challenging (i.e., integration times 
           approaching or exceeding $10^{3}$~hr), although methane could be detected in the 
           spectra of these worlds with integration times below 100~hr out to a distance of 5~pc.
  \item Earth twins are unlikely to be characterized, owing to long integration times and 
          planet-star flux ratios smaller than $10^{-10}$ for a wide range of stellar effective 
          temperatures.
  \item super-Earths could be studied with low-resolution spectra and integration times 
          below $10^{3}$~hr to a distance of 5~pc, but detecting features in higher-resolution 
          spectra will be limited to worlds around a small handful of stars within 3~pc of the Sun.
 \item falling CCD quantum efficiency at red visible wavelengths strongly limits observations 
          of the 0.89~$\mu$m methane band and the 0.94~$\mu$m water vapor band.
\end{itemize}

We note that the majority of the modeled observations of planet types presented here have 
$\Delta t_{\rm{exp}} \propto c_{\rm{D}}/c_{\rm{p}}^{2}$ (to achieve a given SNR).  Thus, 
major gains in driving down requisite integration times could be made by striving for larger 
instrument throughputs, better detector quantum efficiencies, and by using devices with low 
dark current \citep[although achieving lower dark currents in modern detectors can lead to 
trade-offs with other detector properties,][]{hardingetal2015}.  Observational investigations 
will not have control over planetary properties, but we do have control over instrument 
performance requirements.

\appendix
\section{Appendix: Noise Model Expressions}

This Appendix contains a brief discussion of our models of the planet count rate and 
count rates from key noise sources.

\subsection{Planet Count Rate}

The planetary signal is the incident stellar host spectrum, weighted by a wavelength- 
and phase-dependent planetary reflectance, and passed through the optics of a distant 
telescope.  We assume that the host star has radius $R_{\rm{s}}$ and a spectrum of 
a blackbody, $B_{\lambda}\left( T_{\rm{eff}} \right)$, where $T_{\rm{eff}}$ is the stellar 
effective temperature.  Thus, the specific flux density at a distance $d$ from the stellar 
host is, 
\begin{equation}
  F_{\rm{s},\lambda}(d) = \pi B_{\lambda}\!\left( T_{\rm{eff}} \right) \left( \frac{R_{\rm{s}}}{d} \right)^{\!\!2} \ .
\end{equation}
Note that non-blackbody stellar spectra, either from observations or models, can be 
used for $F_{\rm{s},\lambda}(d)$---we simply adopt blackbody spectra for efficiency, 
allowing us to rapidly model large swaths of parameter space.  If the planetary target 
has radius $R_{\rm{p}}$, orbits its host at a distance $r$, is observed at phase angle 
(i.e., star-planet-observer angle) $\alpha$, and has a wavelength-dependent geometric 
albedo $A$, then the specific flux density at a distance $d$ from the planet is,
\begin{equation}
  F_{\rm{p},\lambda}(d) =   A \Phi(\alpha) F_{\rm{s},\lambda}(r) \left( \frac{R_{\rm{p}}}{d} \right)^{\!\!2} 
                                     =  \pi A \Phi(\alpha) B_{\lambda}\!\left( T_{\rm{eff}} \right)  
                                          \left( \frac{R_{\rm{s}}}{r} \right)^{\!\!2} \left( \frac{R_{\rm{p}}}{d} \right)^{\!\!2} \ ,
\label{eqn:Fp}
\end{equation}
where $\Phi(\alpha)$ is the planetary phase function, which is generally a function of 
wavelength.  Thus, the wavelength-dependent planet-star flux ratio is,
\begin{equation}
  \frac{F_{p,\lambda}}{F_{\rm{s},\lambda}} = A \Phi(\alpha)  \left( \frac{R_{\rm{p}}}{r} \right)^{\!\!2} ,
\label{eqn:flxratio}
\end{equation}
which, of course, is only a function of the product of the geometric albedo and 
phase function (i.e., the phase-dependent planetary reflectivity), and the ratio of 
the planetary radius to the orbital separation.

At the telescope, a distance $d$ from the star-planet system, the planetary flux is 
passed through the optical components, with total throughput $\mathcal{T}$ (which 
can, in general, be wavelength-dependent) and spectral bandwidth $\Delta\!\lambda$,
and converted into counts on the detector with quantum efficiency $q$ (in dimensionless 
units of counts per photon).  Thus, the planet count rate is given by,
\begin{equation}
  c_{\rm{p}} = \pi q f_{\rm{pa}} \mathcal{T} \frac{\lambda}{hc} F_{\rm{p},\lambda}(d) \Delta\!\lambda \left( \frac{D}{2} \right)^{\!\!2} \ ,
\end{equation}
which has dimensions of counts per unit time.  Here, $h$ is the Planck constant, 
$c$ is the speed of light, and $f_{\rm{pa}}$ is a factor of order unity (computed 
using $X$, the photometric aperture length) that describes the fraction of light from 
the planet that falls within the photometric aperture.  Inserting Equation~\ref{eqn:Fp} 
and rearranging, we have
\begin{equation}
  c_{\rm{p}} = q f_{\rm{pa}} \mathcal{T} A \Phi(\alpha) B_{\lambda}\!\left( T_{\rm{eff}} \right)  
                     \Delta\!\lambda \frac{\lambda}{hc} \left( \frac{\pi D R_{\rm{s}} R_{\rm{p}}}{2 r d} \right)^{\!\!2}  \ .
\label{eqn:pcnt}
\end{equation}
Furthermore, for a spectrometer with constant resolution, 
$\mathcal{R}=\lambda/\Delta\!\lambda$, we can express the planet count rate as 
\begin{equation}
  c_{\rm{p}} = q f_{\rm{pa}} \mathcal{T} A \Phi(\alpha) B_{\lambda}\!\left( T_{\rm{eff}} \right)  
                      \frac{\lambda^2}{hc\mathcal{R}} \left( \frac{\pi D R_{\rm{s}} R_{\rm{p}}}{2 r d} \right)^{\!\!2}  \ .
\end{equation}

\subsection{Zodiacal and Exozodiacal Light}
\label{subsec:zod}

Local zodiacal light is known to be a function of ecliptic latitude and longitude 
\citep{levasseur-regourd&dumont1980}, although \citet{starketal2014}
demonstrated that a constant V-band surface brightness of 
$M_{\rm{z},V}=23$~mag~arcsec$^{-2}$ was a reasonably accurate representation.  
Thus, we take the wavelength-dependent zodiacal light count rate to be given by 
\begin{equation}
  c_{\rm{z}} = \pi q \mathcal{T} \Omega \Delta\!\lambda \frac{\lambda}{hc} \left( \frac{D}{2} \right)^{\!\!2}  
                     \frac{F_{\odot,\lambda}(1~\rm{AU})}{F_{\odot,V}(1~\rm{AU})} F_{0,V}  10^{-M_{\rm{z},V}/2.5}  \ ,
\label{eqn:cz}
\end{equation}
where $F_{\odot,\lambda}$ is the wavelength-dependent specific solar flux 
density, $F_{\odot,V}$ is the solar flux density in V-band, 
$\Omega$ is the photometry aperture area (with dimensions of arcsec$^{2}$), and 
$F_{0,V}$ is the standard zero-magnitude V-band specific flux density 
($F_{0,V}=3.6\times10^{-8}$~W~m$^{-2}$~$\mu$m$^{-1}$).  We use a square 
aperture with width $X\lambda/D$ along an edge, such that 
$\Omega = \left( X\lambda/D \right)^2$, to most accurately represent a square grid 
of pixels (or spaxels) that would be used to isolate the planet. Note that the ratio 
$F_{\odot,\lambda}/F_{\odot,V}$ in Equation~\ref{eqn:cz} forces the color of the 
zodiacal light spectrum to match that of the Sun, which is a reasonable approximation, 
especially at large elongation angles from the Sun \citep{leinertetal1981}.  Inserting 
$\Omega$, and assuming constant spectral resolution, we have
\begin{equation}
  c_{\rm{z}} = \pi q \mathcal{T} X^{2} \frac{\lambda^{4}}{4hc\mathcal{R}}  
                     \frac{F_{\odot,\lambda}(1~\rm{AU})}{F_{\odot,V}(1~\rm{AU})} F_{0,V}  10^{-M_{\rm{z},V}/2.5}  \ ,
\end{equation}
which is a strong function of wavelength.

Following \citet{starketal2014}, we define a ``zodi'' as the surface brightness of 
an exozodiacal disk at 1~AU from a solar twin, which we label $M_{\rm{ez},V}$.  
Then the wavelength-dependent exozodiacal light count rate is given by,
\begin{equation}
  c_{\rm{ez}} = \pi q \mathcal{T} \Omega \Delta\!\lambda \frac{\lambda}{hc} 
                        \left( \frac{D}{2} \right)^{\!\!2}  \left( \frac{1~\rm{AU}}{r} \right)^{\!\!2} 
                        \frac{F_{\rm{s},\lambda}(1~\rm{AU})}{F_{\rm{s},V}(1~\rm{AU})}
                        \frac{F_{\rm{s},V}(1~\rm{AU})}{F_{\odot,V}(1~\rm{AU})} 
                        N_{\rm{ez}} F_{0,V} 10^{-M_{\rm{ez},V}/2.5}  \ ,
\end{equation}
where $N_{\rm{ez}}$ is the number of exozodis in the disk.  Note that 
the $F_{\rm{s},V}/F_{\odot,V}$ term ensures that exozodiacal light surface 
brightness scales with the intrinsic stellar brightness (at fixed orbital 
distance), and the $(1~\rm{AU}/r)^2$ term causes the disk surface 
brightness to decrease with increasing orbital separation according to 
the $1/r^2$ law.  Again, inserting $\Omega$, and assuming constant spectral 
resolution, we have
\begin{equation}
  c_{\rm{ez}} = \pi q \mathcal{T}  X^{2} \frac{\lambda^4}{4hc\mathcal{R}} 
                        \left( \frac{1~\rm{AU}}{r} \right)^{\!\!2} 
                        \frac{F_{\rm{s},\lambda}(1~\rm{AU})}{F_{\rm{s},V}(1~\rm{AU})}
                        \frac{F_{\rm{s},V}(1~\rm{AU})}{F_{\odot,V}(1~\rm{AU})} 
                        N_{\rm{ez}} F_{0,V} 10^{-M_{\rm{ez},V}/2.5}  \ .
\label{eqn:cez}
\end{equation}
For a discussion of the variety of different exozodiacal light treatments 
and exozodi definitions adopted throughout the literature, see 
\citet{robergeetal2012}.  Note that, while the zodiacal and exozodiacal 
light count rates are independent of telescope diameter, the ratio of the 
planetary signal to the zodiacal and exozodiacal light signals will depend 
on telescope size [$c_{\rm{p}}/\left(c_{\rm{z}}+c_{\rm{ez}}\right) \propto D^{2}$].

\subsection{Leakage, Dark Counts, and Read Noise}

We consider several other key sources of noise.  Stellar light leaked through 
the coronagraph (``leakage'') can contribute a large number of noise counts as 
the host star is many orders of magnitude brighter than the target planet.  The 
leakage count rate is determined from the stellar photon flux passed through the 
observing system, diminished by $C$ (the coronagraph design raw contrast), 
and is given by
\begin{equation}
  c_{\rm{lk}} =  \pi q \mathcal{T} C  \Delta\!\lambda F_{\rm{s},\lambda}(d) \frac{\lambda}{hc} \left( \frac{D}{2} \right)^{\!\!2} 
                     = q \mathcal{T} C  B_{\lambda}\!\left( T_{\rm{eff}} \right) \frac{\lambda^{2}}{hc\mathcal{R}} \left( \frac{\pi D R_{\rm{s}}}{2d} \right)^{\!\!2} \ ,
\label{eqn:clk}
\end{equation}
where the final step assumes constant spectral resolution.  Detections 
are possible for sources fainter than the raw contrast, but coronagraphs 
will have systematic noise floors that limit how far below the raw contrast 
observations can go.  The noise floor limit is due to residuals in the subtracted 
stellar PSF, and is typically a factor of ten below the raw contrast.  This limit is  
important to keep in mind when discussing small, low-albedo, and/or large 
separation planetary companions.

For a given dark current $D_{\rm{e^{-}}}$, in counts per pixel per unit time, the dark 
count rate is
\begin{equation}
  c_{\rm{D}} = D_{\rm{e^{-}}}N_{\rm{pix}} \ ,
\label{eqn:cD}
\end{equation}
where $N_{\rm{pix}}$ is the number of detector pixels contributing to the spectral 
element.  For imaging, the contributing pixels are those that fall within the aperture 
($\Omega$), so that, if a pixel has an angular diameter $\theta_{\rm{pix}}$, then 
\begin{equation}
  N_{\rm{pix,i}} = \frac{4\Omega}{\pi \theta_{\rm{pix}}^{2}} \ ,
\end{equation}
where the sub-script `i' has been introduced for `imaging'.  If the pixel diameter 
is determined by the diffraction limit at some wavelength $\lambda_{0}$, 
giving $\theta_{\rm{pix}} \simeq \lambda_{0}/2D$, then 
the number of contributing pixels is simply
\begin{equation}
  N_{\rm{pix,i}} = \frac{16}{\pi} \left( \frac{X\lambda}{\lambda_{0}} \right)^{2} \ .
\end{equation}

Determining the number of contributing pixels for a spectrometer is more 
complicated.  For an integral field spectrometer (IFS), angularly-resolved 
spectroscopic observations are obtained by placing a grid of small lenses 
(called ``lenslets'') in the focal plane, spectrally dispersing the light from each 
lenslet, and recording the resulting spectra at the detector.  For this setup, the 
number of contributing pixels is
\begin{equation}
  N_{\rm{pix,s}} = n_{\rm{pix}} \Delta \lambda \frac{4\Omega}{\pi \theta_{\rm{lens}}^{2}} \ ,
\end{equation}
where `s' is for `spectroscopy', $n_{\rm{pix}}$ is the number of pixels per unit 
wavelength designed for each lenslet spectrum, and 
$\theta_{\rm{lens}}$ is the angular diameter of an individual lenslet (so that 
$4\Omega/\pi \theta_{\rm{lens}}^{2}$ is the number of lenslets over which the 
aperture is distributed).  For a spectrum sampled at two pixels per spectral resolution 
element, and spread over $\Delta\!N_{\rm{hpix}}$ pixels in the horizontal/spatial 
dimension, we have
\begin{equation}
  n_{\rm{pix}} = \Delta\!N_{\rm{hpix}} \frac{2 \mathcal{R}}{\lambda} = \frac{2\Delta\!N_{\rm{hpix}}}{\Delta\!\lambda} \ ,
\end{equation}
which gives
\begin{equation}
  N_{\rm{pix,s}} = 8 \Delta\!N_{\rm{hpix}} \frac{\Omega}{\pi \theta_{\rm{lens}}^{2}} \ .
\end{equation}
Again, if the lenslets are sized to yield the diffraction limit at some reference 
wavelength, then we would have
\begin{equation}
  N_{\rm{pix,s}} = \Delta\!N_{\rm{hpix}} \frac{32}{\pi} \left( \frac{X\lambda}{\lambda_{0}} \right)^{2} \ .
\end{equation}
Note that the lenslet array typically only covers the inner portion of the overall 
imaging area, with coverage limited by detector space for recording individual lenslet 
spectra.  Thus, lenslet arrays and, more importantly, their associated detectors can 
set a tighter constraint on the OWA than the coronagraph.

Read noise is computed using the number of reads per observation, 
$N_{\rm{read}}$, the number of contributing pixels, and the read noise counts per 
pixel, $R_{\rm{e^{-}}}$, and is given by
\begin{equation}
  c_{\rm{R}} = \frac{N_{\rm{pix}} N_{\rm{read}} }{\Delta t_{\rm{exp}}} R_{\rm{e^{-}}} \ ,
\label{eqn:rn}
\end{equation}
where dividing by the exposure time, $\Delta t_{\rm{exp}}$, permits expressing 
read noise counts as a rate.  The number of reads is determined by assuming 
a maximum exposure time, $\Delta t_{\rm{max}}$ (often limited by cosmic ray 
strikes), with 
\begin{equation}
  N_{\rm{read}} = \frac{\Delta t_{\rm{exp}}}{\Delta t_{\rm{max}}}
\end{equation}
so that Equation~\ref{eqn:rn} can be written as
\begin{equation}
  c_{\rm{R}} = \frac{N_{\rm{pix}} }{\Delta t_{\rm{max}}} R_{\rm{e^{-}}} \ .
\label{eqn:rn_rate}
\end{equation}
Note that integer quantities (e.g., numbers of pixels) can be determined by 
rounding the expressions above, which are, for simplicity, expressed as 
non-integer quantities.


For completeness, internal thermal noise can be included and has a count 
rate given by
\begin{equation}
  c_{\rm{th}} = \pi q \Omega \frac{\lambda}{hc} \epsilon_{\rm{sys}} B_{\lambda}\!\left( T_{\rm{sys}} \right) \Delta\!\lambda \left( \frac{D}{2} \right)^{\!\!2} 
                    = \pi q  \epsilon_{\rm{sys}} B_{\lambda}\!\left( T_{\rm{sys}} \right) X^{2} \frac{\lambda^{4}}{4hc\mathcal{R}} \ ,
\end{equation}
where $\epsilon_{\rm{sys}}$ is the effective emissivity for the observing 
system (of order unity), and $T_{\rm{sys}}$ is the system temperature.  As 
we focus primarily on the visible wavelength range throughout this paper, 
internal thermal contributions are negligible, so we ignore them throughout.

\acknowledgements
TR gratefully acknowledges support from an appointment to the NASA 
Postdoctoral Program at NASA Ames Research Center, administered by 
Oak Ridge Affiliated Universities, and from NASA through the Sagan 
Fellowship Program executed by the NASA Exoplanet Science Institute. 
MM acknowledges support of the NASA Planetary Atmospheres and Origins 
programs.  KS thanks NASA support for the Exo-C mission study through 
the Exoplanet Exploration Program and the Goddard Space Flight Center.
All authors thank M.~Line, and C.~Stark for constructive feedback.  The 
results reported herein benefitted from collaborations and/or information 
exchange within NASA's Nexus for Exoplanet System Science (NExSS) 
research coordination network sponsored by NASA's Science Mission 
Directorate.  Results related to LUVOIR telescopes benefited from 
discussions with S.~Domagal-Goldman and G. Arney, as part of 
collaborative work done within the NASA Astrobiology Institute's Virtual 
Planetary Laboratory, supported by NASA under Cooperative Agreement 
No. NNA13AA93A.
%



\newpage

\section{Tables and Figures}
%
\begin{table}[ht]
  \centering
  \scriptsize
  {\bf Table 1. Symbol Usage} \\
  \vspace{2mm}
  \begin{tabular}{c l}
    \hline
    \hline
    Symbol &  Description  \\
    \hline
          $A$                 &   wavelength-dependent planetary geometric albedo \\
     $\alpha$               &   planet phase angle \\
  $B_{\lambda}$        &   Planck function \\
          $C$                 &  coronagraph design contrast \\
$C_{\rm{noise}}$       &  total number of noise counts on detector for a spectral element \\
    $C_{\rm{b}}$         &  total number of background counts on detector for a spectral element \\
  $C_{\rm{tot}}$         &  total number of counts on detector for a spectral element \\
    $c_{\rm{b}}$          &  total background count rate on detector for a spectral element \\
  $c_{\rm{tot}}$         &  total count rate on detector for a spectral element \\
     $c_{\rm{D}}$        &  dark current count rate on detector \\
     $c_{\rm{ez}}$        &  exozodiacal light count rate on detector \\
     $c_{\rm{p}}$         &  planetary count rate on detector \\
     $c_{\rm{R}}$         &  read noise count rate on detector \\
     $c_{\rm{lk}}$        &  leakage count rate on detector \\
      $c_{\rm{z}}$        &  zodiacal light count rate on detector \\
           $c$                 &  speed of light \\
           $D$                &   telescope diameter    \\
   $D_{\rm{e}^{-}}$     &  dark current \\
           $d$                 &   distance to observed star-planet system \\
$F_{\rm{p},\lambda}$ &  planetary specific flux \\
$F_{\rm{s},\lambda}$ &  stellar specific flux \\
 $F_{\odot,\lambda}$  & solar specific flux \\
     $F_{\odot,V}$        & V-band solar specific flux \\
     $F_{\rm{s},V}$        & V-band stellar specific flux \\
        $F_{0,V}$           & standard zero-magnitude V-band specific flux \\
    $f_{\rm{pa}}$          & fraction of planetary light that falls within photometric aperture \\
          $h$                   &  Planck constant \\
      $\lambda$            &   wavelength \\
 $\Delta\!\lambda$      &  spectral element width \\
      $M_{z,V}$            & V-band zodiacal light surface brightness \\
      $M_{ez,V}$           & V-band exozodiacal light surface brightness \\
   $N_{\rm{ez}}$         &  number of exozodis in exoplanetary disk \\
   $N_{\rm{read}}$     & number of detector reads per observation \\
   $N_{\rm{pix,i/s}}$   & number of contributing pixels for imaging/spectroscopy \\
$\Delta\!N_{\rm{hpix}}$ & number of horizontal/spatial pixels for dispersed spectrum \\
     $n_{\rm{pix}}$      & pixels per unit wavelength at detector for each lenslet spectrum \\
      $\Omega$            &  photometry aperture size, expressed as $4\left(X\lambda/D\right)^{2}$ \\
    $\Phi(\alpha)$        &   wavelength- and phase-dependent planetary phase function \\
           $q$                 &  detector quantum efficiency \\
     $\mathcal{R}$       &  instrument spectral resolution \\
      $R_{\rm{p}}$        &   planetary radius \\
      $R_{\rm{s}}$        &   stellar radius \\
    $R_{\rm{e}^{-}}$     &  read noise counts per pixel \\
             $r$                &   planet-star distance \\
      $\rm{SNR}$         &  signal-to-noise ratio \\
     $\mathcal{T}$        &   telescope and instrument throughput \\
$\theta_{\rm{IWA}}$   &   coronagraph inner working angle \\
$\theta_{\rm{OWA}}$  & coronagraph outer working angle \\
  $\theta_{\rm{pix}}$    & detector pixel angular diameter \\
 $\theta_{\rm{lens}}$   & lenslet angular diameter \\
   $T_{\rm{eff}}$          &    stellar effective temperature \\
 $\Delta t_{\rm{exp}}$ &  exposure time \\
 $\Delta t_{\rm{max}}$ &  detector maximum exposure time \\
            $X$                 &    width of photometric aperture, as multiple of $\lambda/D$ \\
    \hline
  \end{tabular}
\end{table}
\begin{table}[ht]
  \centering
  \small
  {\bf Table 2. Baseline Astrophysical Parameter Values} \\
  \vspace{2mm}
  \begin{tabular}{c l c}
    \hline
    \hline
    Parameter &  Description & Adopted Value  \\
    \hline
     $\alpha$                & planet phase angle &  90$^{\circ}$ \\
        $F_{0,V}$           & standard zero-magnitude V-band specific flux & $3.63\times10^{-8}$~W~m$^{-2}$~$\mu$m$^{-1}$ \\
      $M_{z,V}$            & V-band zodiacal light surface brightness & 23 mag~arcsec$^{-2}$ \\
      $M_{ez,V}$           & V-band exozodiacal light surface brightness & 22 mag~arcsec$^{-2}$ \\
      $N_{\rm{ez}}$      & number of exozodis in exoplanetary disk & 1 \\
    \hline
  \end{tabular}
\end{table}
\begin{table}[ht]
  \centering
  \small
  {\bf Table 3. Baseline Telescope and Instrument Parameter Values} \\
  \vspace{2mm}
  \begin{tabular}{c l c}
    \hline
    \hline
    Parameter & Description & Adopted Value  \\
    \hline
                              $C$                               &   coronagraph design contrast & $10^{-9}$   \\
                     $D_{\rm{e}^{-}}$                    &   dark current & $5 \times 10^{-4}$~sec$^{-1}$  \\
                              $D$                              &    telescope diameter & 2~m  \\
                          $f_{\rm{pa}}$                     &    fraction of planetary light that falls within photometric aperture & 0.87    \\
                 $\Delta\!N_{\rm{hpix}}$               &   number of horizontal/spatial pixels for dispersed spectrum &   3     \\
                               $q$                               &  detector quantum efficiency & Equation~\ref{eqn:quanteff}  \\
                        $\mathcal{R}$                      &  instrument spectral resolution &     70       \\
                       $R_{\rm{e}^{-}}$                    &  read noise counts per pixel & 0.1   \\
                       $\mathcal{T}$                        &  telescope and instrument throughput   &   0.05    \\
       $\theta_{\rm{IWA}}$ ($\lambda/D$)      &   coronagraph inner working angle &  2      \\
       $\theta_{\rm{OWA}}$ ($\lambda/D$)      &  coronagraph outer working angle & 10    \\
                   $\Delta t_{\rm{max}}$                &  detector maximum exposure time  &   1~hr   \\
                                 $X$                             &   width of photometric aperture, as multiple of $\lambda/D$ & 1.5    \\
    \hline
  \end{tabular}
\end{table}
\newpage
\clearpage
\clearpage
\clearpage
\clearpage
\clearpage
\clearpage
\clearpage
\clearpage
\clearpage
\clearpage
\clearpage
\clearpage
\clearpage
\clearpage
\clearpage
\clearpage
%
\end{document}